\documentclass[]{aa}
\usepackage{graphicx}
\usepackage{natbib}
\usepackage{rotating}
\usepackage{epsfig}
\usepackage[toc,page]{appendix}
\usepackage{multirow}
\usepackage{setspace} 
\usepackage{placeins}
\usepackage{multicol}
\usepackage{float}
\usepackage{epsf}
\usepackage{tabularx}
\usepackage{multirow}
\usepackage{balance}
\usepackage{amsmath}
\usepackage{latexsym}
\usepackage[varg]{txfonts}
\usepackage{array}
\usepackage{epsfig}
\usepackage{lscape}
\usepackage{footnote}
\usepackage{graphics}
\usepackage{etoolbox}
\usepackage{leftidx}
\usepackage{footnote}
\usepackage{color}
\usepackage{cleveref}
\usepackage{caption}
\usepackage{subcaption}
\usepackage{natbib}
\usepackage{romannum}
\usepackage{float}

\begin{document}

\title
    { Near infrared spectroscopic observations of high redshift C~{\sc i} absorbers\thanks{The data used in this
paper were collected at the European Southern Observatory under Programmes 084.A-0699, 086.A-0074, 086.A-0643, and 087.A-0548, using X-shooter mounted at the UT2 Cassegrain focus of the Very Large Telescope (VLT). }} 

\author
    {S.~Zou$^1$
   \and P.~Petitjean$^1$ 
   \and P.~Noterdaeme$^1$
   \and C.~Ledoux$^2$ 
   \and J.-K.~Krogager$^{1,6}$
   \and H. Fathivavsari$^3$
   \and R.~Srianand$^4$
   \and S. L\'opez$^5$}
\institute
{Sorbonne Universit\'e, CNRS, UMR 7095, Institut d’Astrophysique de Paris, 98 bis bd Arago, 75014 Paris, France
      \and  European Southern Observatory, Alonso de C\'ordova 3107, Casilla 19001, Vitacura, Santiago, Chile
      \and School of Astronomy, Institute for Research in Fundamental Sciences (IPM), PO Box 19395-5531 Tehran, Iran
      \and  Inter-University Center for Astronomy and Astrophysics,  Post Bag 4, Ganeshkhind, 411 007 Pune, India
     \and Departamento de Astronom\'ia, Universidad de Chile, Casilla 36-D, Santiago, Chile
     \and Dark Cosmology Centre, Niels Bohr Institute, Copenhagen University, Juliane Maries Vej 30, DK-2100 Copenhagen {\O}, Denmark 
 }


\abstract{
We study a sample of 17 $z>1.5$ absorbers selected based on the presence of strong 
C~{\sc i} absorption lines in Sloan Digital Sky Survey (SDSS) spectra and observed with the European Southern Observatory Very Large Telescope (ESO-VLT) spectrograph X-shooter.
We derive metallicities, depletion onto dust, and extinction by dust, and analyse 
the absorption from Mg~{\sc ii}, Mg~{\sc i}, Ca~{\sc ii,} and Na~{\sc i} that are
redshifted into the near infrared wavelength range.
We show that most of these C~{\sc i} absorbers have high metallicity and dust content. 
We detect nine Ca~{\sc ii} absorptions with $W$(Ca~{\sc ii}$\lambda$3934) ~$>$~0.23~\AA~ out of 14 systems
where we have appropriate wavelength coverage. The observed equivalent widths are similar to 
what has been measured in other lower redshift surveys of Ca~{\sc ii} systems.
We detect ten Na~{\sc i} absorptions in the 11 systems where we could observe this absorption.
The median equivalent width ($W$(Na~{\sc i}$\lambda$5891)~=~0.68~\AA) is larger than what
is observed in local clouds with similar H~{\sc i} column densities but also 
in $z<0.7$ Ca~{\sc ii} systems detected in the SDSS. The systematic presence of Na~{\sc i}
absorption in these C~{\sc i} systems strongly suggests that the gas is neutral and cold,
maybe part of the diffuse molecular gas in the interstellar medium (ISM) of high-redshift galaxies.
\par\noindent
Most of the systems (12 out of 17) have $W$(Mg~{\sc ii}$\lambda$2796)~$>$~2.5~\AA~ and six of them
have log~$N$(H~{\sc i})~$<$~20.3, with the extreme case of J1341+1852 that has 
log~$N$(H~{\sc i})~=~18.18. The Mg~{\sc ii} absorptions are spread over more than $\Delta v$ $\sim$ 400~km~s$^{-1}$ 
for half of the systems; three absorbers have $\Delta v$~$>$~500~km~s$^{-1}$.
The kinematics are strongly perturbed for most of these systems, which probably do not arise in quiet 
disks and must be close to regions with intense star-formation activity
and/or are part of interacting objects. All this suggests that a large fraction of the cold gas 
at high redshift arises in disturbed environments.
}
\maketitle


\section{Introduction}
Damped Lyman-$\alpha$ systems (DLA) observed in the spectra of bright
background sources are produced by 
neutral gas \citep{wolf00a} located
in the halo and/or disk of galaxies. The gas in DLAs at high redshift consists 
mostly of a warm neutral phase \citep[e.g.][]{pet00a}, with average temperatures of
the order of several thousand kelvin. Most high-z DLAs are found to have low metallicities  
($\sim$0.1 solar, \citealt{sri12,raf14}) and the cosmological density of the gas in DLAs
is much less than the cosmological density of stars \citep{not12b}.
All this is in line with the idea that most DLAs are part of a transition phase 
intermediate between the intergalactic medium, which is the reservoir
of gas for galaxy formation, and the dense and cold gas that is an ingredient for star formation in the disks of galaxies.
\par\noindent

Understanding the  mechanisms  of  star  formation  at  high  redshifts is central to  
our  knowledge  of  how  galaxies  formed and subsequently evolved chemically. Stars form in molecular 
clouds (e.g. Snow \& McCall 2006)
that are located in the interstellar medium (ISM) of galaxies
whose properties are regulated in turn by radiative and mechanical feedback from stars. 
Deriving the physical properties of the gas in the ISM, in particular in the diffuse
molecular phase of galaxies, is crucial for our understanding of how stars formed in the
early Universe.
It is not easy to detect the cold neutral gas in absorption and attempts to do so have 
been made for many years. In particular, attention has been brought
to molecular hydrogen as it has numerous detectable absorption lines in the
Ultraviolet (UV) \citep{led03,cui05,not08}. 
It has been shown that H$_2$ is detected with overall molecular fractions $>$ 0.1
\% in about 10\% of the DLAs \citep{not08} or possibly less \citep{jor14}.
When detected, the associated gas is usually 
cold with $T_{\rm e}$$\sim$150~K and dense with $n_{\rm H}$$\sim$100~cm$^{-3}$ 
\citep{sri05,bala11}.
The latter systems directly trace the diffuse ISM of high-redshift 
galaxies and are places where it is possible to study the relation between the physical properties of the 
ISM and star-formation activity \citep{not12a}. 
These are also places where one can probe the H~{\sc i}-to-H$_2$ transition in the diffuse 
interstellar medium of high-redshift galaxies \citep{not15,ma15,bala17}.
\par\noindent
It is not that easy to preselect systems where molecular hydrogen can be found.
One way is to concentrate on the presence of neutral carbon.
Indeed, carbon can be ionized by photons of energy smaller than the H~{\sc i} ionization 
limit and its detection should indicate the presence of neutral, cold, and well-shielded gas, thus the presence of H$_2$.
neutral-carbon (C~{\sc i}) at high redshift has not been searched for systematically 
so far (see however \citealt{ge97,ge99,sri05}).
We have thus embarked in a survey of systems selected only on the basis of
the presence of strong C~{\sc i} absorption in quasar spectra from the Sloan Digital Sky Survey (SDSS). 
The selection of systems, the main characteristics of the sample, the detection rate, 
and the H~{\sc i} content have been described in \citet{led15} and the detailed 
analysis of the metallicities will be discussed in Ledoux et al. (in prep).
Out of 66 strong C~{\sc i} systems detected in the SDSS DR7, 17 have been observed with the ESO spectrograph X-shooter, which 
offers the unique opportunity to study the associated metal lines redshifted in the infrared region of the
spectrum and in particular the Ca~{\sc ii}, Na~{\sc i,}
and Ti~{\sc ii} lines that are frequently used in the local universe to study the interstellar medium.
The large spectral range of X-shooter is
also ideal for deriving the dust attenuation of the quasar spectrum induced by the
presence of a DLA along the line of sight. One of these systems has been 
singled out as a "ghostly" DLA and is described in details in \cite{fat17}.
\par\noindent
The presence of Ca~{\sc ii} absorption 
is of particular interest for studying the properties of the interstellar medium of high-$z$ galaxies.
It is important to bear in mind that Ca is usually highly depleted 
onto dust and that the ionization energy of Ca~{\sc ii} being 11.87~eV, 
Ca~{\sc ii} may not be the dominant ionization stage of Ca even in H~{\sc i} dominated gas.
Even though the presence of strong Ca~{\sc ii} absorption is therefore not a 
characteristic of cold gas, its presence or absence can yield interesting information on the 
amount of dust and on the radiation density below the Lyman limit. 
In turn, Na~{\sc i} is  ionized above 5.14~eV and is therefore associated with cold and neutral gas.
In the local universe a tight correlation is seen between $N$(Na~{\sc i}) and $N$(H~{\sc i})
\citep{fer85,wak00}. It will be interesting to verify if such correlation
holds at high redshift as well. In addition the ratio of the Na~{\sc i} and Ca~{\sc ii} column densities
is a useful indicator of the physical state of the gas \citep[]{rou52,wel96}.
\par\noindent
In our Galaxy, when observed at high spectral resolution, the
Na~{\sc i} complexes break into sub-components with a median Doppler parameter of $\sim$0.73~km~s$^{-1}$
and a median separation between adjacent components of $\sim$2~km~s$^{-1}$. The typical temperature
of the gas is 80~K \citep{wel94}. 
The velocity distribution of the corresponding Ca~{\sc ii} absorption is broader, due to outlying components 
at higher temperature ($T>6000$~K) seen only 
in Ca~{\sc ii}. Even individual Ca~{\sc ii} components are broader than the corresponding 
Na~{\sc i} or K~{\sc i} components implying that the Ca~{\sc ii} absorption 
arises predominantly in
warmer and more diffuse gas occupying a larger volume \citep{hob74,hob75,wel96,wel01}.
Absorption of Ca~{\sc ii}  can arise both from cold, relatively dense gas,
where Ca is typically heavily depleted onto grains and Ca~{\sc ii} is its dominant
ionization state, and also in warmer, lower density gas, where Ca is less depleted
but Ca~{\sc ii} is a trace ionization state.
\par\noindent
It is found that Ca~{\sc ii} (Na~{\sc i}) absorbers at intermediate and high radial velocities 
are present in 40–55\% (20–35\%) of the sightlines through the halo of our Galaxy
\citep{bek08,benb12}. The Ca~{\sc ii}/Na~{\sc i} ratio is
found to be smaller in halo gas compared to what is observed in the disk of the Galaxy
\citep{kee83,fer85,vla93,sem93,benb12}.
\par\noindent
Previous studies of Ca~{\sc ii} and Na~{\sc i} absorption outside the Galaxy were carried out 
at low redshift so that the absorption lines are redshifted in the optical
window (\citep[see][]{bla88} \cite{bow91} 
The impact parameters of the Ca {\sc ii}--associated galaxies have also been studied previously \citep{kun84,wom90,pet00b,hall02,wang05,che11,zyc07,rah16}.
%
\par\noindent
\cite{wild05} were the first to search SDSS spectra systematically for Ca~{\sc ii} absorbers
and to show that they induce a reddening of the quasar spectrum with an average $E$(B-V)~=~0.06
\citep{wild06} and that they are more evolved than the overall population of DLAs. 
\citet{sar14,sar15} used SDSS-Data Release 9 data and showed that the equivalent width distribution
reveals two populations of Ca~{\sc ii} systems with 
$W_{\rm r}$(Mg~{\sc ii}$\lambda$2796)/$W_{\rm r}$(Ca~{\sc ii}$\lambda$3934)~less and greater than 1.8, 
respectively. These authors show as well that the systems with 
$W_{\rm r}$(Ca~{\sc ii}$\lambda$3934) smaller and 
larger than $\sim$0.7~\AA~ have properties, respectively, consistent with those of halo 
gas and intermediate between halo-type and disk-type gas \citep[see][]{zyc09}.
\cite{gub16} study the dust depletion of Ca~{\sc ii} and Ti~{\sc ii}
in 34 systems at $z<0.4$ and conclude that these lines trace predominantly neutral gas
in the disks and inner halo regions of galaxies 
 (see also \citealt{cox07}). 
Finally, Na~{\sc i} and Ca~{\sc ii}
have been detected in a DLA at $z\sim 1$ towards QSO~APM~08279+5255 
\citep{pet00b}. In this system, it is clear that Na~{\sc i} absorption is
confined to narrow components whereas Ca~{\sc ii} has a shallow and broad profile
very much consist with what have been found in the Galaxy and its halo.
\par\noindent
The paper is organised as follows.
We describe the sample and the observations in Section~2. Section~3 gives
the Na~{\sc i}, Ca~{\sc ii,} and $E(B-V)$ measurements. We discuss the results 
in Section~4 before concluding in Section~5.

\section{Sample and observations}
We  systematically  searched the SDSS-DR7 \citep{aba09} 
quasar spectra \citep{sch10} for C~{\sc i} absorption systems
and found 66 of them at $z>1.5$ with $W_{\rm r}$(C {\sc i}$ \lambda$1560)~$>$~0.12~\AA. The sample is complete 
for $W_{\rm r}$(C~{\sc i}$\lambda$1560)~$>$~0.40~\AA~ \citep{led15}.
Follow-up observations have been performed with the ESO Ultraviolet and Visual Echelle Spectrograph (UVES) spectrograph for 27 systems and 
the ESO/X-shooter spectrograph 
for 17 systems (see Ledoux et al. in prep). In this paper we concentrate on the latter systems.  
\par\noindent 
The instrument X-shooter \citep{ver11} covers the full wavelength range from 300~nm 
to 2.5~$\mu$m at intermediate spectral resolution using three spectroscopic arms (UV-Blue (UVB),
Visible (VIS) and Near-IR (NIR)). We observed the quasars in slit mode for slightly more than one hour each. 
To optimise the sky subtraction in the NIR, telescope nodding was performed following an ABBA 
scheme with a nod throw of 5 arcsec and a jitter box of 1 arcsec.  
The two-dimensional (2D) and one-dimensional (1D) spectra were extracted using 
the X-shooter pipeline in its version 2.5.2 \citep{mod10}.  
\par\noindent
Generally the final spectra are close to the nominal resolving power of R~= 4350, 
7450, and 5300 in the UVB, VIS, and NIR arms, for slit widths 
of 1.0, 0.9, and 0.9 arcsec, respectively. For some objects however, the seeing was better than
the widths of the slits, and resolutions were higher.  
We estimated the actual resolution following the method described in \cite{fyn11}.
Flux calibration has been performed using observations of standard stars provided
by the ESO. We note that observations have been performed when the Atmospheric Dispersion Corrector (ADC) was
still in use. 
\par\noindent
There is no clear distinction between the X-shooter sub-sample and 
the overall sample of C~{\sc i} absorbers. A comparison between sub-samples will be 
performed in Ledoux et al. (in prep).
Names of the objects and characteristics of the C~{\sc i} absorption systems
 derived from X-shooter data are given in Tables~1 and 2. 
Spectra and absorption profiles are shown in the Appendices.
\par\noindent

\section{Measurements}
\subsection{Equivalent widths}
\par\noindent
For isolated transitions we measure the equivalent width directly from the spectrum,
integrating the observed normalized flux over the absorption profile. In the case of C~{\sc i}, the C~{\sc i}, C~{\sc i}*, and C~{\sc i}** transitions are blended together
so that we used the program VPFIT \citep{car14} to disentangle the absorptions. Results of the fits are shown
in Appendix~\ref{comments}. Equivalent widths of the overall C~{\sc i} 
$\lambda\lambda$1560,1656 absorption with errors measured at the 1$\sigma$ level are given in 
columns 2 and 3 of Table \ref{naca}. These values are derived by integrating the 
observed normalized flux
in the spectrum, except
 for QSO J0917+0154; in this case, C~{\sc i}$\lambda$1560 is strongly blended 
with other metal lines, thus we used the value obtained from the fit.  
We follow \citet{vol06} to estimate the errors.
\par\noindent
When Na~{\sc i} and/or Ca~{\sc ii} absorption lines are detected, we fit the 
absorption feature with a Gaussian function and derive the 
equivalent widths from the fit. The results are shown in Appendix~\ref{comments}. 
When no line is detected, we derive an upper limit on the equivalent width as 
$EW_{\rm lim}$~=~3.2$\times FWHM$/$SNR$ in 3$\sigma$ where $SNR$ is the signal-to-noise ratio at the 
expected position of the line, and full width at half maximum ($FWHM$) corresponds to twice the width of an unresolved 
spectral feature. The absorption redshift is determined 
by the position of the strongest C~{\sc i} component and is used as the zero of the velocity
scale for the figures in Appendix~\ref{comments}.  

\subsection{Metallicities and extinction}\label{ebv}

\begin{figure}
  \resizebox{\hsize}{!}{\includegraphics[clip=true, width=15cm]{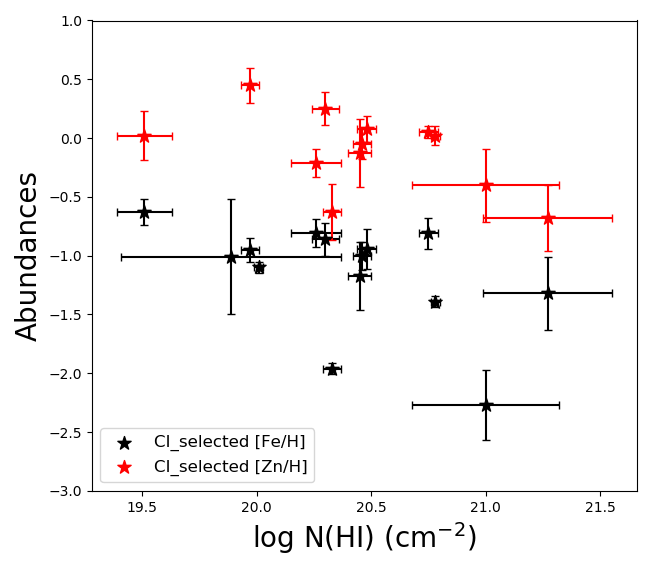}}
\caption{\small{\label{zcompare}Metal abundances versus $N$(H~{\sc i}) column density
 in our sample, red stars are for [Zn/H] and black stars are for [Fe/H]. We use the gas-phase 
metal abundances and H~{\sc i} column densities from Table \ref{tableZ}. 
}}
\end{figure}

\begin{figure}
\centering
\resizebox{\hsize}{!}{\includegraphics[clip=true, width= 50cm]{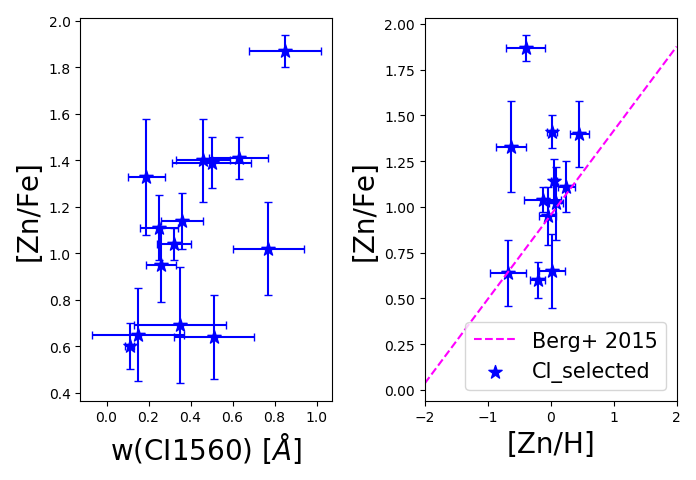}}
\caption[]{\small{\label{dust_CI} Left panel: Depletion factor [Zn/Fe] versus W(C~{\sc i}$\lambda$1560).
Right panel: Depletion factor [Zn/Fe] versus metallicity, [Zn/H].
The purple dashed line in the right panel is the mean relation observed for typical DLAs \citep{ber15}
}}
\end{figure}

The spectral resolution of X-shooter is not high enough to obtain trustful Doppler 
parameters and column densities except when either the absorption is optically thin 
(e.g. Zn~{\sc ii}) or several transitions of the same species
are observed. This is the case for Fe~{\sc ii} and Si~{\sc ii} for which we derive column 
densities from Voigt profile fitting of all observed transitions together (see Table~\ref{tableZ}).
 We used VPFIT \citep{car14} to model the absorption profiles. The redshift and
Doppler parameter of the components were imposed to be the same in all profiles. 
When  estimating  best-fit  parameters, VPFIT takes  as  input  the
normalised  spectrum  and  the  resolution  provided  by  the  user.
This means that continuum placement uncertainties are not reflected in the error estimates.
We therefore estimated the errors in column densities by varying slightly the continuum
for lines that are not fully saturated. When one of the lines was optically thin, we derived
the error in the column density from the error in the equivalent width (see Section 3.1). 
For other species we measured only the equivalent width without trying to derive a column density.
\par\noindent
Gas-phase metal abundances derived from Fe~{\sc ii}, Si~{\sc ii,} and Zn~{\sc ii} absorptions are listed 
in Table \ref{tableZ}. We added the molecular contribution to the total hydrogen column density
whenever H$_2$ was detected \citep[see][]{not18}. This contribution is non-negligible
in the cases of J1237+0647 and J0917+0154.
We plot in Fig.~\ref{zcompare} the metal abundances versus H~{\sc i} column density
(red and black stars for, respectively, Zn and Fe). It is apparent that the
C~{\sc i} systems have large metallicities (around solar) and in any case larger
than what is measured in typical DLAs at similar redshifts (e.g. \citealt{raf14}). 
Depletion of iron onto dust measured as [Zn/Fe] is significant. 
What is striking also is that not only are H~{\sc i} column densities not very large 
(several systems do not qualify as standard DLAs as they show log~$N$(H~{\sc i})~$<$~20.3), 
but also there is a tendency for metallicity to decrease as column density increases.
This may well be a consequence of the usual dust bias, which implies that
systems with both high H~{\sc i} column densities and metallicities drop out of the sample, 
because the quasar flux is attenuated below the flux limit of the quasar survey \citep{boi95}.
It could be as well that the quasars colours are affected by the presence of dust, shifting 
the quasar out of colour selection of the quasar survey.
%
%

\par\noindent
The dust depletion indicator [Zn/Fe] is plotted versus $W$(C~{\sc i}$\lambda$1560) 
and metallicity (as [Zn/H]) in the two panels of Fig.~\ref{dust_CI}. 
It is apparent in the left panel that dust depletion is significant. There is, however, no strong correlation between the amount of dust depletion and the amount of neutral carbon.
The relation seems to hold even for the high metallicities measured in our sample although
a large scatter is observed in the measured depletion factors.
Although these considerations are useful for the following discussion, 
it is clear that these properties should be studied in the complete sample of C~{\sc i} systems, 
not only in the sub-sample of quasars observed with X-shooter. This will be done in a companion 
paper (Ledoux et al. in prep).
\par\noindent
%
%
Given the metallicities and dust depletions measured in our sample, the quasar
spectra are significantly attenuated by the presence of dust in the absorption systems.
We estimate this attenuation, measured as $A_{\rm V}$ and $E(B-V)$, by fitting a quasar spectral 
template to the data assuming a set of fixed extinction 
curves (Small Magellanic Cloud(SMC), Large Magellanic Cloud(LMC), and LMC2) parametrized by \citet{gor03}. The dust reddening is assumed to be caused solely by the foreground C~{\sc i}
absorber. We use the quasar template derived by \citet{sel16} and fit only the data in bona fide continuum regions, regions 
which are not strongly affected by absorption or broad emission lines. This template has been obtained
by combining seven spectra of bright quasars taken with X-shooter. Signal-to-noise ratio varies from 50 in the UV
to 200 in the optical and 100 in the near infrared.
Before fitting, the template is smoothed with a Gaussian kernel ($\sigma$ = 7~pixels) 
to prevent the noise in the template from falsely fiting noise peaks in the real data. In order to take into account the 
uncertainty in the template, we subsequently convolve the errors on the observed data with the uncertainty estimate 
for the template.
In the near infrared, we perform a 5-$\sigma$ clipping in 
order to discard outlying pixels introduced by the removal of skylines during data reduction.
We furthermore allow for variations in the iron pseudo-continuum in the rest-frame UV by including 
the template derived by \citet{ves01}. We separate the contributions from \ion{Fe}{ii} and \ion{Fe}{iii} into 
different templates and allow each to vary independently.

\par\noindent During the observations of the target J1302+2111, 
the atmospheric dispersion correction (ADC) malfunctioned leading to strong chromatic slit losses for the UVB and VIS 
arms (the NIR arm does not have an ADC unit and hence is not affected). Similar artefacts from malfunctioning ADCs of 
X-shooter have been reported by \citet{lop16}.
By comparing the X-shooter spectrum to the available SDSS spectrum, we conclude that the UVB arm is not strongly 
affected by chromatic slit loss, however, the VIS arm is heavily affected. We therefore do not use the VIS arm for 
the extinction analysis.

\par\noindent We then fit the template to the data using four free parameters: 
the attenuation $A_{\rm V}$, 
the two scaling parameters for the \ion{Fe}{ii} and \ion{Fe}{iii} contributions, and an arbitrary flux scale 
that corresponds to the flux in the IR as the attenuation is usually negligible at these wavelengths.
We fit the template to the data for each of the three extinction 
curves considered, SMC, LMC, and LMC2, and we assign the best fit as the solution with the lowest $\chi^2$. 
We note that for the target J2340--0052, the 
lowest $\chi^2$ is obtained with an LMC2 extinction law, however, upon visual inspection it is clear that the fit is not good 
as the best-fit iron contribution over-estimates the actual fluctuations in the data. Thus, when removing the iron contribution 
the spectrum is fitted better with the SMC extinction curve. The best-fit reddening and the associated best-fit extinction curve 
are given in Table~\ref{tableebv}. In cases where the reddening is too small to distinguish between various extinction curves, we list 
the best-fit extinction curve as `N/A' in Table~\ref{tableebv} and give the $R_{\rm V}$ value for the SMC curve for simplicity. We note that 
for such small values of $A_{\rm V}$, the value of $R_{\rm V}$ does not change the resulting $E(B-V)$ significantly. The main uncertainty 
for the dust fitting comes from intrinsic variations to the template. We have modelled this by assuming a distribution of the 
relative intrinsic spectral slope ($\Delta\beta$) as modelled by a Gaussian function with a width of 0.2~dex \citep{sel16}. This yields 
an estimated 0.07~mag uncertainty on the best-fit $A_{\rm V}$ , which by far dominates the total uncertainty as the statistical uncertainty 
from the fit is of the order of 0.01~mag.

\par\noindent For the cases where an SMC extinction curve is preferred by the fit, the resulting $A_{\rm V}$ should be 
regarded as an upper limit, since for this extinction curve it is very difficult to disentangle dust in the quasar 
and dust in the absorber \citep[see discussion in][]{kro15}. In Figure~\ref{J0216ebv}, we show the best-fit model 
for one target (the other spectra are shown in Appendix~\ref{appendixebv}).

\begin{figure}
\centering
\resizebox{\hsize}{!}{\includegraphics[clip=true, width=13cm]{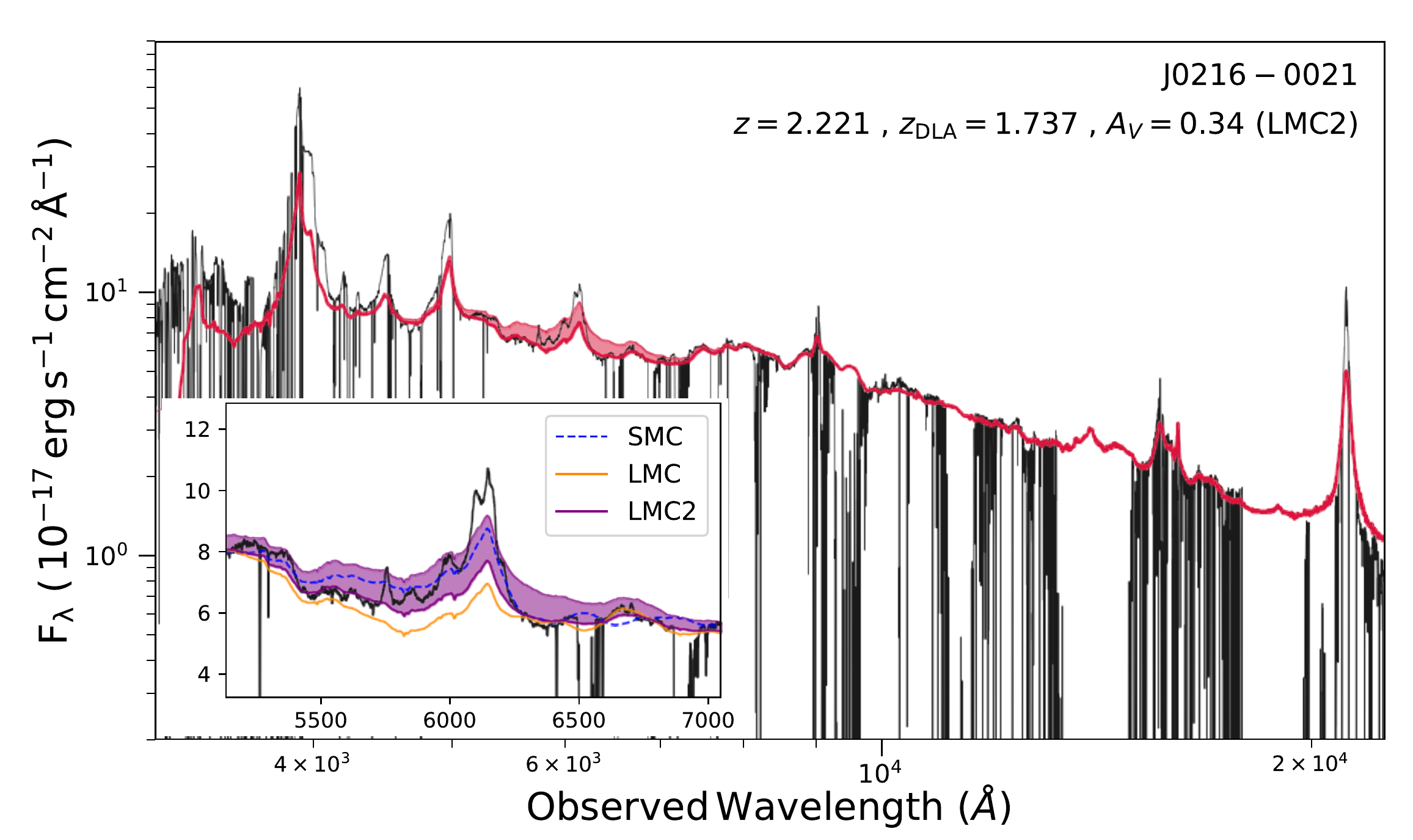}}
\caption[]{\small{\label{J0216ebv} X-shooter spectrum of J0216-0021. The red curve
corresponds to the quasar template of \citet{sel16} reddened 
by different extinction curves (see insert).
}}
\end{figure}

\section{Results}

\subsection{MgII} \label{mg2}

\begin{figure}
\centering
\resizebox{\hsize}{!}{\includegraphics[clip=true, width=12cm]{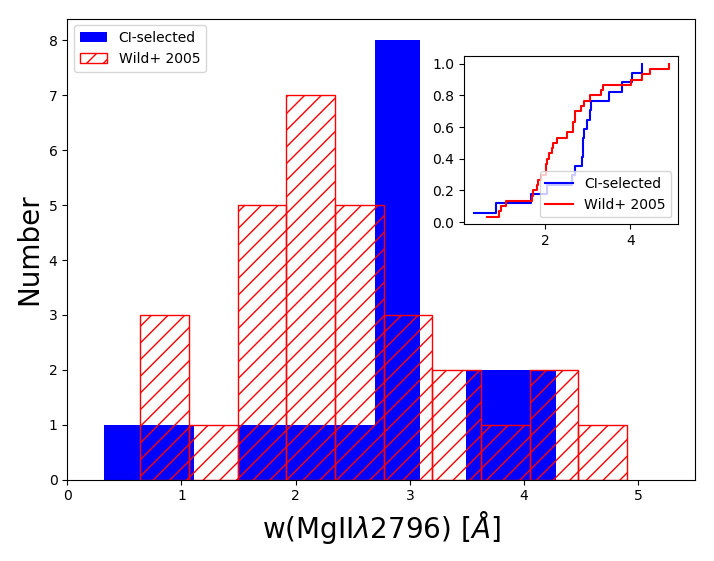}}
\caption[]{\small{\label{mg_hist}$W$(Mg~{\sc ii}$\lambda$2796) distributions. 
The blue histogram is for our sample of C~{\sc i} absorbers. The red hashed histogram is for the Ca~{\sc ii}
sample of \cite{wild05} at $0.84 < z < 1.3$. The lines in the insert are the cumulative distributions of the two samples.
}}
\end{figure}

The $W$(Mg~{\sc ii}$\lambda$2796) observed in our C~{\sc i} systems are large;
13 out of 17 (82\%) of these systems have $W$(Mg~{\sc ii}$\lambda$2796)~$>$~2.5~\AA ,~
while such strong systems are rare in Mg~{\sc ii} surveys
even at high redshifts (see Rao et al. 2005).
As can be seen in Fig.~\ref{mg_hist}, this is systematically larger than what is observed in systems selected on 
the basis of the presence of Ca~{\sc ii} in SDSS spectra (\citealt{wild05}; red histogram in Fig.~\ref{mg_hist}). 
The sample of \citet{wild05} includes 31 Ca~{\sc ii} absorbing systems in the redshift range 0.84 $<$ $z$ $<$1.3. 
This number of strong systems can be compared with what is observed in high-redshift DLAs.  \citet{ber16}
study the 36 blindly selected DLAs with 2$<$$z$$<$4 detected in the XQ-100 legacy survey \citep{lop16}. Only three of their DLAs have $W$(MgII$\lambda$2696)~$>$~2.5~\AA~
and much lower metallicities. 
\par
\begin{figure}
\centering
\resizebox{\hsize}{!}{\includegraphics[clip=true, width=15cm]{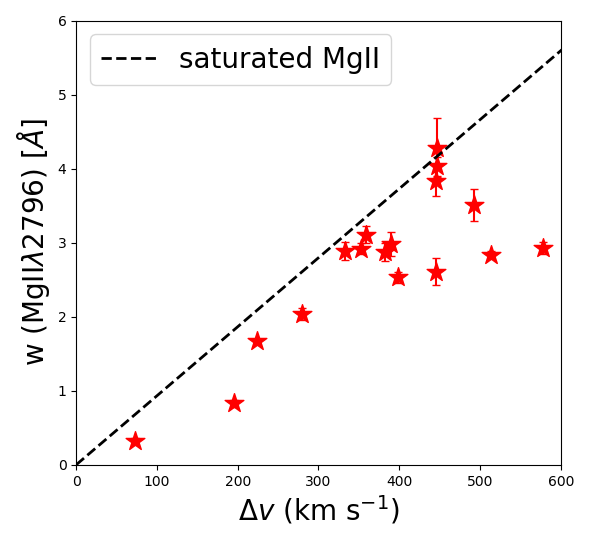}}
\caption[]{\small{}
$W$(Mg~{\sc ii}$\lambda$2796) versus the width of the absorption profile (see text for definition).
The black dashed line is the relation expected when the Mg~{\sc ii} line is totally saturated. 
\label{mgv90}}
\end{figure}

\noindent
We have measured the velocity spread of the Mg~{\sc ii} absorption, $\Delta v$.
Since, the SNR in the infrared is not optimal, we measure $\Delta v$ as the velocity
separation between the two extreme pixels where $\tau<0.1$. This is similar to the standard 
$\Delta v_{90}$ definition and more robust for our data. The idea is to include all
satellite absorption and to have a good representation of the kinematical extent of the absorption.
In Fig.~\ref{mgv90}, it can be seen that 
the $W$(Mg~{\sc ii}$\lambda$2796) equivalent width (EW) is strongly correlated with $\Delta$ $v$. 
The correlation we see is quite similar to what is seen for 
typical DLAs \citep{ell06}. The difference resides again in the presence
in our sample of a high fraction of large ($>$300~km~s$^{-1}$) $\Delta v$ values.
The median value of $\Delta v$ in our sample is $\sim$400~km~s$^{-1}$ ;
three absorbers have $\Delta v$$>$500~km~s$^{-1}$. Few such extreme systems are known in the
literature at high redshift \citep{led06} and
are found to be associated with molecular hydrogen \citep{pet02,led02}. Only three DLAs out of 36 have  a
kinematical extention larger 
than 200~km~s$^{-1}$ in the sample of \citet{ber16}, when seven of our systems
have $\Delta v$$>$400~km~s$^{-1}$. 
It is therefore surprising that 
our C{\sc i}-selected systems show such disturbed kinematics.
The dashed line in Fig.~\ref{mgv90} is the relation one expects if the lines are 
completely saturated over the whole absorption profile. It is therefore clear
that above 300~km~s$^{-1}$, the velocity spread of the absorption is dominated by satellite
components.
\par\noindent
These large kinematical spreads could be due to strong winds or the consequences of interactions
between several galaxies. Interestingly, we found three systems 
in the spectra of J1047+2057, J1133-0057, and J2350-0052 where 
the Mg~{\sc ii} absorption shows two distinct saturated sub-systems separated by
more than 200~km~s$^{-1}$ in the C~{\sc i} and Mg~{\sc ii} absorption profiles. 
In J1133-0057, we detect also two distinct Na~{\sc i} components.
Therefore probably both processes can be invoked to explain the 
large velocity spreads of Mg~{\sc ii} absorptions in our sample.
\par\noindent

\subsection{Ca~{\sc ii}}\label{para_naca}

\par\noindent 
Due to the absorber redshifts, the Na~{\sc i} and Ca~{\sc ii} lines in our sample are redshifted to the 
near-infrared wavelength range. The data quality in the NIR arm is not good enough to derive 
robust column densities. Thus we decided to use the equivalent width only 
to discuss the observations of Na~{\sc i} and Ca~{\sc ii}. The results are listed in 
Table \ref{naca}. We detect Ca~{\sc ii} in nine systems out of 14 where we could obtain an 
equivalent width limit. The other spectra are spoiled by atmospheric features.
\par\noindent
\citet{wild05} and \citet{sar14} searched for Ca~{\sc ii} systems
in the SDSS data and therefore at $z_{\rm abs}<1.3$. 
\citet{nes08} searched for Ca~{\sc ii} in 16 known DLAs with 0.6~$<$~$z_{\rm abs}$~$<$~1.3
and detected Ca~{\sc ii} in 12 of them. They warn, however, that their sample is biased towards
strong Mg~{\sc ii} systems. The Mg~{\sc ii} mean equivalent width in their sample 
is 1.9~\AA~ compared to 1.35~\AA~ in an unbiased sample.  
The sample by \citet{rah16} consists of nine DLAs at $z\sim 0.6$. In five of the seven observed fields
they could detect associated galaxies for which they estimate a metallicity of 0.2 to 0.9 solar when
the gas has a metallicity in the range 0.05 to 0.6 solar.
%
\par
\begin{figure}
\centering
\resizebox{\hsize}{!}{\includegraphics[clip=true, width=12cm]{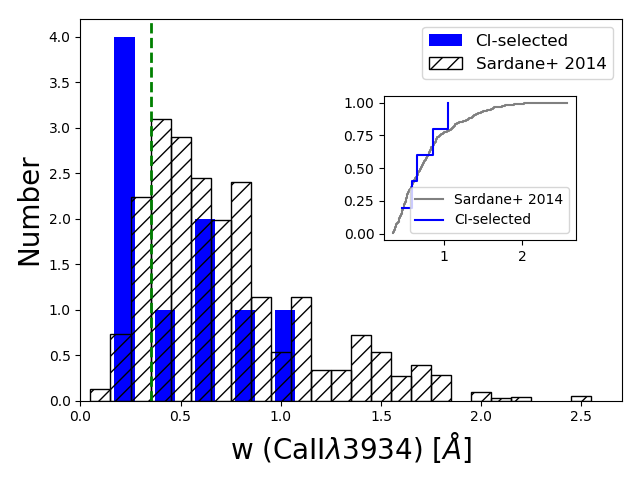}}
\caption[]{\small{\label{ca_hist}}Blue histogram is the rest equivalent width distribution of the
Ca~{\sc ii} K lines detected in our sample; 
grey-hashed histogram is the scaled equivalent width distribution of Ca~{\sc ii} K lines
detected in the SDSS by \citet{sar14}. The two cumulative distributions are shown in the insert; considering the detection limit of SDSS, we applied the Kolmogorov–Smirnov test when $W$(Ca~{\sc ii}$\lambda$3934) > 0.35 $\AA$.}
\end{figure}

\noindent
\cite{sar14} found 435 Ca~{\sc ii} doublets in the SDSS DR7 and DR9 databases, with $z<1.34$.
In Fig.~\ref{ca_hist} we plot their equivalent width distribution (scaled for convenience and shown in black histograms) 
together with ours (blue histogram). There is no obvious difference between the two distributions.
The discrepancy in the first bin is probably due to the higher detection limit of the SDSS study.  Assuming the SDSS detection limit is $\sim$ 0.35 \AA, we applied a Kolmogorov–Smirnov test above this limit. A $p$ value of 0.96 indicates
that the two samples are drawn from a similar population. 

\subsection{Na~{\sc i}}  
%
%
\par
\begin{figure}
\centering
\resizebox{\hsize}{!}{\includegraphics[clip=true, width= 60cm]{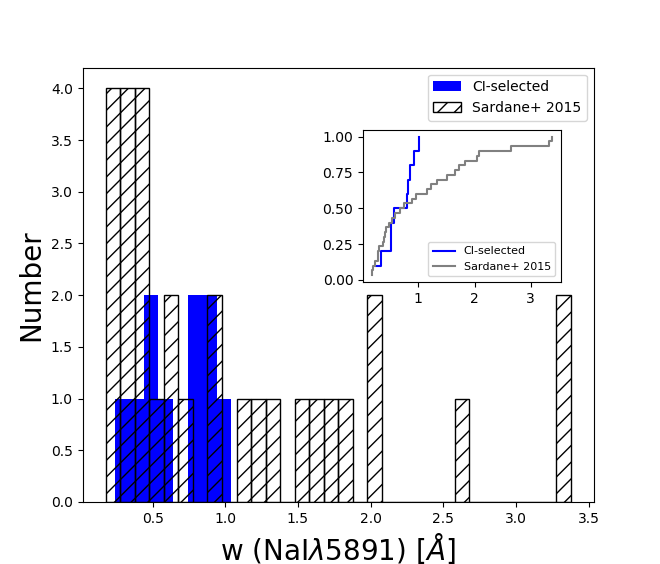}}
\caption[]{\small{\label{nahist}
Blue histogram is the rest equivalent width distribution of the
Na~{\sc i} lines detected in our sample; 
grey-hashed histogram is the equivalent width distribution of Na~{\sc i} lines
detected in the SDSS by \citet{sar15}. The two cumulative distributions are shown in the insert. 
}}
\end{figure}

\noindent
We detect Na~{\sc i} in ten systems out of 11 systems for which the spectral range has a good signal-to-noise.
In other spectra, the wavelength range where the Na~{\sc i} absorption is expected to be redshifted
is spoiled by atmospheric features. Results are listed in column 4 of Table~\ref{naca}.
We compare in Fig.~\ref{nahist} the Na~{\sc i} rest equivalent width distribution in our sample to the distribution in the sample
of 30 $z<0.7$ Ca~{\sc ii} systems detected in SDSS spectra by \citet{sar15}.
It is apparent that the sample by Sardane et al. contains more systems with large Na~{\sc i} equivalent widths.
We caution, however, that a few of these systems have inconsistent doublet ratios. 
We therefore checked the absorption in the SDSS spectra and noticed that some of the strong lines are affected by noise.
A few strong Na~{\sc i} systems seem to be real, however.
It seems that in the sample of C~{\sc i}-selected systems the strong Na~{\sc i} systems are missing.
This is somewhat surprising as we would expect some correlation between Na~{\sc i} and C~{\sc i}
absorptions (see Fig.~\ref{ewnaca}). This possibly can be explained by the fact that strong Na~{\sc i} systems drop out
of our sample because of additional extinction, as the corresponding extinction does affect the
spectrum more strongly at higher redshift. Alternatively there could be some evolution with redshift, 
strong Na~{\sc i} systems being absent at high redshift. This could be related to the possibility that
strong winds are more frequent at lower redshift.  


\begin{table*}
  \small
  \caption{\label{tableZ}Column density measurements using Voigt profile fitting}
\begin{tabular}{lccccccccc}
\hline
QSO   &$z_{abs}$  &   log~$N$(H {\sc i})  &   log~$N$(Fe {\sc ii})  &  log~$N$(Zn {\sc ii})  & log~ $N$(Si {\sc ii}) &   [Fe/Zn] & [Fe/H]$^c$ &  [Si/H]$^c$   & [Zn/H]$^c$ \\
\hline
J0216-0021 &  1.7359&  20.01$\pm$0.04 & 14.41$\pm$0.05 & -                          & 15.21$\pm$0.09  &  -                          &  -1.10$\pm$0.07 &  -0.31$\pm$ 0.10 &                  - \\
J0815+2640 & 1.6798&  20.78$\pm$0.04 & 14.89$\pm$0.05 & 13.36$\pm$0.08 & 15.52$\pm$0.07  & -1.41$\pm$0.09   &  -1.39$\pm$0.07 &  -0.77$\pm$ 0.07 &   0.02$\pm$0.08  \\
J0854+0317 & 1.5663&  20.46$\pm$0.06 & 14.96$\pm$0.11 & 12.97$\pm$0.12 & 15.58$\pm$0.11   & -0.95$\pm$0.16   &  -1.00$\pm$0.12 &  -0.39$\pm$ 0.12 &  -0.05$\pm$0.13    \\
J0917+0154 & 2.1059&  20.75$\pm$0.04 & 15.44$\pm$0.11 & 13.64$\pm$0.05 & 16.17$\pm$0.06  & -1.14$\pm$0.12   &  -0.97$\pm$0.12 &  -0.25$\pm$ 0.08 &  0.17$\pm$0.07    \\
J1047+2057 & 1.7740&  20.48$\pm$0.04 & 15.04$\pm$0.17 & 13.12$\pm$0.11 & 15.68$\pm$0.08  & -1.02$\pm$0.20   &  -0.94$\pm$0.18 &  -0.31$\pm$ 0.09 &   0.08$\pm$0.12   \\
J1122+1437 & 1.5538&  20.33$\pm$0.04 &  13.87$\pm$0.05  & 12.26$\pm$0.24 & 14.72$\pm$0.09 & -1.33$\pm$0.25    &  -1.96$\pm$0.08 &  -1.12$\pm$ 0.10 & -0.63$\pm$0.25   \\
J1133-0057 &  1.7045&  21.00$\pm$0.32$^a$ & 14.23$\pm$0.03& 13.16$\pm$0.06 & 15.71$\pm$0.06 & -1.87$\pm$0.07&  -2.27$\pm$0.32&-0.80$\pm$ 0.33 &  -0.40$\pm$0.33\\
                    &  1.7063 &&&&&&&&\\
J1237+0647 &  2.6896& 19.89$\pm$0.48 & 14.53$\pm$0.06 & -                         & 15.26$\pm$0.22 &  -                           &  -1.01$\pm$0.49  & -0.29$\pm$ 0.53 &  -    \\
J1248+2848 &  1.5124&   -                        & 15.07$\pm$0.17 & 12.82$\pm$0.19 & 15.75$\pm$0.06  & -0.69$\pm$0.25   &       - &  -               &            -   \\
J1302+2111 &  1.6556&  21.27$\pm$0.28 &15.45$\pm$0.16  & 13.15$\pm$0.09& 16.01$\pm$0.05 & -0.64$\pm$0.18    &  -1.32$\pm$0.32 &  -0.77$\pm$ 0.29 &  -0.68$\pm$0.30\\     
J1314+0543 &  1.5829& 19.97$\pm$0.04 & 14.52$\pm$0.10 & 12.98$\pm$0.15& 15.36$\pm$0.11   & -1.40 $\pm$0.18    &  -0.95$\pm$0.11 &  -0.12$\pm$ 0.12 &   0.45$\pm$0.16 \\
J1341+1852 &  1.5442& 18.18$\pm$0.07 & 12.98$\pm$0.05 & -                          & 13.92$\pm$0.11  &  -                         &   $^b$      &   $^b$      &   -   \\      
J1346+0644 &  1.5120&  -                         & 14.70$\pm$0.07 & 13.15$\pm$0.08 & 15.51$\pm$0.08 & -1.39$\pm$0.11   &   -             &  -             &   -    \\ 
J2229+1414 &  1.5854& 19.51$\pm$0.12 & 14.38$\pm$0.05 & 12.09$\pm$0.19 & 15.29$\pm$0.12 & -0.65$\pm$0.20   &  -0.63$\pm$0.14 &   0.27$\pm$ 0.17 &  0.02$\pm$0.23    \\
J2336-1058 &   1.8287& 20.30$\pm$0.06 & 14.98$\pm$0.06 & 13.15$\pm$0.13 & 15.54$\pm$0.07 & -1.11$\pm$0.14   &  -0.86$\pm$0.09 &  -0.31$\pm$ 0.09 &   0.25$\pm$0.15     \\
J2340-0053 &   2.0546& 20.26$\pm$0.11 & 14.96$\pm$0.06 & 12.62$\pm$0.08 & 15.14$\pm$0.08 & -0.60$\pm$0.10   &  -0.81$\pm$0.13 &  -0.64$\pm$ 0.14 &  -0.21$\pm$0.14     \\
J2350-0052 &   2.4265& 20.45$\pm$0.05 & 14.79$\pm$0.03 & 12.89$\pm$0.06 & 15.36$\pm$0.05 & -1.04$\pm$0.07   &  -1.17$\pm$0.07 &  -0.61$\pm$ 0.08 &  -0.13$\pm$0.08   \\    
\hline
\multicolumn{10}{l}{$^a$Value from the fit by \cite{fat17}.}\\
\multicolumn{10}{l}{$^b$Ionization correction should be taken into account.}\\
\multicolumn{10}{l}{$^c$We added the molecular contribution to the total hydrogen column density
whenever H$_2$ is detected \citep[see][]{not18}.
}\\
\end{tabular}
\end{table*}

\begin{table*}
  \small
  \caption{\label{naca}Equivalent widths (\AA) of absorption features}
\centering
\begin{tabular}{lcccccccccc}
\hline
QSO &    C~{\sc i}$\lambda$1560 & C~{\sc i}$\lambda$1656  &   Na~{\sc i}$\lambda$5891 &  Na~{\sc i}$\lambda$5897 &  Ca~{\sc ii}$\lambda$3934 & Ca~{\sc ii}$\lambda$3969 &  Mg~{\sc ii}$\lambda$2798 & Mg~{\sc ii}$\lambda$2803 & Mg~{\sc i}$\lambda$2852 \\
\hline
J0216-0021  &0.32$\pm$0.14  &0.84$\pm$0.11   &0.34$\pm$0.05   &0.09$\pm$0.03 &0.23 $\pm$0.09 & $<$0.22              &2.62$\pm$0.04   &2.21$\pm$0.05   &0.78$\pm$0.04\\
J0815+2640  &0.63$\pm$0.14  &0.99$\pm$0.13  &0.82$\pm$0.21   &0.45$\pm$0.14 &0.47 $\pm$0.16 & $<$0.31              &2.89$\pm$0.06   &2.69$\pm$0.05   &-            \\
J0854+0317  &0.26$\pm$0.07  &0.46$\pm$0.08  &$<$0.23              &$<$0.23            &0.66 $\pm$0.28 & $<$0.12              &2.89$\pm$0.04   &2.73$\pm$0.03   &0.79$\pm$0.06\\
J0917+0154  &0.36$\pm$0.10  &0.61$\pm$0.32  &   -                       &         -               &$<$0.53            & $<$0.53               &3.81$\pm$0.09   &3.73$\pm$0.08   &1.59$\pm$0.11\\
J1047+2057  &0.77$\pm$0.17  &1.17$\pm$0.13  &0.86$\pm$0.15   &0.75$\pm$0.14  &0.59 $\pm$0.14&0.50 $\pm$0.15    &3.50$\pm$0.06   &3.68$\pm$0.04   &1.33$\pm$0.05\\
J1122+1437  &0.19$\pm$0.09  &0.26$\pm$0.08  &0.56$\pm$0.10   &0.53$\pm$0.09  &0.21 $\pm$0.24 & $<$0.28              &0.83$\pm$0.03   &0.65$\pm$0.03   &0.21$\pm$0.03\\
J1133-0057  &0.85$\pm$0.17   &1.42$\pm$0.17  &1.00                    &0.88                   &0.17 $\pm$0.07 &0.07 $\pm$0.10   &2.91$\pm$0.07   &2.80$\pm$0.09   &1.20$\pm$0.06\\
                     &             -        &   -                          & 0.88                   &0.67                   &- &- &- &-    &-\\                                                                  
J1237+0647  &0.39$\pm$0.22  &0.60$\pm$0.17  &0.51$\pm$0.21   &0.34$\pm$0.23 & -                          & $<$0.16              &4.28$\pm$0.16   &3.84$\pm$0.21   &0.99$\pm$0.24\\
J1248+2848  &0.35$\pm$0.13  &0.55$\pm$0.17  &  -                       &       -                  &0.87 $\pm$  0.10 & 0.75 $\pm$0.20  &4.03$\pm$0.07   &3.68$\pm$0.07   &1.21$\pm$0.10\\
J1302+2111  &0.51$\pm$0.19  &0.59$\pm$0.17  &0.80$\pm$0.21   &$<$0.40             &1.06 $\pm$  0.30 & 0.48 $\pm$0.20  &2.87$\pm$0.07   &2.73$\pm$0.06   &1.30$\pm$0.09\\
J1314+0543  &0.46$\pm$0.13  &0.65$\pm$0.15  &0.92$\pm$0.13   &1.00$\pm$0.12 &$<$0.35                & $<$0.35             &2.04$\pm$0.08   &1.79$\pm$0.09   &0.43$\pm$0.08\\
J1341+1852  &0.13$\pm$0.06  &0.14$\pm$0.06  &  -                        &       -                 &$<$0.14                & $<$0.14            &0.32$\pm$0.01   &0.25$\pm$0.02   &0.03$\pm$0.01\\
J1346+0644  &0.50$\pm$0.19  &0.80$\pm$0.18  &  -                        &       -                 &$<$0.31                & $<$0.31            &2.98$\pm$0.07   &2.69$\pm$0.07   &0.84$\pm$0.11\\
J2229+1414  &0.15$\pm$0.22  &0.21$\pm$0.24  &0.51$\pm$0.14   &0.37$\pm$0.12 &$<$0.50                & $<$0.50             &2.70$\pm$0.13   &2.32$\pm$0.11   &0.51$\pm$0.11\\
J2336-1058   &0.25$\pm$0.09  &0.35$\pm$0.06  &0.24$\pm$0.11   &0.17$\pm$0.07 &    -                        &   -                       &3.08$\pm$0.05   &2.68$\pm$0.04   &1.23$\pm$0.07\\
J2340-0053   &0.11$\pm$0.02  &0.17$\pm$0.03  &      -                   &       -                  &0.25 $\pm$  0.02 & 0.10 $\pm$0.10  &1.66$\pm$0.01   &1.53$\pm$0.01   &0.47$\pm$0.01\\
J2350-0052   &0.32$\pm$0.08  &0.51$\pm$0.03  &      -                   &       -                  &    -                      &   -                        &3.06$\pm$0.05   &2.66$\pm$0.05   &0.80$\pm$0.08\\

\hline
\end{tabular}
\end{table*}   

\begin{table*}
  \small
  \caption{\label{tableebv}Dust extinction in the sample; second column is the extinction law used when fitting the dust attenuation.
The mean SNRs in each of the X-shooter arms are given in the last three columns.
}
\centering
\begin{tabular}{lccccccc}
\hline

QSO&    Ext law& $E(B-V)$ &     A(V)&   Rv     & UVB SNR & VIS SNR  & NIR SNR   \\        
 \hline                   
J0216-0021&    LMC2  &  0.123 & 0.34 & 2.76   &49    &   66   &55   \\ 
J0815+2640&    LMC   &  0.138 & 0.47 & 3.41   &35    &   39   &18\\
J0854+0317&    SMC   &  0.099 & 0.27 & 2.74   &52    &   66   &50\\
J0917+0154&    SMC   &  0.135 & 0.37 & 2.74   &22    &   25   &8 \\
J1047+2057&    LMC2  &  0.174 & 0.47 & 2.76   &60    &   69    &27 \\
J1122+1437&    N/A   &  0.000 & 0.00 & 2.74       &35    &   76    & 35\\   
J1133-0057&    SMC   &  0.226 & 0.62 & 2.74       &45    &   63    &50\\
J1237+0647&    LMC2  &  0.152 & 0.42 & 2.76   &40    &   55  &21 \\
J1248+2848&    N/A   &  0.000 & 0.00 & 2.74       &28    &   33   &28 \\
J1302+2111&    SMC   &  0.051 & 0.14 & 2.74   &30    &  33  &16 \\
J1314+0543&    SMC   &  0.036 & 0.10 & 2.74   &42    & 45   &20    \\
J1341+1852&    SMC   &  0.033 & 0.09 & 2.74   &75    & 80    &22\\
J1346+0644&    N/A   &  0.018 & 0.05 & 2.74       &29     &32   &30\\
J2229+1414&    N/A   &  0.000 & 0.00 & 2.74       &25     &28   &11\\
J2336-1058&    N/A   &  0.007 & 0.02 & 2.74       &62     &65   &40\\
J2340-0053&    SMC   &  0.066 & 0.18 & 2.74      &150   &180   &115\\
J2350-0052&    LMC   &  0.038 & 0.13 & 3.41       &50   &70   &35 \\

\hline
\end{tabular}
\end{table*}   

\section{Properties of C~{\sc i} systems at high redshift}

\subsection{Dust dimming}
\begin{figure*}
\centering
\includegraphics[clip=true, width =12cm]{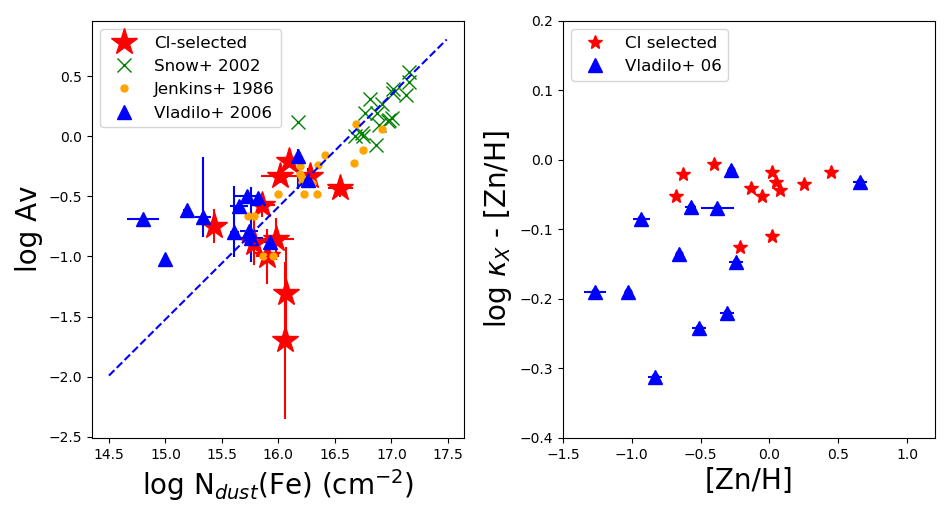}
\caption[]{\small{\label{na_depletion} Left panel:
Attenuation $A_{\rm V}$ versus the Fe~{\sc ii} column density into dust. 
Red stars are for the C~{\sc i}-selected sample, blue triangles correspond to the data from \citet{vla06}, 
green crosses and orange points are  
for samples of local interstellar clouds from \citet{snow02} and \citet{jen86}, respectively.
The blue dashed line is a linear regression of the Milky Way(MW) data with fixed unit slope \citep[see][]{vla06}. 
Right panel: The dust-to-gas ratio minus metallicity is plotted against metallicity.  
}}
\end{figure*}

\begin{figure*}
\centering
\includegraphics[clip=true, width= 15cm]{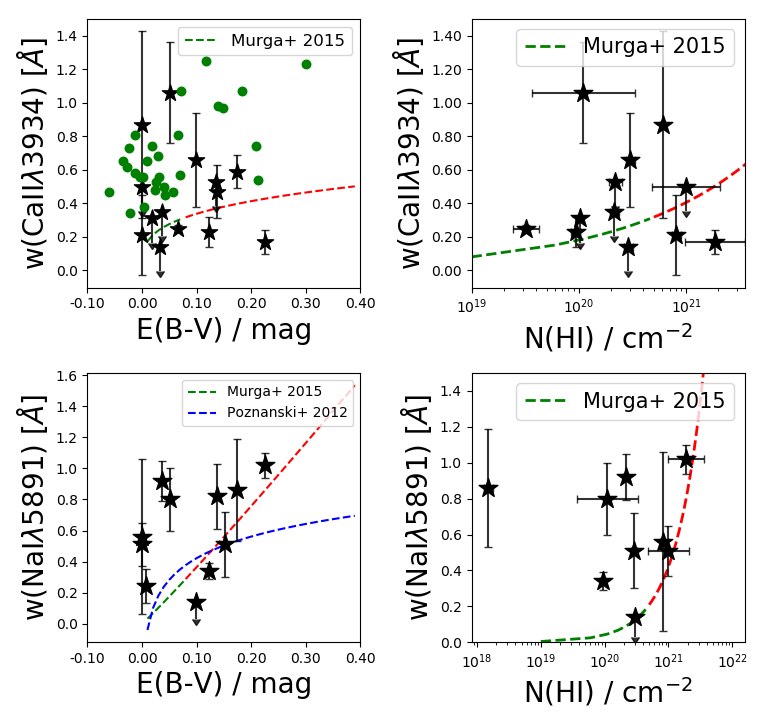}
\caption[]{\small{\label{webv}Left panels: $W$(Ca~{\sc ii}$\lambda$3934) in the upper
panel and $W$(Na~{\sc i}$\lambda$5891) in the lower panel as a function of the colour excess $E(B-V)$. The green dots 
are the data from \citet{wild05}.
In all panels the green dashed lines correspond to the observations of \citet{mur15} in the interstellar and 
circumgalactic media of the Milky Way.
The red dashed lines are extrapolations of these relations. 
The blue line in the $W$(Na~{\sc i}) versus $E(B-V)$ plot is an empirical relation given by \cite{poz12}}:
log($E(B-V)$)~=~2.16$\times$$W - 1.91$.
Right panels: Same as left panels with log~$N$(H~{\sc i}) on the x-axis.
}
\end{figure*}

\citet{vla06} studied the extinction induced by DLAs on the quasar flux.
They found that the extinction $A_{\rm V}$  increases with the column density of iron locked into dust
calculated as follows:
\begin{equation}\label{eq.fe}
  \begin{split}
  N({\rm Fe})_{\rm dust}& = N({\rm Fe})- N({\rm Fe})_{\rm gas}\\
  &= N({\rm Zn})_{\rm gas}\times (10^{[Fe/Zn]_{\odot}} - 10^{[Fe/Zn]_{gas}} ).
  \end{split}
 \end{equation}
In Fig.~\ref{na_depletion} we plot the attenuation $A_{\rm V}$ versus the  
column density of iron into dust measured in the C~{\sc i} systems. Points by \citet{vla06} are
indicated as blue triangles. Apart from a few points 
with low $A_{\rm V}$, the trend is found to be the same for the different samples. 
We note that these low values are quite uncertain. We note also that most
of our systems have log~$N_{\rm dust}$(Fe)~$\sim$~16.
It has been shown that for a column density larger than this, the molecular hydrogen
content of DLAs is high \citep{led03,not08}. 
We calculated the dust-to-gas ratio $\kappa_{\rm X}$ based on the metallicity [X/H] and 
iron to metal ratio [Fe/X], according to the definition in \cite{led03},
\begin{align}
     \kappa_{\rm X}=10^{\rm [X/H]}(1-10^{\rm [Fe/X]}).
      \end{align}
It is clear that this ratio is directly proportional to the metallicity. We therefore plot
this number minus the metallicity as a function of zinc metallicity in the right panel 
of Fig.~\ref{na_depletion}.
It can be seen this number is larger at higher metallicities (see also \citealt{wis17}).
It is intriguing as well to see that the scatter is much less as soon as metallicity
gets closer to the solar value. This may be a consequence of these systems being
somehow chemically mature.
\citet{led03} indicate that the detection probability of H$_2$ in DLAs increases as soon as
log$\kappa >~-1.5$. It is clear that all our systems fulfil this condition.
Actually \citet{not18} show that H$_2$ is present in all systems where 
the H$_2$ molecular transitions are covered by our spectra.
\par\noindent
\citet{wild05} indicated that quasars with strong Ca~{\sc ii} systems along their 
line of sight tend to have larger colour excess.
In the left panels of Fig.~\ref{webv} we plot the $W$(Ca~{\sc ii}$\lambda$3934) and 
$W$(Na~{\sc i}$\lambda$5891) as a function of $E(B-V)$. The mean value of \citet{wild05} is
$E(B-V)$~$\sim$~0.06 for $W$(Ca~{\sc ii}$\lambda$3934)$\sim$0.55~\AA. It can be seen
in the figure that our systems have slightly smaller equivalent widths, albeit
errors are large, for the same colour excess.
\par\noindent
Correlation between dust reddening and the presence of Ca~{\sc ii} and/or Na~{\sc i} absorption lines 
in our Galaxy has been investigated by \citet{poz12} and more recently by \citet{mur15}.
The Na~{\sc i} doublet absorption strength in particular is generally expected to correlate with
the amount of dust along the line of sight. \citet{rich94} have shown, using 57 high-resolution 
stellar spectra taken  by \citet{sam93}, that  the  equivalent width of individual Na~{\sc i} 
components correlates with the colour excess measured for these stars, albeit with a noticeable scatter. 
At lower resolution, the correlation survives although the scatter is even larger due to 
blending of individual components. \citet{poz12} combined two samples of Na~{\sc i} lines observed at high and low resolutions 
from, respectively, the High Resolution Spectrograph (HIRES) and Echellette Spectrograph and Imager (ESI) spectrographs on the Keck telescope and SDSS and derived an empirical
relation between $E(B-V)$ and $W$(Na~{\sc i}). This relation is over-plotted in the left-bottom panel
of Fig.~\ref{webv}. It can be seen that our points are not inconsistent with this relation,
although $W$(Na~{\sc i}$\lambda$5891) in C~{\sc i} systems seem to be larger than what is expected.
\par\noindent
\citet{mur15} used SDSS spectra to obtain mean Na~{\sc i} and Ca~{\sc ii} absorptions on the sky
and correlate them with $N$(H~{\sc i}) and extinction maps.
Both Ca~{\sc ii} and Na~{\sc i} absorption strengths correlate strongly with $N$(H~{\sc i}) and
$E(B-V)$, increasing linearly towards higher values until the saturation effect becomes significant at
$N$(H~{\sc i})$\sim$5$\times$10$^{20}$~cm$^{-2}$ and $E(B-V) \sim 0.08$~mag. Their relations are
over-plotted in Fig.~\ref{webv}.
\par
\begin{figure}
  \resizebox{\hsize}{!}{\includegraphics{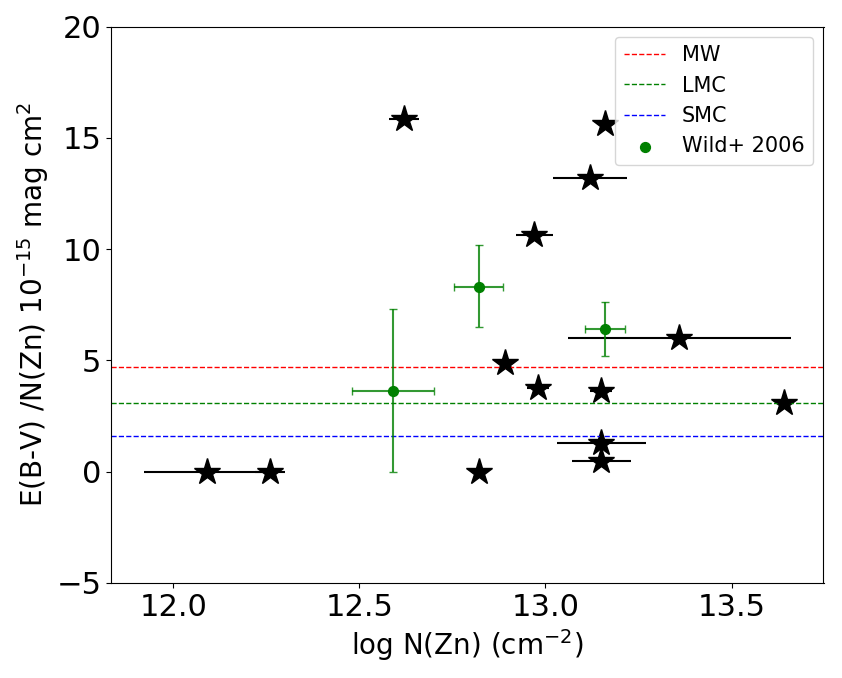}}
  \caption{\small{Dust-to-metal ratio against column density of Zn~{\sc ii}. Dashed lines indicate
values measured in the MW, LMC, and SMC. Green points correspond to values derived in, respectively,
all Ca~{\sc ii} systems and sub-samples of high-$W$ value and low-$W$ Ca~{\sc ii} systems by \citet{wild06}.
}}
  \label{Rdm} 
\end{figure}

\noindent 
The dust-to-metal ratio defined as $E(B-V)$/$N$(Zn~{\sc ii}) is plotted versus the 
Zn~{\sc ii} column density in Fig.~\ref{Rdm}. Values measured in the MW, LMC, and SMC are
indicated as horizontal dashed lines. Green points indicate values measured by \citet{wild06}
in Ca~{\sc ii} systems and sub-samples of strong and weak systems.
It is noticeable that most of the C~{\sc i} systems have dust-to-metal ratios of the order
of what is seen in the MW, LMC, and SMC, with the exception of four systems,
J1133-0057, J2340-0053, J0854+0317, and J1047+2057, which show much larger 
(by a factor of two to three) values than in the MW. \citet{wild06}
have already noticed that the values of the dust-to-metal ratios determined
for the Ca~{\sc ii} absorbers were in the range 
$R_{DM}$ =(4–8)$\times$ 10$^{-15}$~mag~cm$^2$, 
which is often higher than the values derived for the Milky Way.
This was particularly true for their strong-system sample including all
systems with $W$(Ca~{\sc ii}$\lambda$3934)~$>$~0.7~\AA. Out of the four C~{\sc i} systems with
large values, none qualifies for this denomination. The equivalent widths are
observed to be in the range 0.17-0.6~\AA. 
We already noticed that the C~{\sc i} system towards J1133-0057 is very peculiar
as it is at the redshift of the quasar and arises from a very small cloud
\citep{fat17}. Errors are quite large for the system towards
J2340-0053 since $E(B-V)$ and $N$(Zn~{\sc ii}) are small. For the two 
other systems, there is no apparent explanation. It may be conjectured that 
dust composition is quite different in these systems and high resolution data
may help investigate this issue.

\subsection{Nature of the systems}
\begin{figure}
  \resizebox{\hsize}{!}{\includegraphics{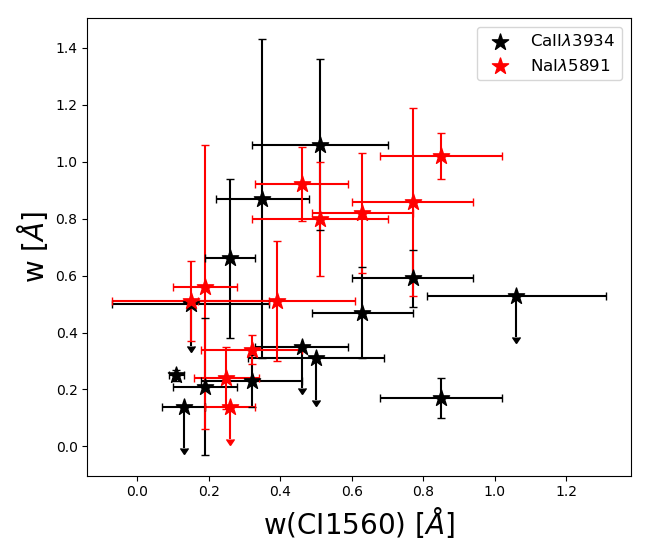}}
  \caption{\small{$W$(Ca~{\sc ii}$\lambda$3934) (black stars) and
$W$(Na~{\sc i}$\lambda$5891) (red stars) as a function of the $W$(C~{\sc i}$\lambda$1560).
}}
  \label{ewnaca}
\end{figure}
\begin{figure}
\centering
\resizebox{\hsize}{!}{\includegraphics[clip=true, width=15cm]{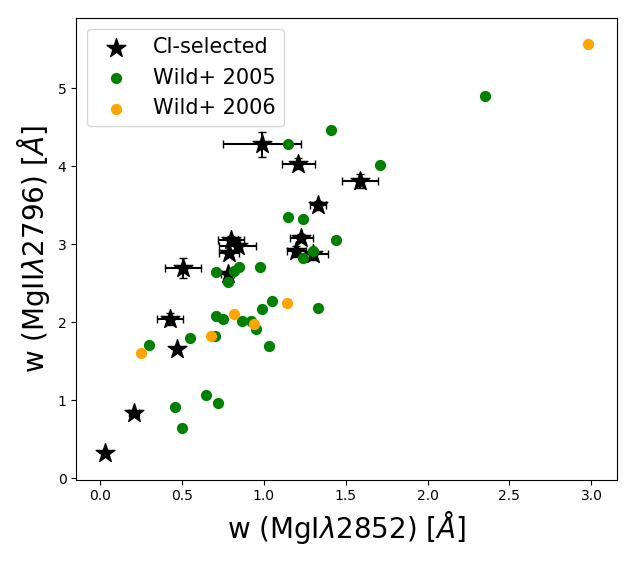}}
\caption[]{\small{} $W$(Mg~{\sc ii}$\lambda$2796) versus $W$(Mg~{\sc i}$\lambda$2852)
equivalent width. The green dots are the data from \citet{wild05}, the orange dots are from \citet{wild06}. 
\label{mgIImgI}}
\end{figure}
We plot in Fig.~\ref{ewnaca} the $W$(Ca~{\sc ii}$\lambda$3934) (black stars) and
$W$(Na~{\sc i}$\lambda$5891) (red stars) as a function of the $W$(C~{\sc i}$\lambda$1560). There is a positive correlation
for Na~{\sc i} with a linear correlation coefficient of 0.80. This is expected as Na~{\sc i} is seen in cold gas 
as traced by C~{\sc i}.
 We note that the correlation can be interpreted as a consequence of an increasing number of components 
as the equivalent widths increase. 
If there is a correlation for Ca~{\sc ii}, it is positive, but  there is no strong evidence for it,
the linear correlation coefficient being only 0.36.
This may indicate that depletion of Ca varies strongly from one system to the other.
\par\noindent
\par\noindent
The correlation between equivalent widths is seen also between Mg~{\sc ii} and Mg~{\sc i}
(Fig.~\ref{mgIImgI}; see also \citealt{wild06}). This is also probably related to the increase of the 
kinematical extension of the absorptions with increasing rest equivalent width (see Fig.~\ref{mgv90}). 

\par\noindent



From the right panels of Fig.~\ref{webv}, we have seen that there is no simple relation between
$W$(Ca~{\sc ii}$\lambda$3934) and $N$(H~{\sc i}). This has been already emphasized by \citet{nes08}
at lower redshift.
The latter authors conclude that systems with $W > 0.25$~\AA~ should be DLAs. This 
conclusion is verified in our sample.
\par\noindent

%
\subsubsection{
Association with galaxies}
\begin{figure}
\centering
\resizebox{\hsize}{!}{\includegraphics[clip=true, width= 60cm]{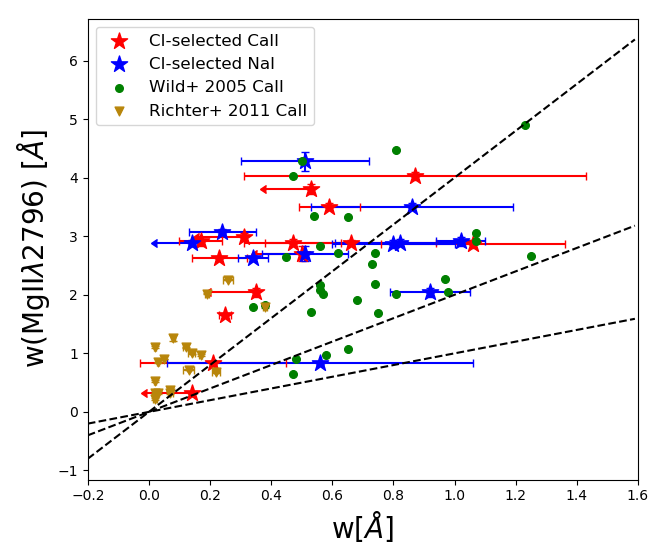}}
\caption[]{\small{\label{wild_mg} $W$(Mg~{\sc ii}$\lambda$2796) as a function
of Ca~{\sc ii}$\lambda$3934 (red stars) and Na~{\sc i}$\lambda$5891 (blue stars) equivalent widths. Green points
are from \citet{wild05} for systems with 0.6~$<$~$z_{\rm abs}$~$<$~1.3. The orange triangles are the sample from \citet{ric11}.
The three dashed lines are for $W$(MgII) / $W$(CaII) = 1, 2, and 4 respectively.  
}}
\end{figure}
%
%
Two characteristics of our C~{\sc i} systems are clearly specific, namely that the velocity spread
of the Mg~{\sc ii} absorption is unusually large (nine out of 17 of our
systems have $\Delta v$~$>$~400~km~s$^{-1}$; see Fig.~5) and that ten out of 11 
of our systems where we could investigate the presence of Na~{\sc i} absorptions
have a detection. In Fig.~\ref{wild_mg} we plot the $W$(Mg~{\sc ii}$\lambda$2796) versus the Ca~{\sc ii} (red stars) and Na~{\sc i} (blue stars) equivalent widths. Data by 
\citet{wild05} at lower redshift are over-plotted as green points. 
It is apparent that the $W$(Mg~{\sc ii}$\lambda$2796) measured in our systems are amongst the largest.
A trend can be seen in the sense of larger $W$(Mg~{\sc ii}$\lambda$2796) for
larger $W$(Ca~{\sc ii}$\lambda$3934), although the scatter is very large. 
This may be an indication that large $W$(Mg~{\sc ii}$\lambda$2896) can be found
in a wide variety of situations and in particular at any distance from the associated
galaxy. 
We should note that in the sample of  star-forming galaxies in 
the redshift range 0.4$-$1.4 observed by \citet{mar12}, 
about 12\% have $W$(Mg~{\sc ii}$\lambda$2796)~$>$~2.5~\AA~ and  
there is an apparent correlation between the stellar mass and the $W$(Mg~{\sc ii}$\lambda$2796).
%
\par\noindent
It is well known that DLAs are not always detected in the disk of galaxies
and that the DLA phase is extended towards the halo and even
the intergalactic filaments (e.g. \citealt{pon08,bou13}).
This may well be the case as well for the C~{\sc i} systems in our sample.
\par\noindent
Several studies have been dedicated to the link between Ca~{\sc ii} systems and associated galaxies
at low redshift \citep{bow91,hew07,ric11,zyc07}.
Galaxies are detected with impact parameters from 5~kpc \citep{pet96} to more than 30~kpc \citep{rah16}.
In turn, cold gas detected by the presence of molecular hydrogen or very high H~{\sc i} DLA systems 
seems to be located in the disks of galaxies \citep[see][]{not14}.
The C~{\sc i} systems studied here are associated with cold gas and Ca~{\sc ii} absorptions
but are not always DLAs. It is probable that they arise in different environments.
It would therefore be of particular interest to search for the associated galaxies.

\subsubsection{C~{\sc i} systems and outflows}
We have seen that the C~{\sc i} systems in our sample show very large Mg~{\sc ii} velocity extents.
Very large Mg~{\sc ii} absorbers are relatively rare although the large statistics
of the SDSS imply that a large number of systems with $W$(Mg~{\sc ii}$\lambda$2796)~$>$~3~\AA~ is known 
\citep{nes05,pro06,qui10}. By stacking thousands of relatively shallow SDSS images of the fields around strong  Mg~{\sc ii}
absorption systems, \citet{zib07} demonstrated that the strongest
systems are associated with bluer galaxies closer to the sightline
to the background quasar compared to weaker systems. 
This suggests a relation between star forming galaxies and strong Mg~{\sc ii} systems,
the velocity extent of the latter possibly arising from a super wind.
This is strengthened by the correlation between $W$(Mg~{\sc ii}) and the mean [OII] luminosity in the
associated galaxy detected by \citet{men09} and \citet{not10b}. \citet{jos17} also found that the [OII] luminosity of the galaxy associated
with Mg~{\sc ii} systems increases with $W$(Mg~{\sc ii}). 
\par\noindent
On the other hand, \citet{nes11}
studied two fields of ultra-strong Mg~{\sc ii} systems
(J0747+305, $z_{\rm abs}$~=~0.7646, $W$(Mg~{\sc ii}$\lambda$2796)~=~3.6~\AA~
and J1417+011, $z_{\rm abs}$~=~0.669, $W$(Mg~{\sc ii}$\lambda$2796)~=~5.6~\AA).
They detected two galaxies in each field at the same redshift as the absorption at 36 and 61~kpc 
for J0747+305 and 29 and 58~kpc for J1417+011 from the line of sight to the quasar. 
This means that either the associated
galaxy is at a very small impact parameter and is not seen in the ground-based images
or the system can arise at large distances from the host galaxy as suggested in the previous section.
\par\noindent
\citet{bou07} surveyed 21 fields around $z_{\rm abs}$~$\sim$~1, $W$(Mg~{\sc ii}$\lambda$2796)~$>$~2~\AA~ 
systems and detected strong H$\alpha$ emission in 14 of the fields.
The corresponding star-formation rate is in the range 1–20~M$_{\odot}$~yr$^{-1}$. The impact
parameter is $<$10~kpc for only two of the galaxies and between 11 and 54~kpc for the 12 others.
\par\noindent
In our sample, 13 out of 17 systems have $W$(Mg~{\sc ii}$\lambda$2796)~$>$~2~\AA~ (see Fig.~5).
Out of these systems, eight depart from the relation corresponding to a saturated line.
This shows that the kinematics is strongly perturbed for most of these systems.
It would be surprising if they arose in quiet disks and they must be located close to regions of star-formation 
activity and/or be part of objects in interaction. 
However, \citet{bou12} searched the fields of 20 strong ($W$(Mg~{\sc ii}$\lambda$2796)~$>$~2~\AA)
$z\sim 2$ Mg~{\sc ii} systems for star-formation activity and detected only four of them. This may 
indicate that not all strong Mg~{\sc ii} systems are related to strong star-formation activity.
Since our systems contain cold gas, 
it would be most interesting to search for the counterparts of star-formation activity around
the systems in our sample.
As noticed before, we found three systems 
in the spectra of J1047+2057, J1133-0057, and J2350-0052 where 
the Mg~{\sc ii} absorptions show two distinct saturated sub-systems separated by
more than 200~km~s$^{-1}$ in the C~{\sc i} and Mg~{\sc ii} absorption profiles. 
In J1133-0057, we detected also two distinct Na~{\sc i} components. This is a small but
significant fraction of the sample where it is reasonable to believe that the
systems arise in interacting objects. We should note that in the case of J2350-0052, \citet{kro17}
detect Ly-$\alpha$ emission at an impact parameter of $\sim$ 6 kpc. In any case, all this suggests that a large fraction of the cold gas at high redshift arises in 
disturbed environments probably in places with strong star-formation activity. 

\section{Conclusion}
%
In this paper we have studied a sample of 17, $z_{\rm abs} > 1.5$, absorption systems selected only 
on the basis of the presence of C~{\sc i} absorption in the SDSS spectrum of background quasars
\citep{led15} and observed with the ESO spectrograph X-shooter.
The $W$(C~{\sc i}$\lambda$1656) are in the range 0.17 to 1.42~\AA.
Thanks to the large wavelength coverage of the
X-shooter instrument, we can study, for the first time at these high redshifts, 
the Ca~{\sc ii}$\lambda\lambda$3934,3969 and Na~{\sc i}$\lambda\lambda$5891,5897 absorption 
together with Mg~{\sc ii} and Mg~{\sc i} absorptions.
\par\noindent 
We show that most of these systems have high metallicities and dust content compared
to standard DLAs at these redshifts. 
We detect nine Ca~{\sc ii} absorptions with $W$(Ca~{\sc ii}$\lambda$3934)~$>$~0.23~\AA~ out of 13 systems
where we could have observed the line. The observed equivalent widths are similar to 
what is observed in other lower redshift surveys.
We detect ten Na~{\sc i} absorptions out of 11 systems where we could observe this species.
No trend is seen between either $W$(Ca~{\sc ii}) or $W$(Na~{\sc i}) and metallicity.
While most of the systems have dust-to-metal ratios of the order
of what is seen in the Milky Way, Large, and Small Magellanic Clouds, four of the systems have values
two or three times larger than what is observed in the Milky Way.
Although the errors affecting our estimates are still large, this may be an indication
that these C~{\sc i} systems are at an advanced stage of chemical evolution. There is,
however, no indication that the systems with large dust-to-metal ratios are a peculiar subset of the 
overall sample.
\par\noindent
The systematic presence of Na~{\sc i} in these C~{\sc i} systems indicates that they
probably probe the cold gas in the ISM of high-redshift galaxies.
The characteristics of the systems are such that
most of the systems should show molecular hydrogen. This will be confirmed in an
associated paper \citep{not18}.
\par\noindent
Most of the systems (12 out of 17) have $W$(Mg~{\sc ii}$\lambda$2796)~$>$~2.5~\AA.
The Mg~{\sc ii} absorptions are spread over more than $\sim$400~km~s$^{-1}$ for half of the
systems; three absorbers have an extension larger than 500~km~s$^{-1}$. This is reminiscent of 
the detection of molecular hydrogen (log~$N$(H$_2$)~=~17.4 and 16.5) in Q0551$-$366 and Q0013-004,
where the Mg~{\sc ii} absorptions are spread over $\Delta v$~$\sim$~700 and 1000~km~s$^{-1}$, 
respectively, for a metallicity relative to solar [Zn/H]~=~$-$0.13 and $-$0.02 and 
$W$(C~{\sc i})~$\sim$~0.25~
and 0.33~\AA~ \citep{led02,pet02}. These large velocities can be the consequence of either interaction 
or star-formation activity in the associated galaxy or cold flow disks \citep{ste13}. 
\citet{led06} have interpreted the observed correlation between $\Delta v$ and
metallicity in DLAs as a tracer of an underlying mass-metallicity relation. 
This is based on the assumption that $\Delta v$ is a tracer of the virial velocity
and thus the halo mass. Others have interpreted large velocities as the consequences
of strong winds powered by star formation (e.g. \citealt{bou16}).
We note that both interpretations would imply that if the kinematics is strongly perturbed for most of 
these systems, the winds probably do not arise in quiet disks and are located close to regions of intense star-formation 
activity and/or are part of objects in interaction. 
However, \citet{bou12} searched the fields of 20 strong ($W$(Mg~{\sc ii}$\lambda$2796)~$>$~2~\AA)
$z\sim 2$ Mg~{\sc ii} systems for star-formation activity and detected only four of them. This may 
indicate that not all strong Mg~{\sc ii} systems are related to strong star-formation activity.
Since our systems contain cold gas, 
it would be most interesting to search for their optical counterparts and associated star-formation activity in emission.
\par\noindent

\begin{acknowledgements}
PN, RS, and PPJ gratefully acknowledge support from the Indo-French
Centre for the Promotion of Advanced Research (Centre Franco-Indien
pour la Promotion de la Recherche Avanc\'ee) under contract
No. 5504-B. 
 
SZ acknowledges support from Universit\'e Pierre et Marie Curie under contrat doctoral No.2124/2015.

\par\noindent JK acknowledges support from EU-FP7 under the Marie-Curie grant agreement no. 600207 with reference DFF-MOBILEX—5051-00115.

\par\noindent S.L. has been supported by FONDECYT grant number 1140838 and partially by PFB-06 CATA.

\par\noindent We thank T. Kr\"uhler for help with the X-shooter data analysis. We thank the referee for a thorough reading of the manuscript and constructive comments. 
\end{acknowledgements}

\bibliographystyle{aa}

\bibliography{./phd}



\begin{appendix}

\section{Comments on individual systems}\label{comments}
We use here the notations Ca~{\sc ii} K and H for Ca~{\sc ii}$\lambda$3934$\AA$
and Ca~{\sc ii}$\lambda$3969$\AA$, respectively,  Na~{\sc i}~D for the Na~{\sc i}$\lambda\lambda$5891,5897
doublet, and $W$ for the rest equivalent width.
The neutral carbon ground base fine structure levels are denoted as C~{\sc i}, C~{\sc i}*, and C~{\sc i}** for
the  2s$^2$2p$^2$$^3$P$_J$$^e$ levels where J = 0,1,2.


\subsection{J0216-0021 $-$ $z_{\rm abs}$ = 1.735888}

\begin{figure*}
  \begin{minipage}[c][10cm][t]{.5\textwidth}
  \vspace*{\fill}
  \centering
  \includegraphics[width=7cm]{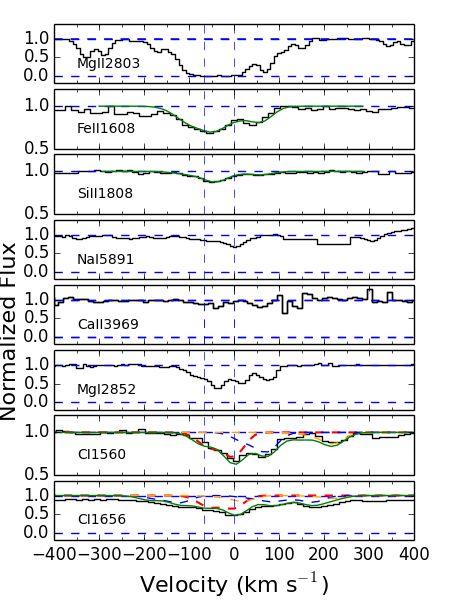}
  \subcaption{}
  \label{0216v}
\end{minipage}%
\begin{minipage}[c][10cm][t]{.5\textwidth}
  \vspace*{\fill}
  \centering
  \includegraphics[width=6.2cm]{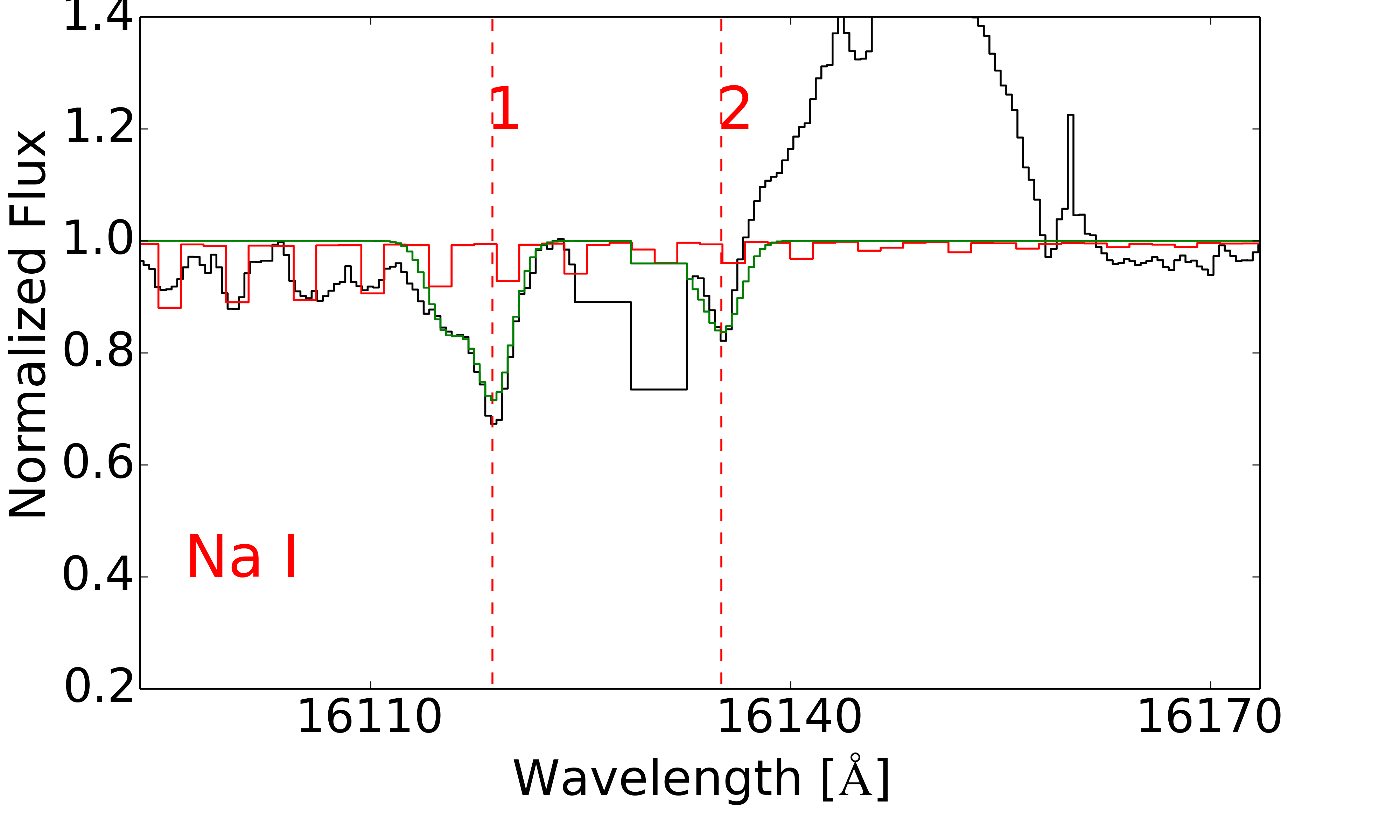}
  \label{0216na}\par\vfill
  \includegraphics[width=6.4cm]{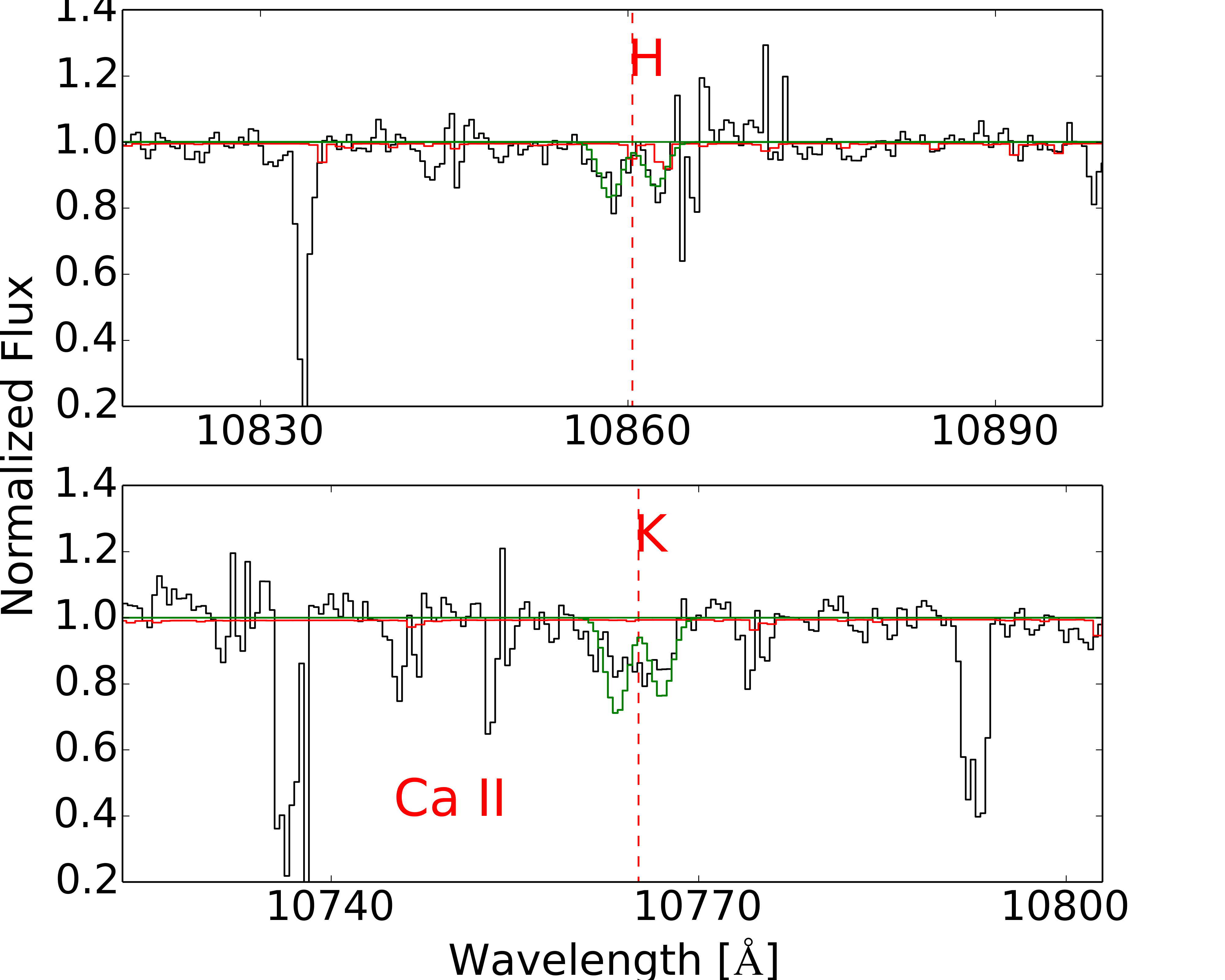}
  \subcaption{}
  \label{0216ca}
\end{minipage}
\caption{\label{0216}J0216-0021 (a) the velocity plot of a sub-set of the
absorbing species. The vertical dashed lines indicate the positions of the C~{\sc i} components. 
Fits to the absorption profiles are over-plotted with the data.
For C~{\sc i} transitions, the red dashed line corresponds to the true ground state, the
blue dashed line is for the C~{\sc i}* absorption, and the orange dashed line is for 
the C~{\sc i}** absorption. The redshift is taken in Table \ref{tableZ} to give the zero velocity. Right panels (b):  The upper panel is  the spectrum at the expected position 
of Na~{\sc i}$\lambda\lambda$5891,5897, the 1 and 2 indicate the Na{\sc i}$\lambda$5891 and Na{\sc i}$\lambda$5897 lines respectively; the lower panel is the spectrum at the expected positions of the Ca~{\sc ii}$\lambda\lambda$3934,3969 lines, the H and K notations indicate the Ca{\sc ii}$\lambda$3969 and Ca{\sc ii}$\lambda$3934 lines respectively. The red curve is the telluric spectrum template of X-shooter.}
\end{figure*}

In this system the Ca{~\sc ii} K line is clearly detected but we indicate an upper limit 
for the H line since the data around the H line is noisy (see Fig.~\ref{0216ca} lower panel). 
The Na~{\sc i}~D is detected clearly even though there are strong spikes close to
the two absorption lines (see Fig.~\ref{0216ca} upper panel). We subtracted the equivalent width of the weak telluric contamination. 
The C~{\sc i}$\lambda$1560 line is not used in the fit of the C~{\sc i} absorption as it is strongly blended. 
There are two components at $z =1.735283$ and 1.735888, while the strongest component 
of Fe~{\sc ii}, Si~{\sc ii,} and Mg~{\sc i} is seen in-between these two components (see also J1248+2848). 
This could be an artefact due to the spectral resolution. To discuss the details of the 
structure of the absorbing cloud, higher resolution data are needed.

\subsection{J0815+2640 $-$ $z_{\rm abs}$=1.679778}
The Ca~{\sc ii} H and Na~{\sc i}~D lines are clearly detected. We conservatively give an upper limit
for the Ca~{\sc ii} H line in Table~\ref{naca}. The Na~{\sc i}~D line is somehow blended with telluric absorptions.
We therefore have subtracted the sky contamination from the EW.
The Mg~{\sc i}$\lambda$2852 feature is strongly blended with absorption from the sky. 
Weak lines from the Ni~{\sc ii} triplet are detected. 

\begin{figure*}
  \begin{minipage}[c][10cm][t]{.5\textwidth}
  \vspace*{\fill}
  \centering
  \includegraphics[width=7cm]{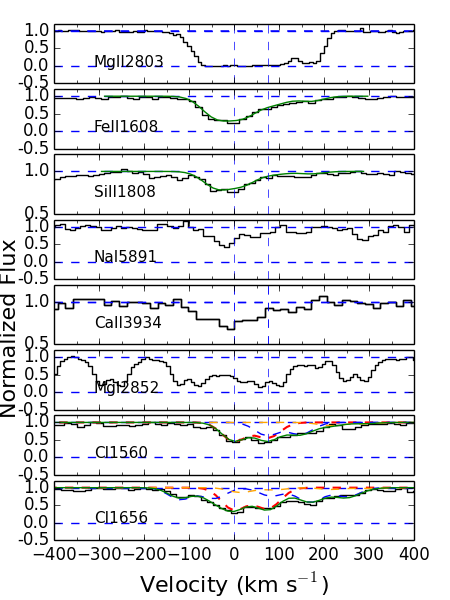}
  \subcaption{}
  \label{0815v}
\end{minipage}%
\begin{minipage}[c][10cm][t]{.5\textwidth}
  \vspace*{\fill}
  \centering
  \includegraphics[width=6.2cm]{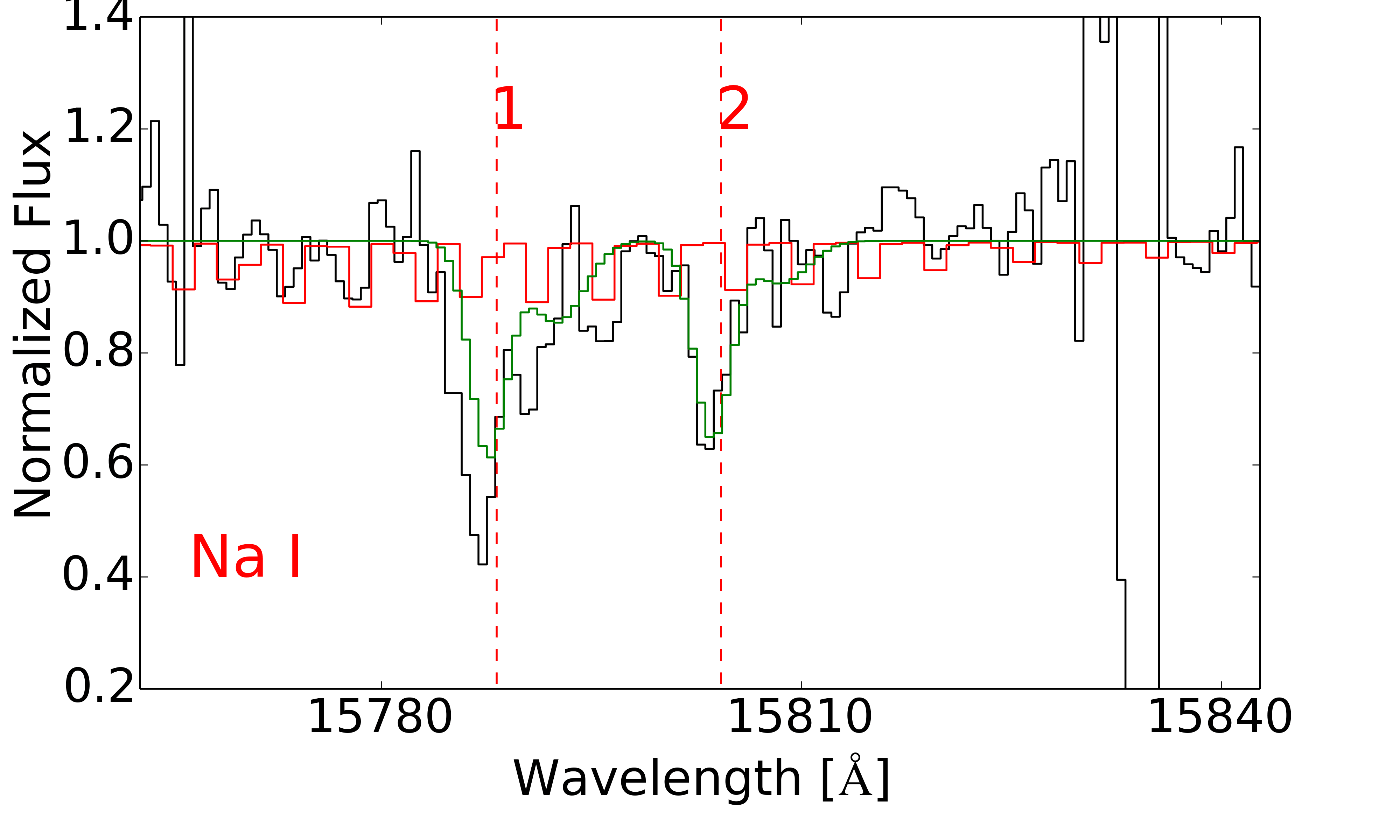}
  \label{0815na}\par\vfill
  \includegraphics[width=6.4cm]{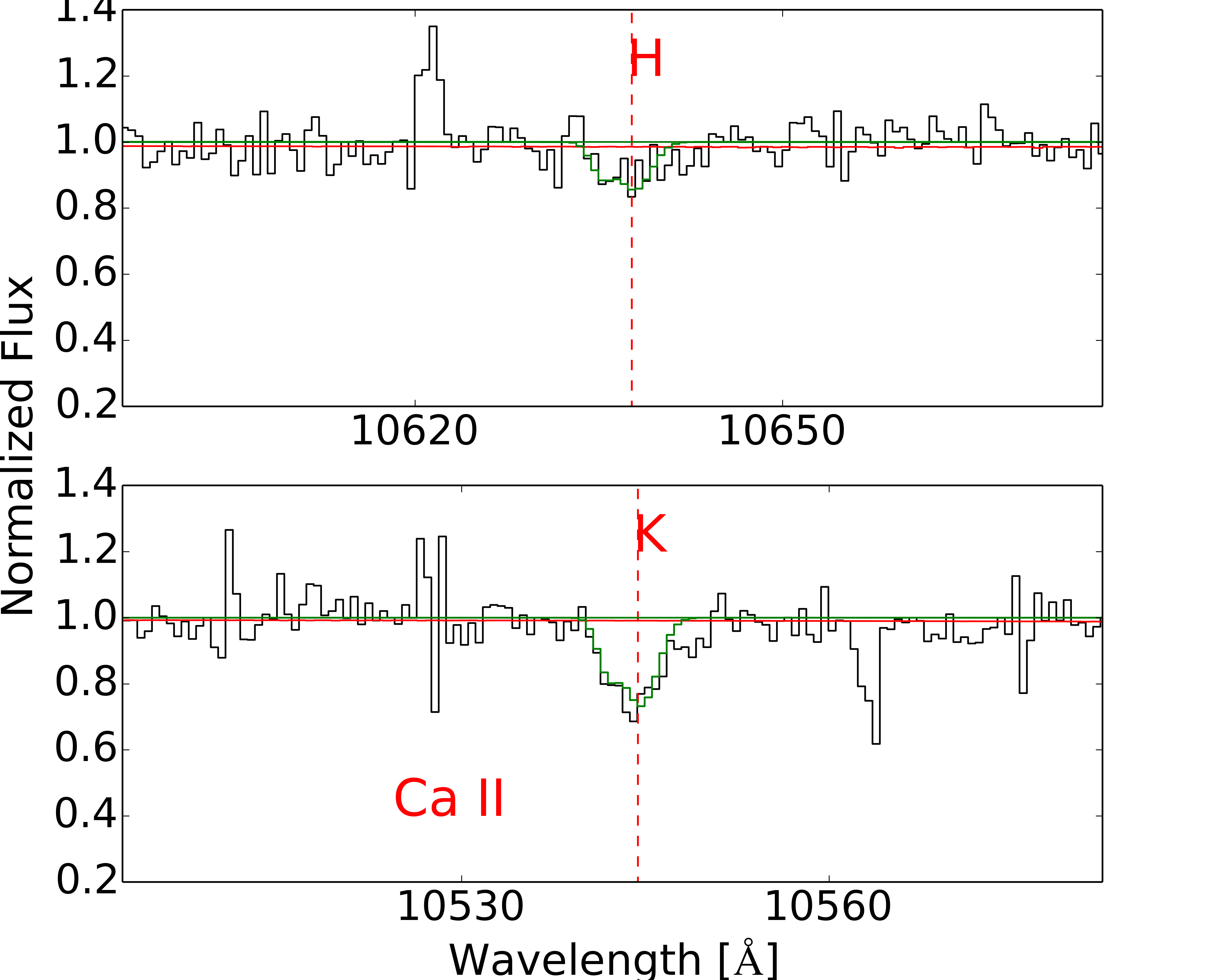}
  \subcaption{}
  \label{0815ca}
\end{minipage}
\caption{J0815+2640 $-$ Same as Fig.\ref{0216}
}
\end{figure*}

\subsection{J0854+0317 - $z_{\rm abs}$ = 1.566320}
The Ca~{\sc ii} K line is detected in this system, when the Ca~{\sc ii} H line is lost in the noise.
The spectrum around the expected position of Na~{\sc i} D is relatively good but no
absorption is seen down to 0.23~\AA. 
There is a slight shift of $\sim$-25~km~s$^{-1}$ between the strongest component 
of Fe~{\sc ii} and Si~{\sc ii} and the strongest C~{\sc i}$\lambda$1656 component. 
The C~{\sc i}$\lambda$1560 line is blended with other lines at $\sim$ -125 km/s  and 400 km/s. 
Weak lines of Zn~{\sc ii} and Mn~{\sc ii} are detected.

\begin{figure*}
  \begin{minipage}[c][10cm][t]{.5\textwidth}
  \vspace*{\fill}
  \centering
  \includegraphics[width=6.6cm]{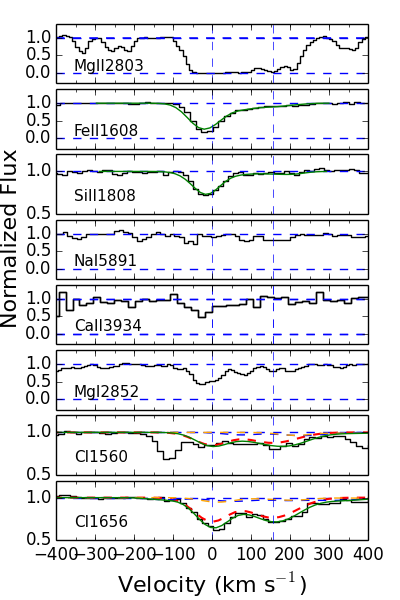}
  \subcaption{}
  \label{0854v}
\end{minipage}%
\begin{minipage}[c][10cm][t]{.5\textwidth}
  \vspace*{\fill}
  \centering
  \includegraphics[width=6.2cm]{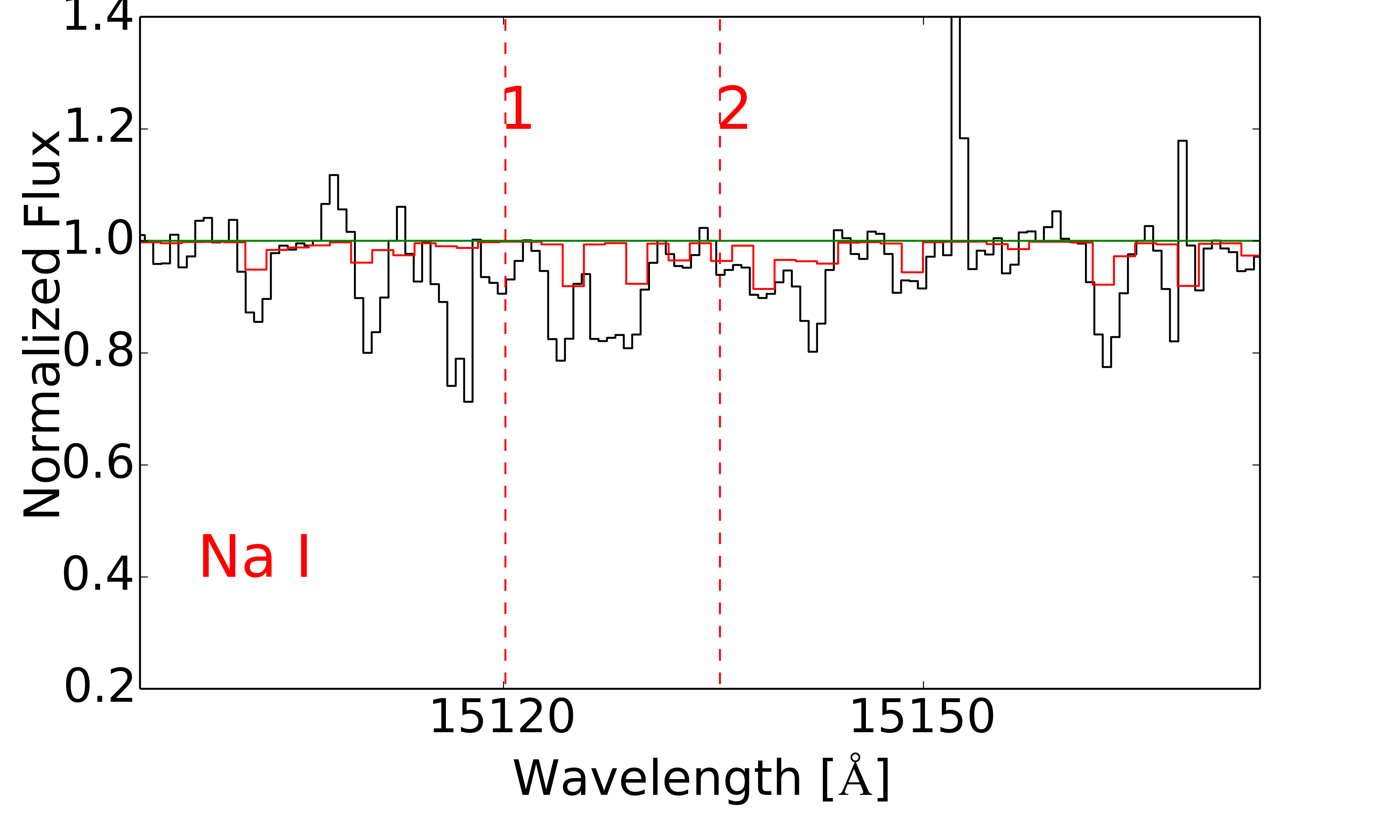}
  \label{0854na}\par\vfill
  \includegraphics[width=6.4cm]{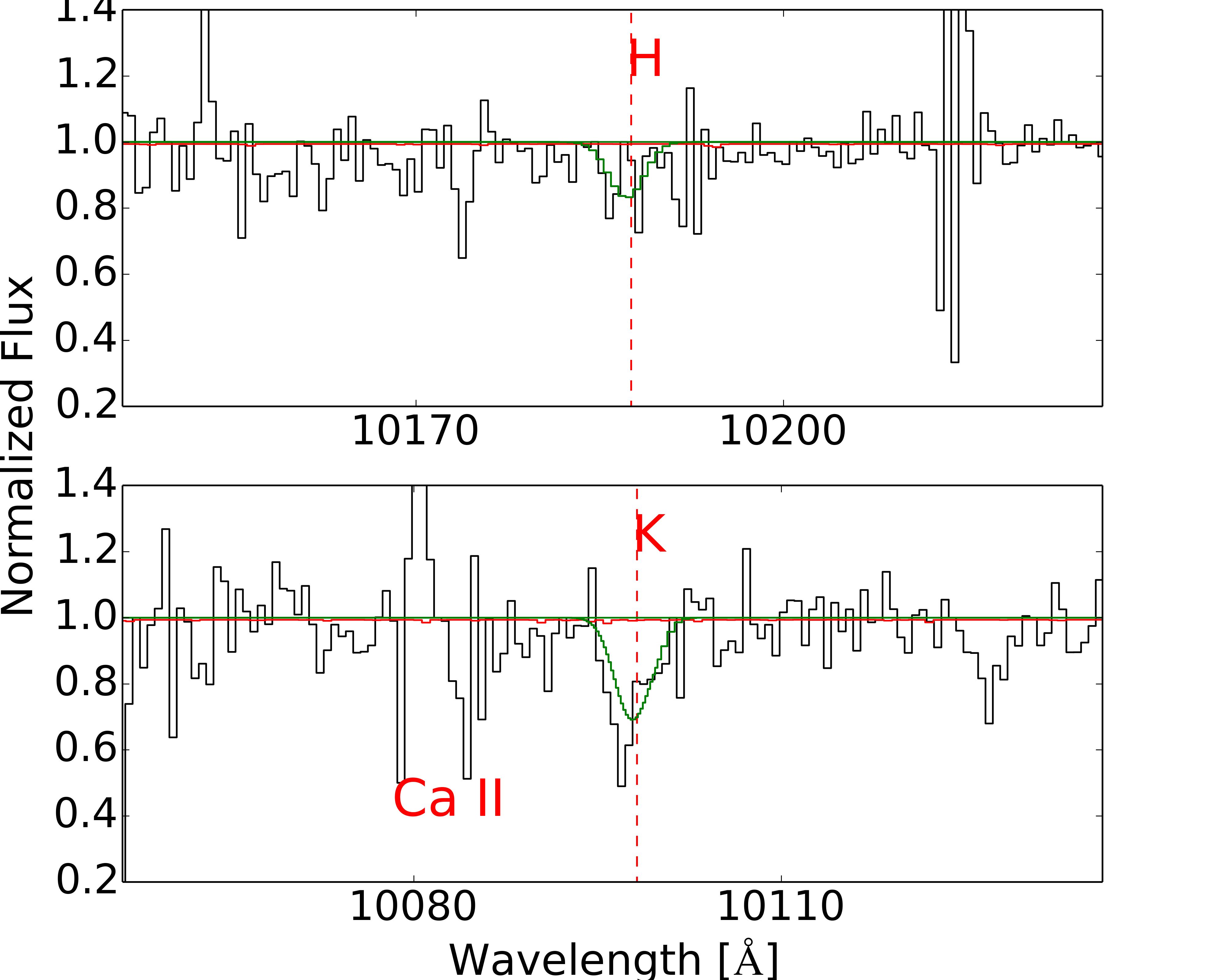}
  \subcaption{}
  \label{0854ca}
\end{minipage}
\caption{J0854+0317 $-$ Same as \ref{0216}.
}
\end{figure*}

\subsection{J0917+0154 $-$ z$_{abs}$ = 2.105934} 
Data are noisy at the expected positions of Ca~{\sc ii} H\&K.
The Na~{\sc i} D is redshifted  in the gap between the H and K bands.
The 
Ni~{\sc ii} and Zn~{\sc ii} lines are also detected. There is no obvious CO line detection in the 
X-shooter spectrum. The C~{\sc i}$\lambda$1560 line is strongly blended, therefore we did 
not include it in our fit.

\begin{figure*}
  \begin{minipage}[c][10cm][t]{.5\textwidth}
  \vspace*{\fill}
  \centering
  \includegraphics[width=6.5cm]{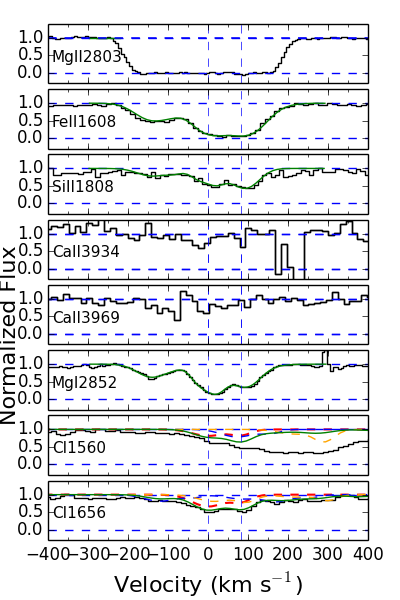}
  \subcaption{}
  \label{0917v}
\end{minipage}%
\begin{minipage}[c][10cm][t]{.5\textwidth}
  \vspace*{\fill}
  \centering
  \includegraphics[width=6.4cm]{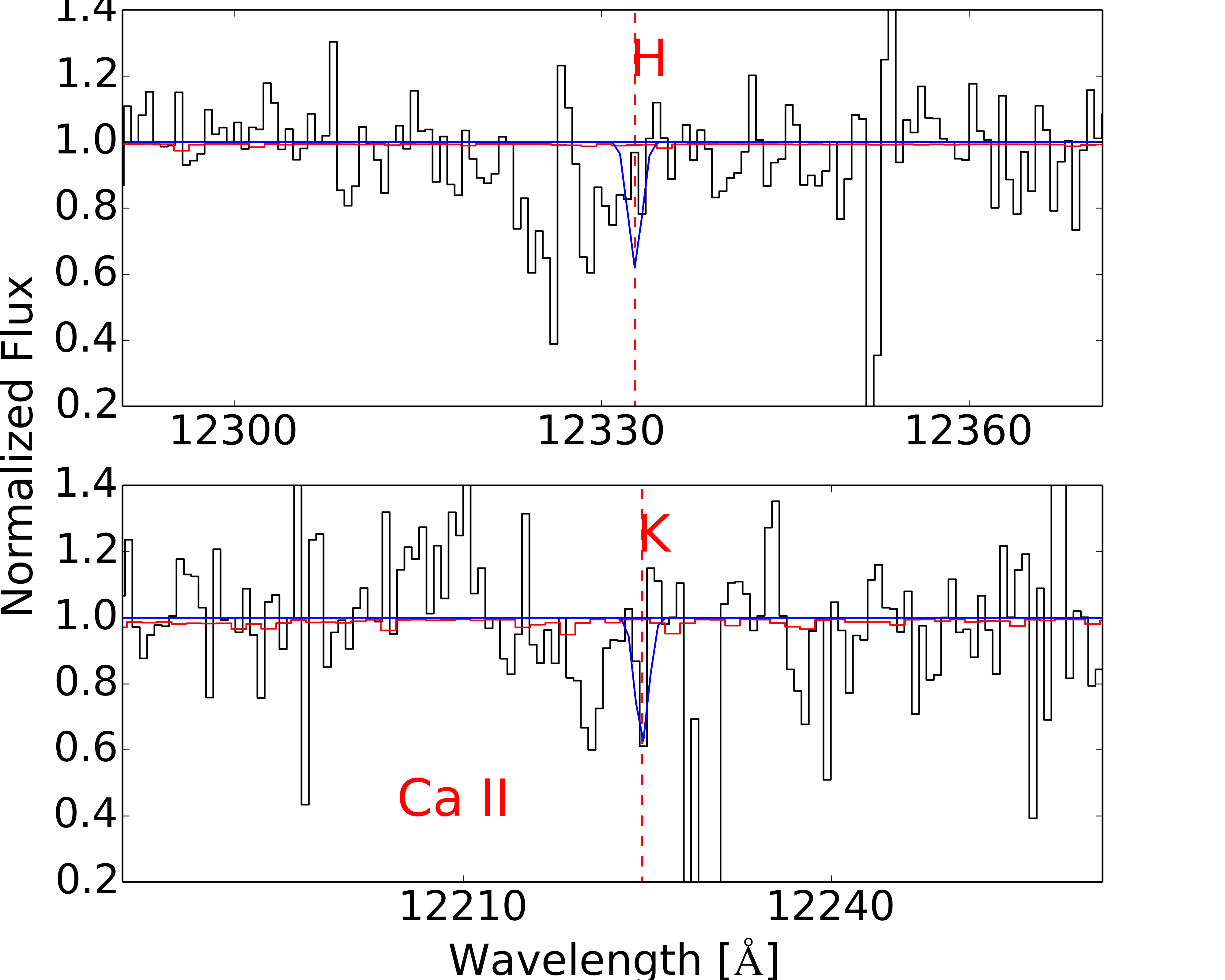}
  \subcaption{}
  \label{0917ca}
\end{minipage}
\caption{J0917+0154 $-$ Same as \ref{0216}.
}
\end{figure*}

\subsection{J1047+2057 $-$ $z_{\rm abs}$ = 1.773960}
Both Ca~{\sc ii} H\&K and Na~{\sc i} D are detected and strong. 
The spectrum around Na~{\sc i} D is noisy however, thus the error on EW is relatively large. 
The redshift defined by the strongest C~{\sc i} component does not correspond to the exact 
centre of the Na~{\sc i} D lines. The strongest component of Na~{\sc i} D is shifted by around 25 km/s
compared to C~{\sc i}.
This may be an artefact of the intermediate resolution of X-shooter as the 
C~{\sc i} fit is complex and needs at least three components. Higher resolution data are needed
to study in more detail the structure of the absorbing cloud.
The elements H$_2$ and CO are detected. Mg~{\sc ii} for this system is uncommonly strong and spans around 
800~km~s$^{-1}$ including components at -350~km~s$^{-1}$ and +400~km~s$^{-1}$.
\cite{not10} first detected CO in this system 
and used it to  measure the cosmic microwave background (CMB) temperature at the corresponding redshift.
They obtain a column density of log~$N$(CO)~=~14.74$\pm$0.07. From the X-shooter data
we derive log~$N$(CO)~=~14.56$\pm$0.92, which is in agreement with the result by \cite{not11}. 
\cite{dap16} used the CO absorption from this system to constrain the cosmological 
variation of the proto-to-electron mass ratio. 
NiII and ZnII are detected.

\begin{figure*}
  \begin{minipage}[c][10cm][t]{.5\textwidth}
  \vspace*{\fill}
  \centering
  \includegraphics[width=6.5cm]{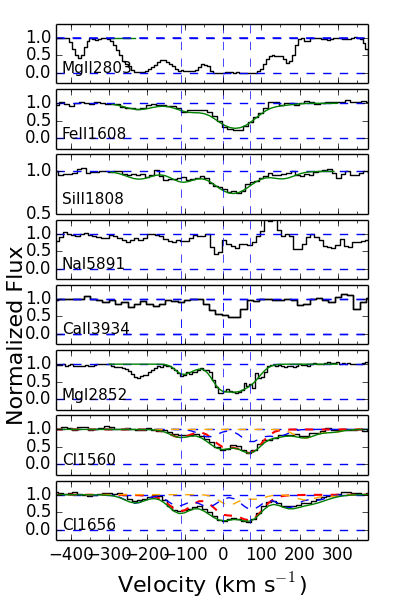}
  \subcaption{}
  \label{1047v}
\end{minipage}%
\begin{minipage}[c][10cm][t]{.5\textwidth}
  \vspace*{\fill}
  \centering
  \includegraphics[width=6.2cm]{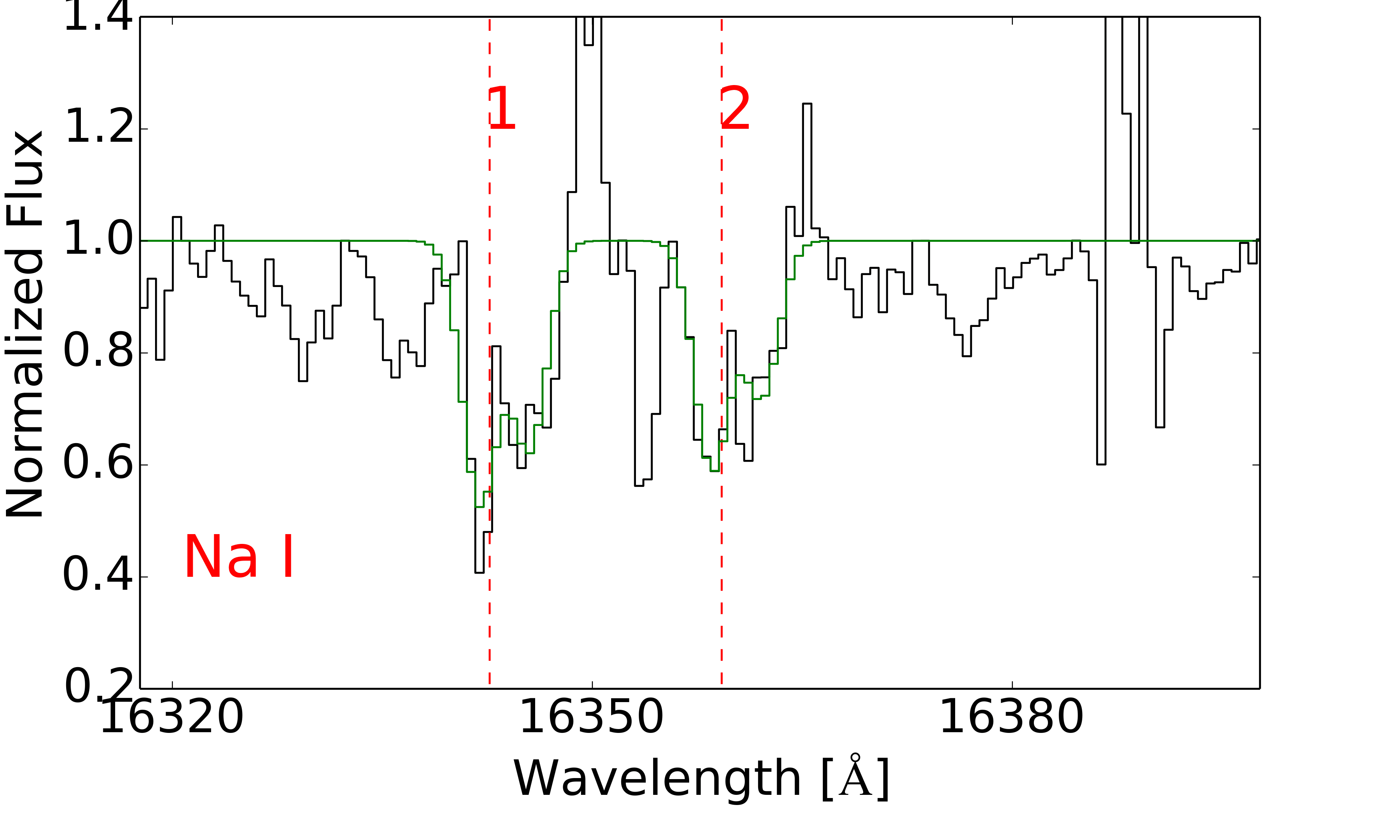}
  \label{1047na}\par\vfill
  \includegraphics[width=6.4cm]{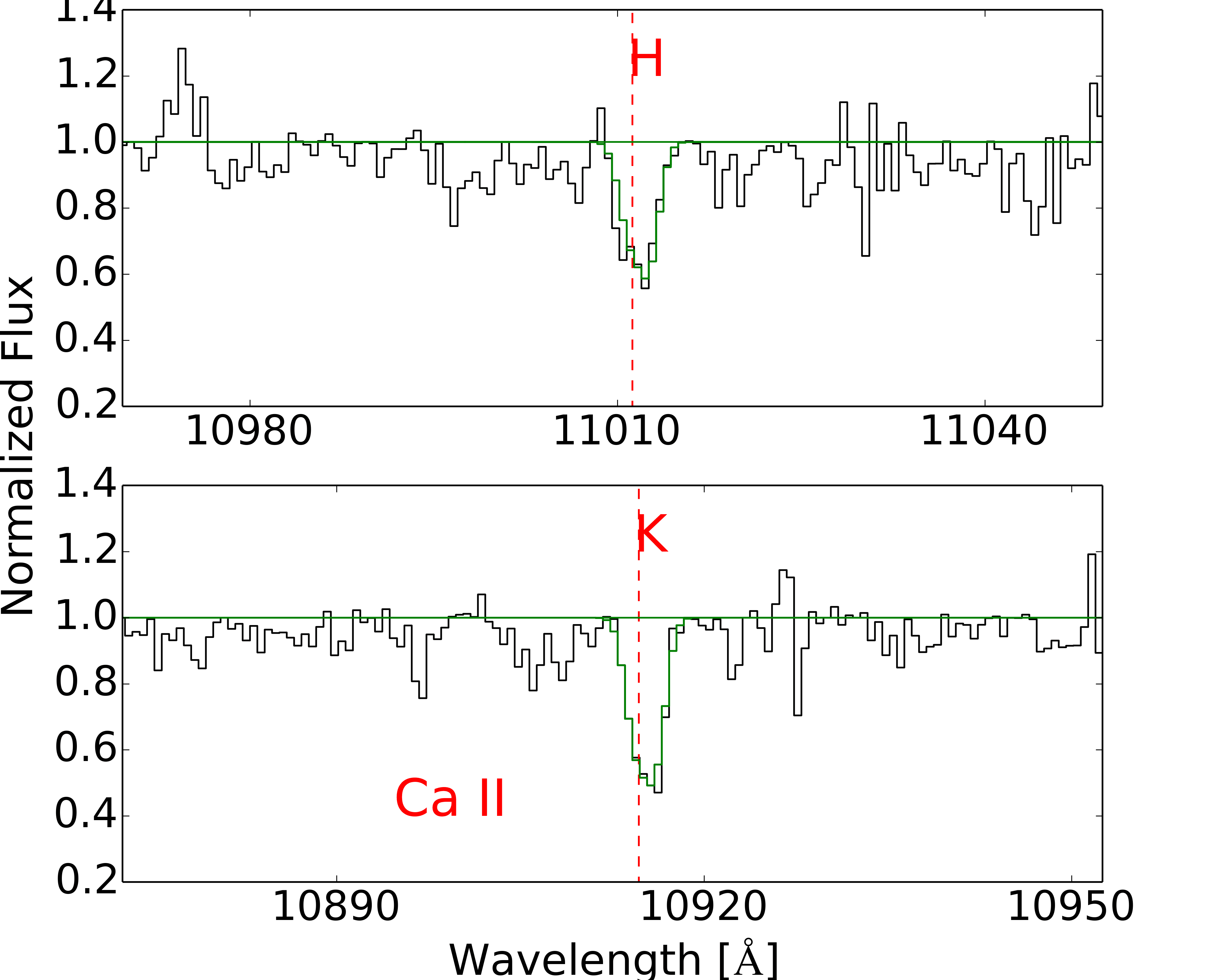}
  \subcaption{}
  \label{1047ca}
\end{minipage}
\caption{J1047+2057 $-$ Same as \ref{0216}. 
}
\end{figure*}

\subsection{J1122+1437 $-$ $z_{\rm abs}$ = 1.553779}
\begin{figure*}
  \begin{minipage}[c][10cm][t]{.5\textwidth}
  \vspace*{\fill}
  \centering
  \includegraphics[width=7cm]{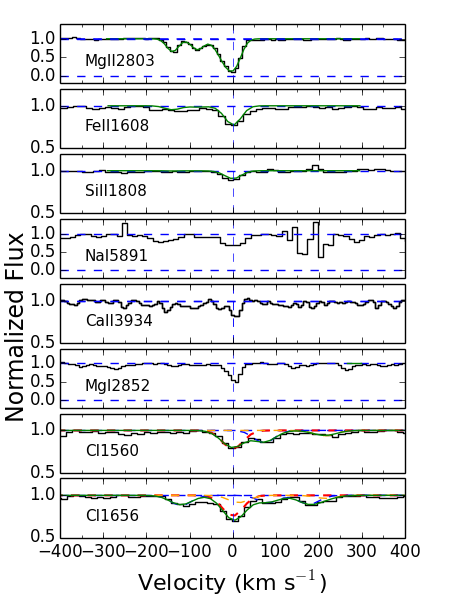}
  \subcaption{}
  \label{1122v}
\end{minipage}%
\begin{minipage}[c][10cm][t]{.5\textwidth}
  \vspace*{\fill}
  \centering
  \includegraphics[width=6.2cm]{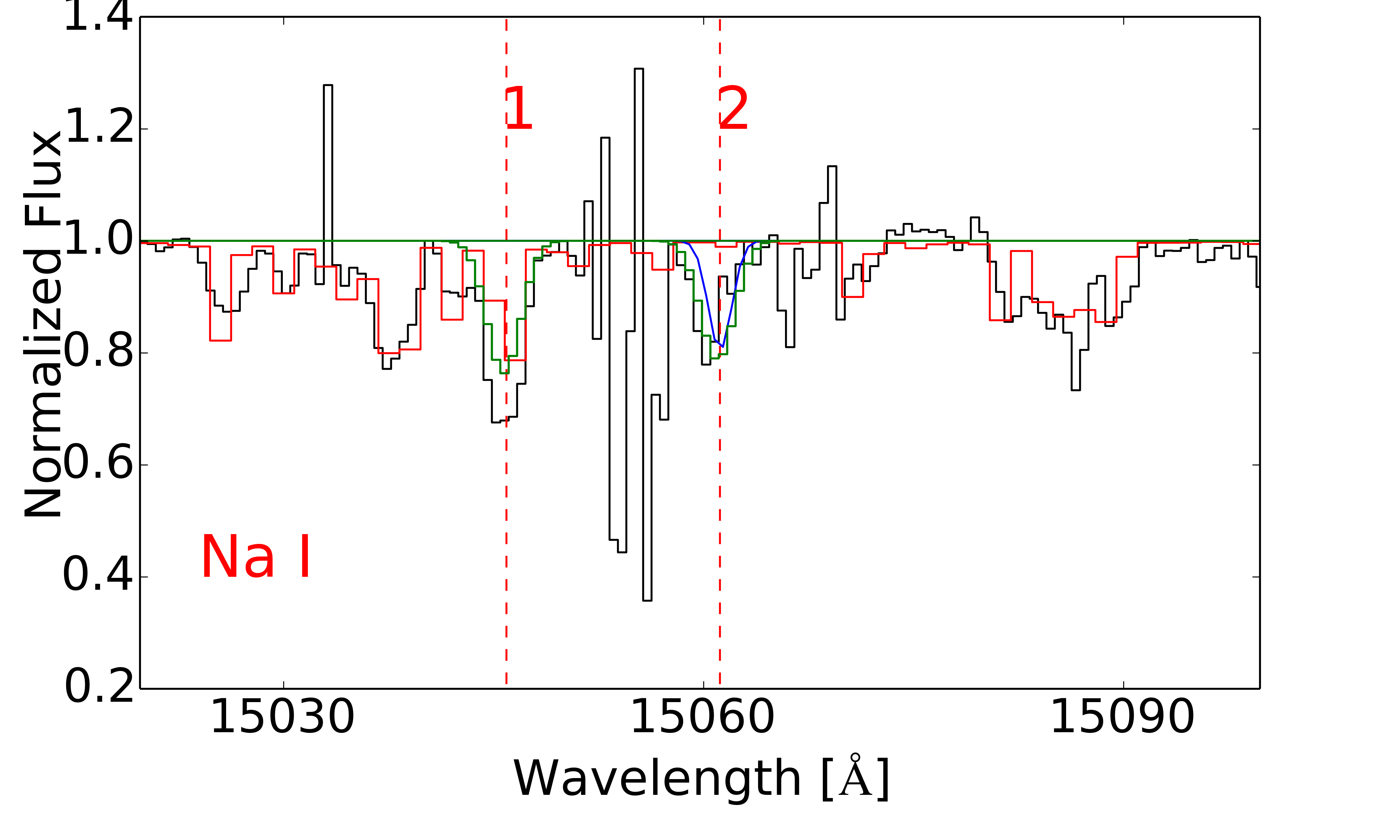}
  \label{1122na}\par\vfill
  \includegraphics[width=6.4cm]{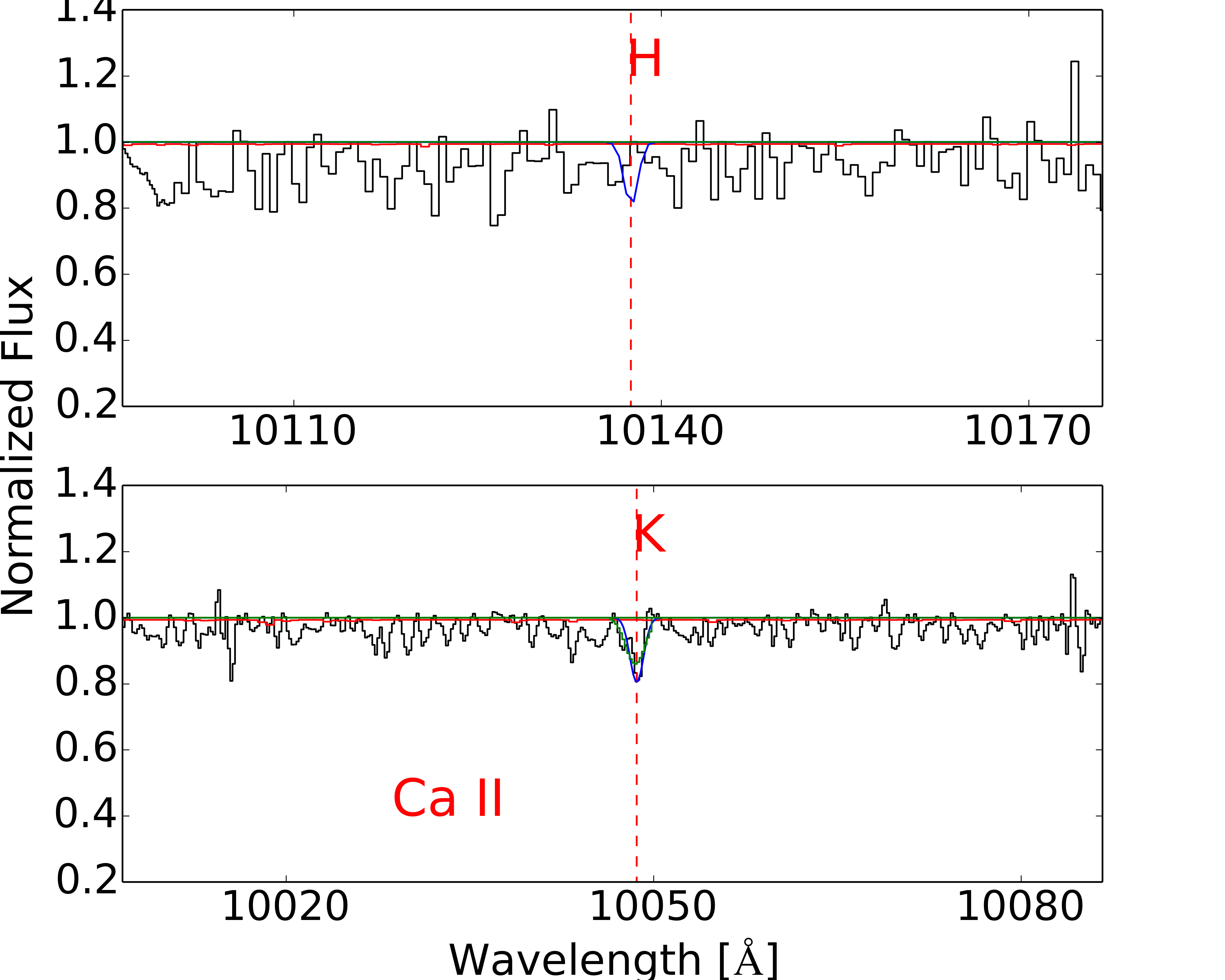}
  \subcaption{}
  \label{1122ca}
\end{minipage}
\caption{J1122+1437 $-$ Same as \ref{0216}.   
}
\end{figure*}

In this spectrum Ca~{\sc ii} and Na~{\sc i} are both detected. The Ca~{\sc ii} K line is clearly detected but not
Ca~{\sc ii} H (see Table \ref{naca}).
For the Na~{\sc i} D doublet, both lines are detected. 
However, Na~{\sc i}$\lambda$5891 is strongly blended with a telluric feature but there is 
an apparent excess that can be estimated and subtracted (Fig. \ref{1122ca} upper panel).
Metal lines and C~{\sc i} absorptions are detected in one strong component plus several weaker components
blueshifted by -170,-100, and -20 km s$^{-1}$ relative to the main component. 

\subsection{J1133-0057 $-$ z$_{abs}$ = 1.704536}
Both Ca~{\sc ii} H\&K and Na~{\sc i} D are clearly detected in this good spectrum. This system is a 
peculiar but interesting case for which the absorbing cloud is small and located at a short distance to the quasar.
Indeed, the broad line region (BLR) is only partially covered by the cloud. This system has been analysed by 
\cite{fat17} who derived an H~{\sc i} column density of 21.00$\pm$0.30. 
There are two main components separated by $\sim$200~km/s. Although Ca~{\sc ii} is detected
only in the strongest system, Na~{\sc i} is detected in both. Ca must be highly depleted
into dust, which is consistent with the corresponding large attenuation of the quasar (see Table 3). 

\begin{figure*}
  \begin{minipage}[c][10cm][t]{.5\textwidth}
  \vspace*{\fill}
  \centering
  \includegraphics[width=6.8cm]{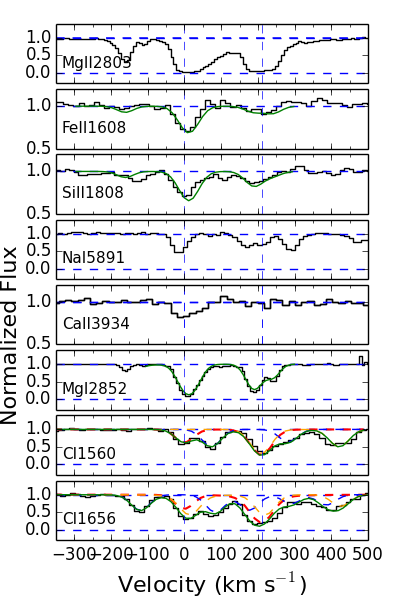}
  \subcaption{}
  \label{1133v}
\end{minipage}%
\begin{minipage}[c][10cm][t]{.5\textwidth}
  \vspace*{\fill}
  \centering
  \includegraphics[width=6.2cm]{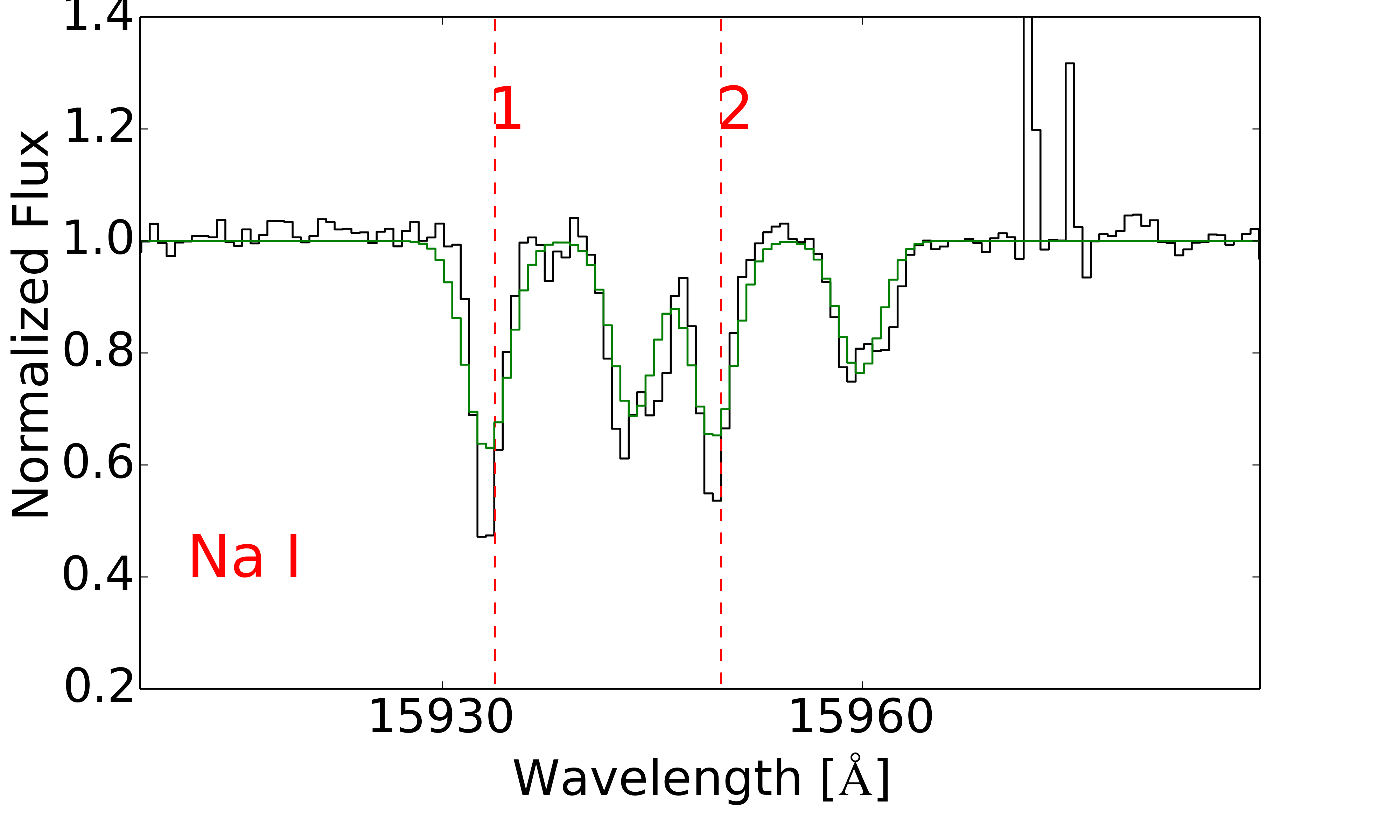}
  \label{1133na}\par\vfill
  \includegraphics[width=6.4cm]{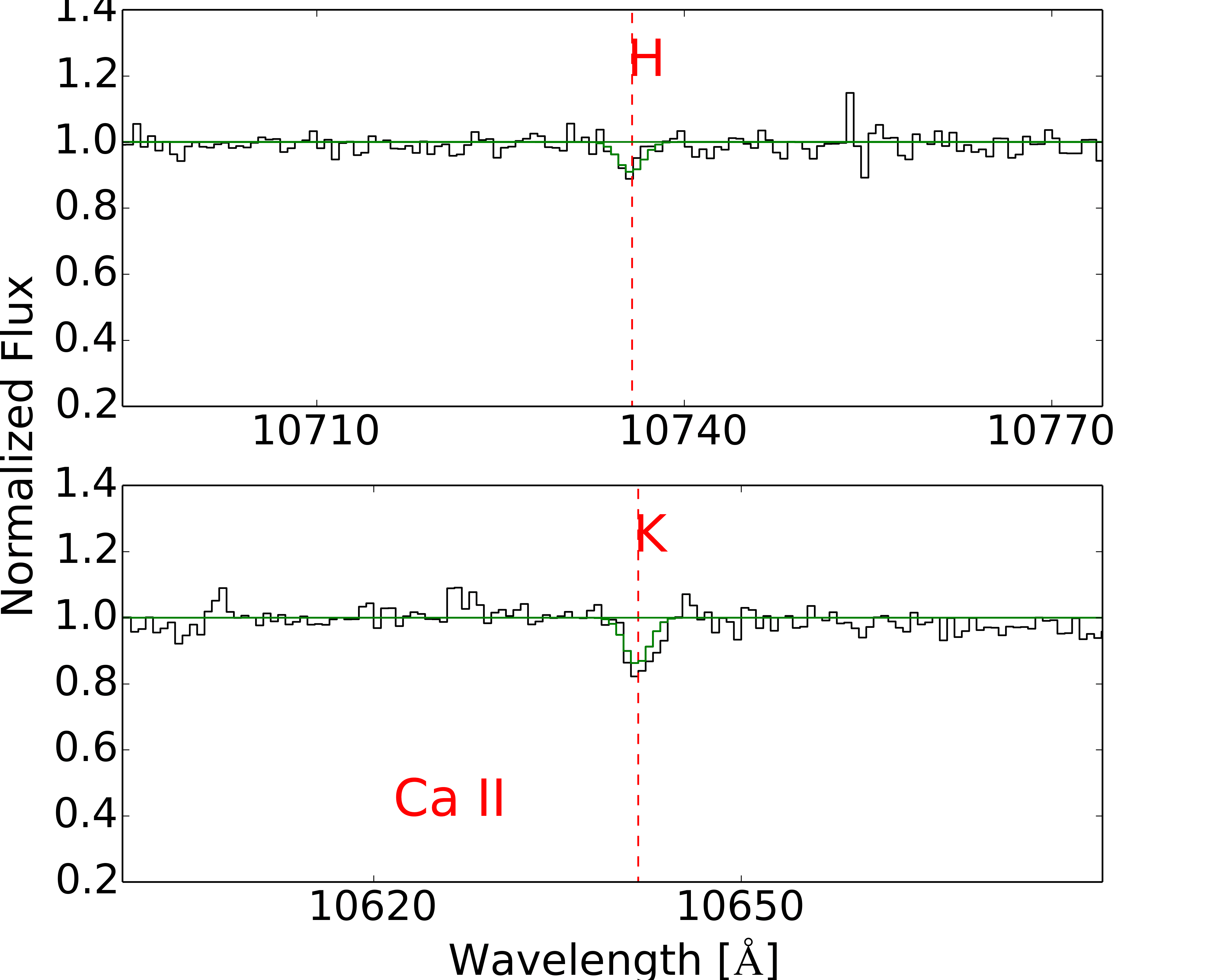}
  \subcaption{}
  \label{1133ca}
\end{minipage}
\caption{J1133-0057 $-$ Same as \ref{0216}. 
}
\end{figure*}

\subsection{J1237+0647 - $z_{\rm abs}$ = 2.689602}
In this spectrum Na~{\sc i} D is clearly detected while Ca~{\sc ii} H\& K are strongly blended with telluric absorptions, 
so there is no EW upper limit for Ca~{\sc ii} D. 
Detection of H$_2$ and CO in this system has been published by \cite{not10}. 
The excitation of the CO rotational levels is used to measure $T_{\rm CMB}$.
The spectrum used by \cite{not10} was obtained with UVES and the authors derive 
log~$N$(HI) = 20.0$\pm$0.15, to be compared with our measurement 19.89$\pm$0.47. The system
is also analysed by \cite{dap16}. Three components are used to fit the absorptions from C~{\sc i} 
and its fine structure lines. The components of Fe~{\sc ii} and C~{\sc i} are in exactly the same place.
%

\begin{figure*}
  \begin{minipage}[c][10cm][t]{.5\textwidth}
  \vspace*{\fill}
  \centering
  \includegraphics[width=6.6cm]{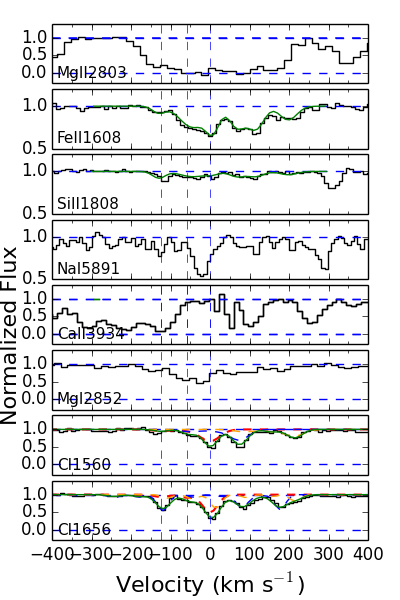}
  \subcaption{}
  \label{1237v}
\end{minipage}%
\begin{minipage}[c][10cm][t]{.5\textwidth}
  \vspace*{\fill}
  \centering
  \includegraphics[width=6.2cm]{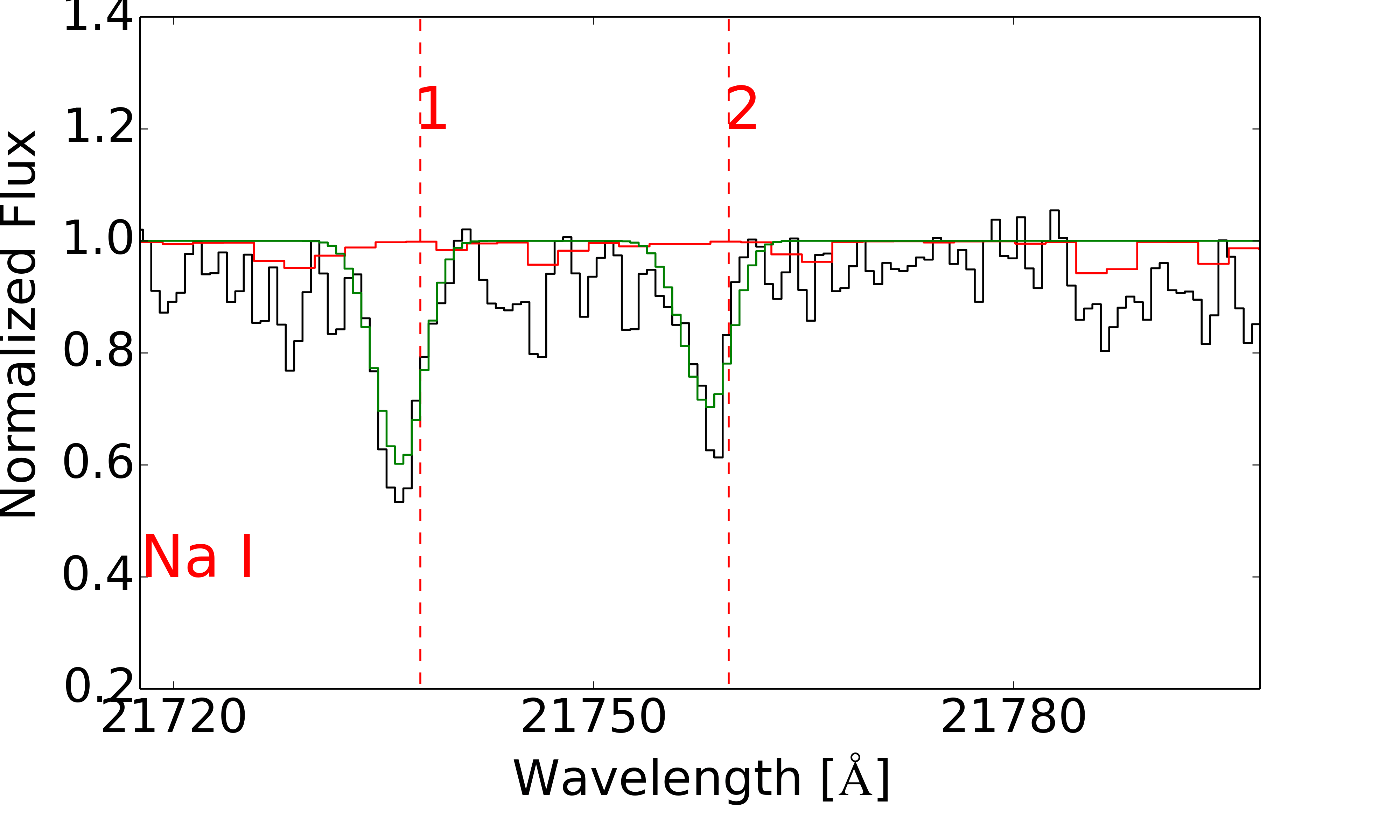}
  \label{1237na}\par\vfill
  \includegraphics[width=6.4cm]{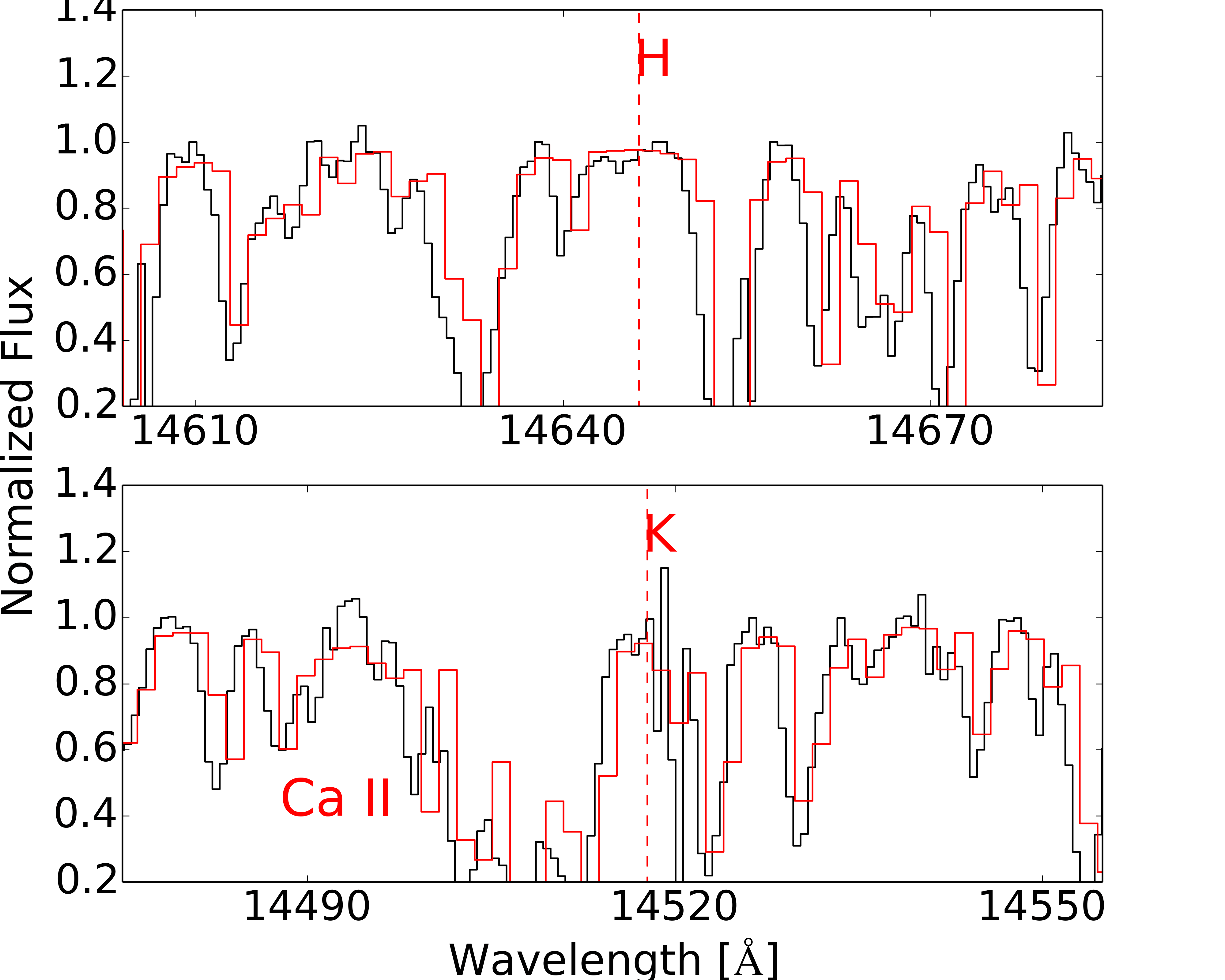}
  \subcaption{}
  \label{1237ca}
\end{minipage}
\caption{J1237+0647 : Same as \ref{0216}.
}
\end{figure*}

\subsection{J1248+2848 $-$ $z_{\rm abs}$ = 1.512373}
Both Ca~{\sc ii} K \& H lines are detected for this system. 
The Ca~{\sc ii} H line is, however, blended with a sky feature, thus the value given 
in Table \ref{naca} for the $EW$ Ca~{\sc ii} H line is somewhat tentative. As in the case of J1346+0644, 
Na~{\sc i} D lines are badly blended with strong telluric absorptions. This is why we do not
give an $EW$ upper limit for Na~{\sc i} D. The component at $v\sim$ -250 km s$^{-1}$ 
is included into the fit and the measurement of $W$(Mg~{\sc i}$\lambda$2852). The width of the 
metal line absorption profile is around 575 km s$^{-1}$. 

\begin{figure*} 
  \begin{minipage}[c][10cm][t]{.5\textwidth}
  \vspace*{\fill}
  \centering
  \includegraphics[width=6.6cm]{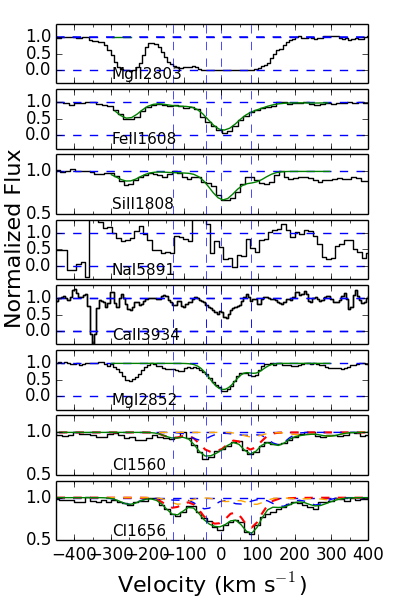}
  \subcaption{}
  \label{1248v}
\end{minipage}%
\begin{minipage}[c][10cm][t]{.5\textwidth}
  \vspace*{\fill}
  \centering
  \includegraphics[width=6.2cm]{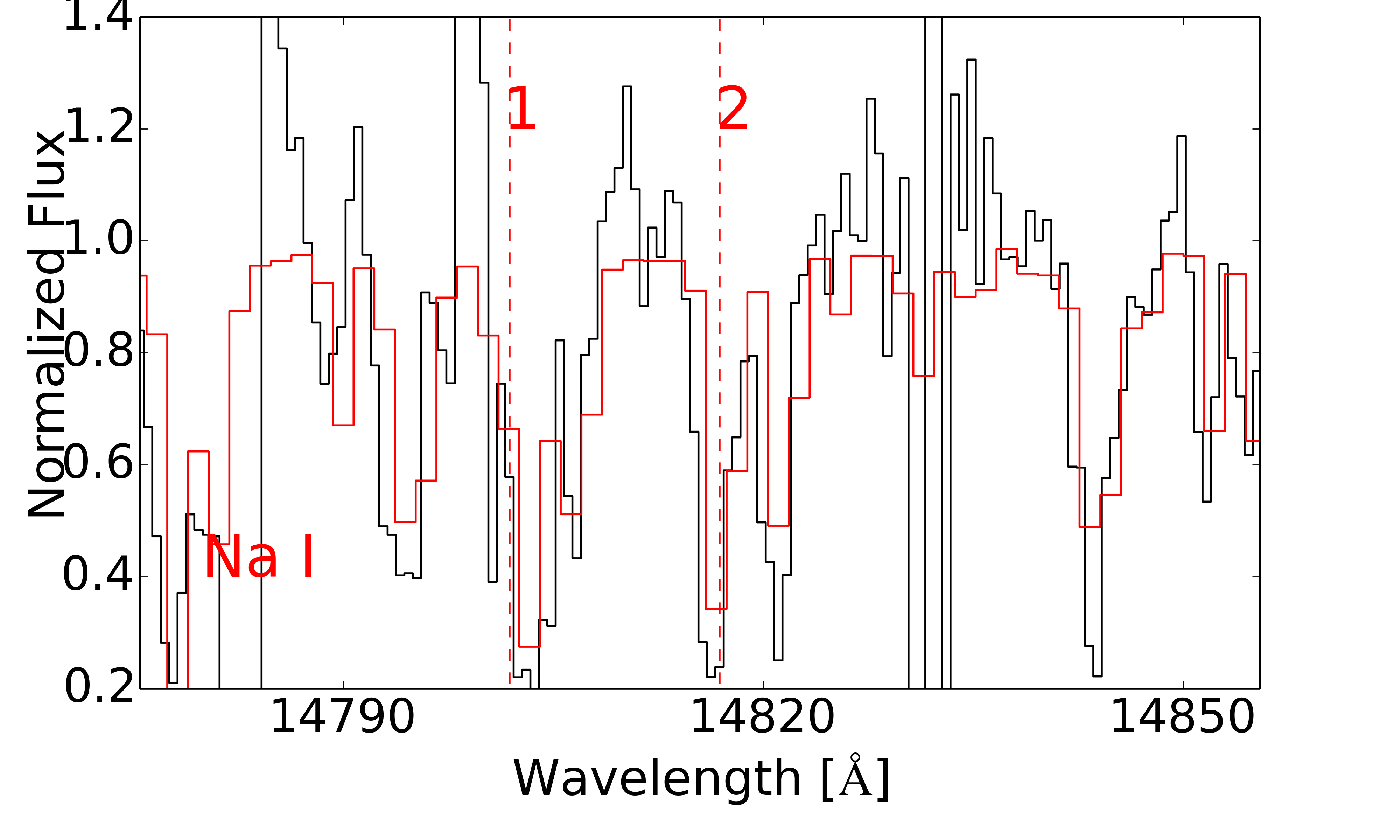}
  \label{1248na}\par\vfill
  \includegraphics[width=6.4cm]{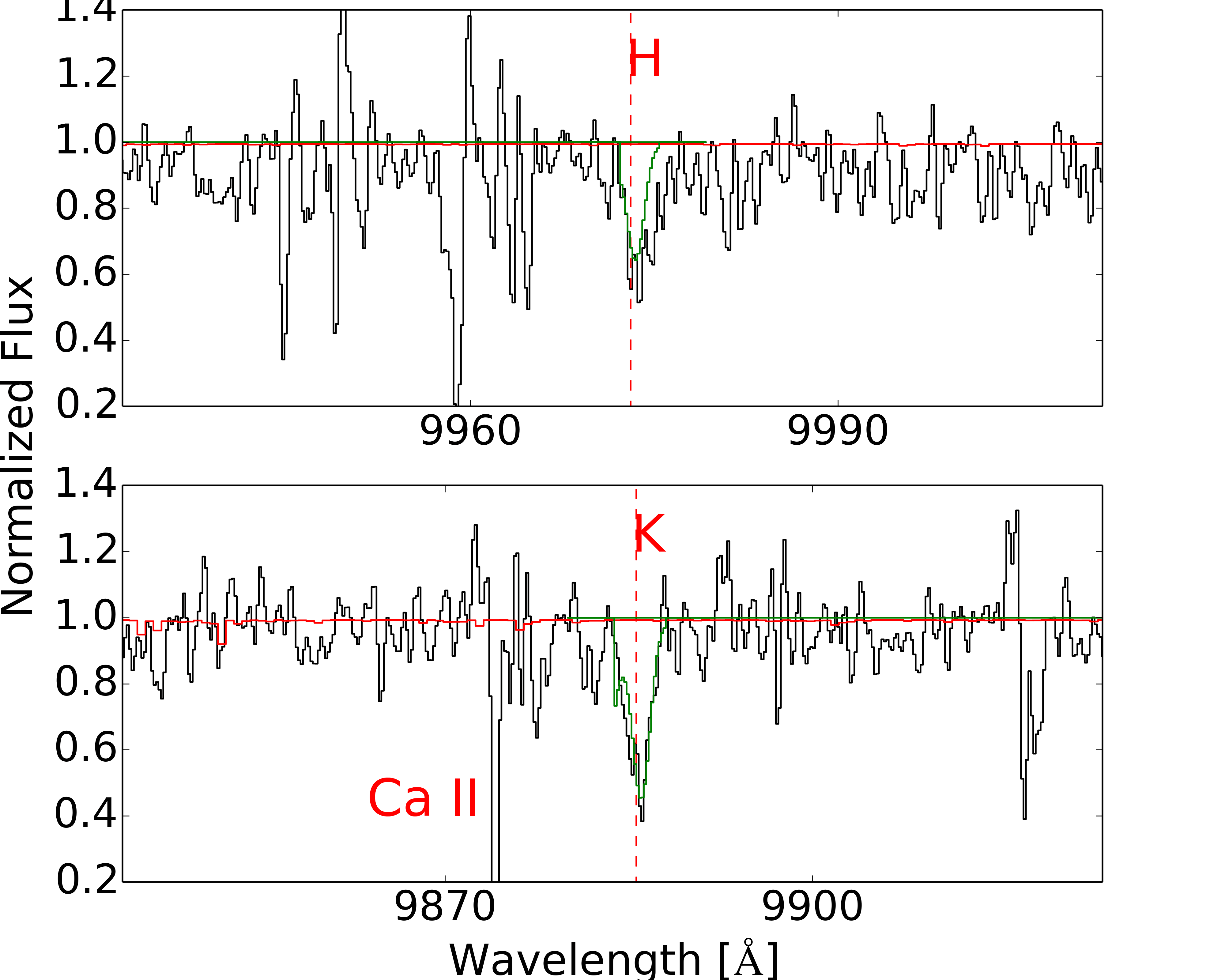}
  \subcaption{}
  \label{1248ca}
\end{minipage}
\caption{J1248+2848: Same as \ref{0216}.
}
\end{figure*}

\subsection{J1302+2111 $-$ $z_{\rm abs}$ = 1.655602}
Both Na~{\sc i} and Ca~{\sc ii} are detected for this system. The Na~{\sc i}$\lambda$5891 line is clean, 
However, there is a spike close to Na~{\sc i}$\lambda$5897 that prevents a direct measurement of the $EW$ .
We thus fit the doublet using VPFIT to derive the $EW$ for Na~{\sc i}$\lambda$5897. 
The Fe~{\sc ii}$\lambda$2586 and Fe~{\sc ii}$\lambda$2600 lines are strongly blended with sky absorptions, 
so we did not use them to fit the Fe~{\sc ii} lines. 
C~{\sc i} is well fitted with one component.

\begin{figure*}
  \begin{minipage}[c][10cm][t]{.5\textwidth}
  \vspace*{\fill}
  \centering
  \includegraphics[width=6.5cm]{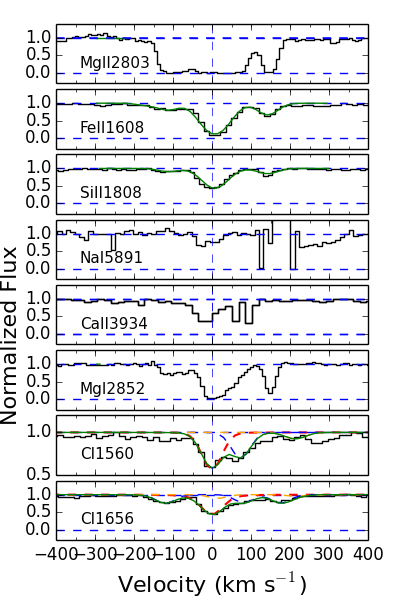}
  \subcaption{}
  \label{1302v}
\end{minipage}%
\begin{minipage}[c][10cm][t]{.5\textwidth}
  \vspace*{\fill}
  \centering
  \includegraphics[width=6.2cm]{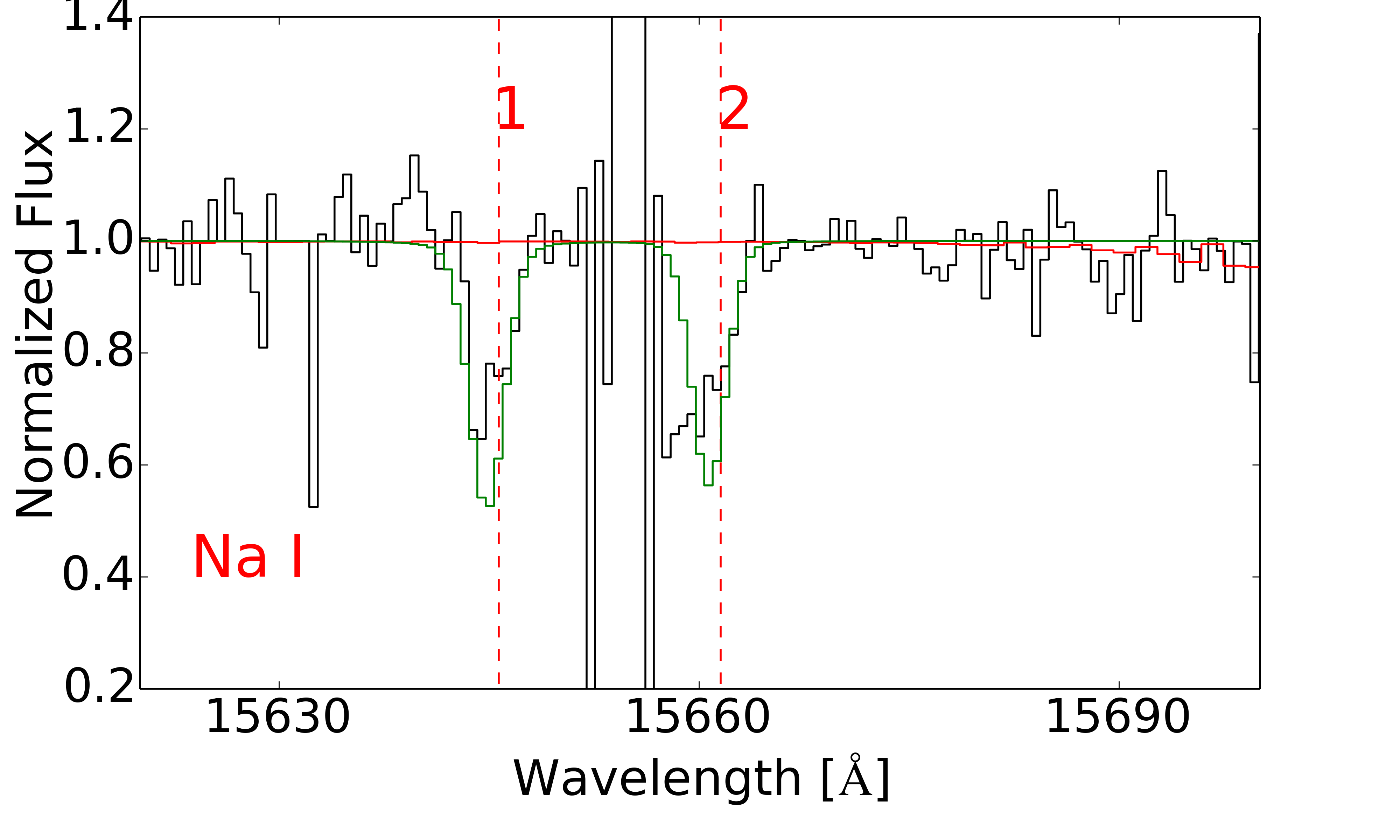}
  \label{1302na}\par\vfill
  \includegraphics[width=6.4cm]{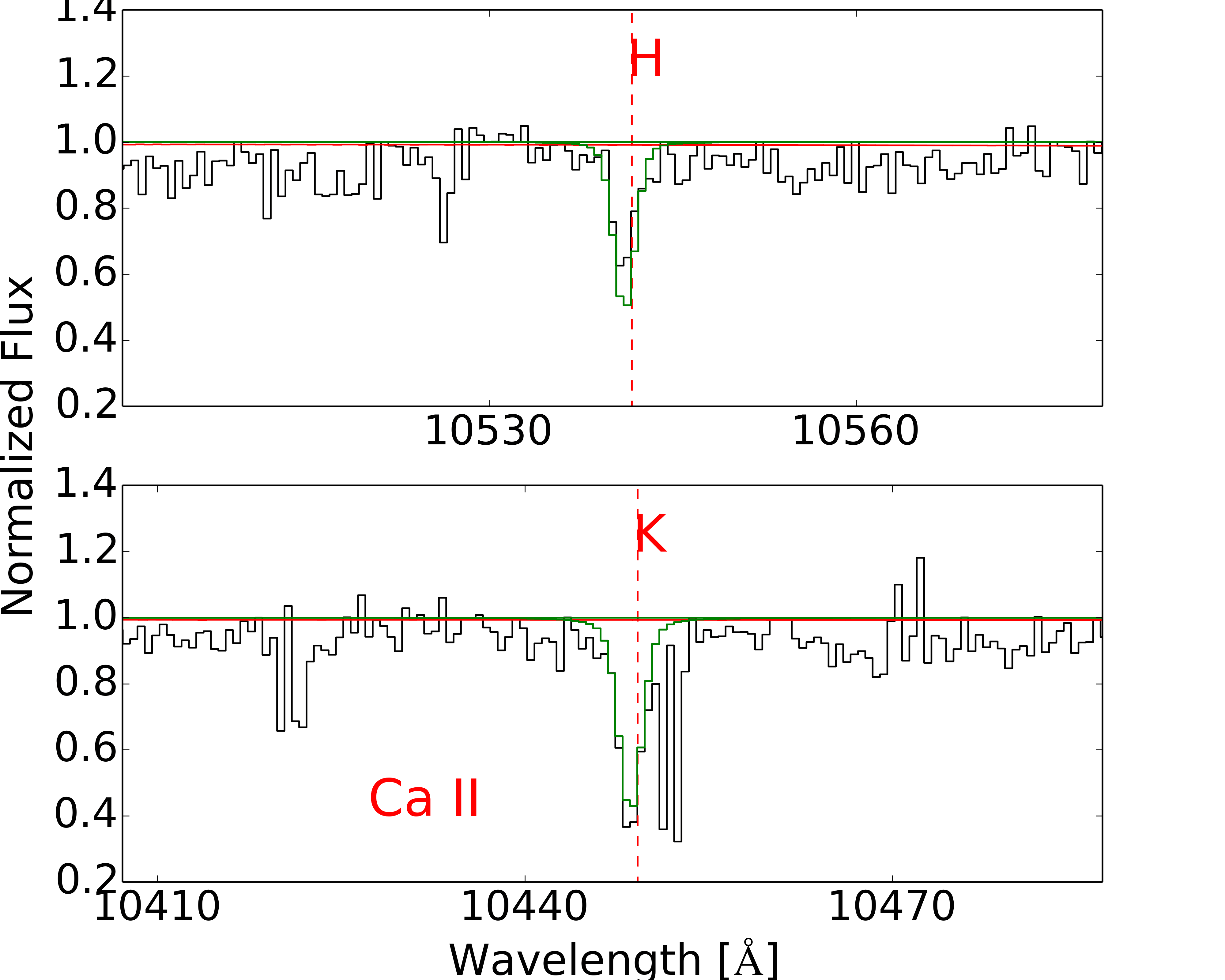}
  \subcaption{}
  \label{1302ca}
\end{minipage}
\caption{J1302+2111 : Same as \ref{0216}.
}
\end{figure*}


\subsection{J1314+0543 $-$ $z_{\rm abs}$ = 1.5828}
There is a clear detection of Na~{\sc i} D in this system. 
The spectrum is affected by a spike near Na~{\sc i}$\lambda$5897, thus we fit the Na~{\sc i}$\lambda$5891 
line and show the corresponding Na~{\sc i}$\lambda$5897 line.
The $EW$ values in Table \ref{naca} are from the VPFIT fit. 
There is not apparent detection of Ca~{\sc ii}  and we give upper limits on $EW$. There are two strong 
components at $v=0$, 160 km/s in the Fe~{\sc ii}, Si~{\sc ii}, C~{\sc i,} and Mg~{\sc i} absorption profiles. 
The spike in the spectrum could be due at least partly to Na~{\sc i}$\lambda$5891 in the second component, the corresponding Na~{\sc i}$\lambda$5897 being affected by a residual from sky subtraction.
We detect Zn~{\sc ii}, Mn~{\sc ii} and Ni~{\sc ii} , with depletion factor [Fe/Zn]~=~$-$0.95$\pm$0.05.

\begin{figure*}
  \begin{minipage}[c][10cm][t]{.5\textwidth}
  \vspace*{\fill}
  \centering
  \includegraphics[width=7cm]{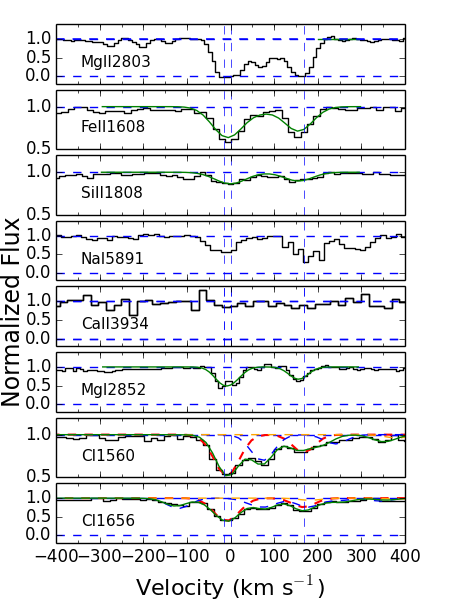}
  \subcaption{}
  \label{1314v}
\end{minipage}%
\begin{minipage}[c][10cm][t]{.5\textwidth}
  \vspace*{\fill}
  \centering
  \includegraphics[width=6.2cm]{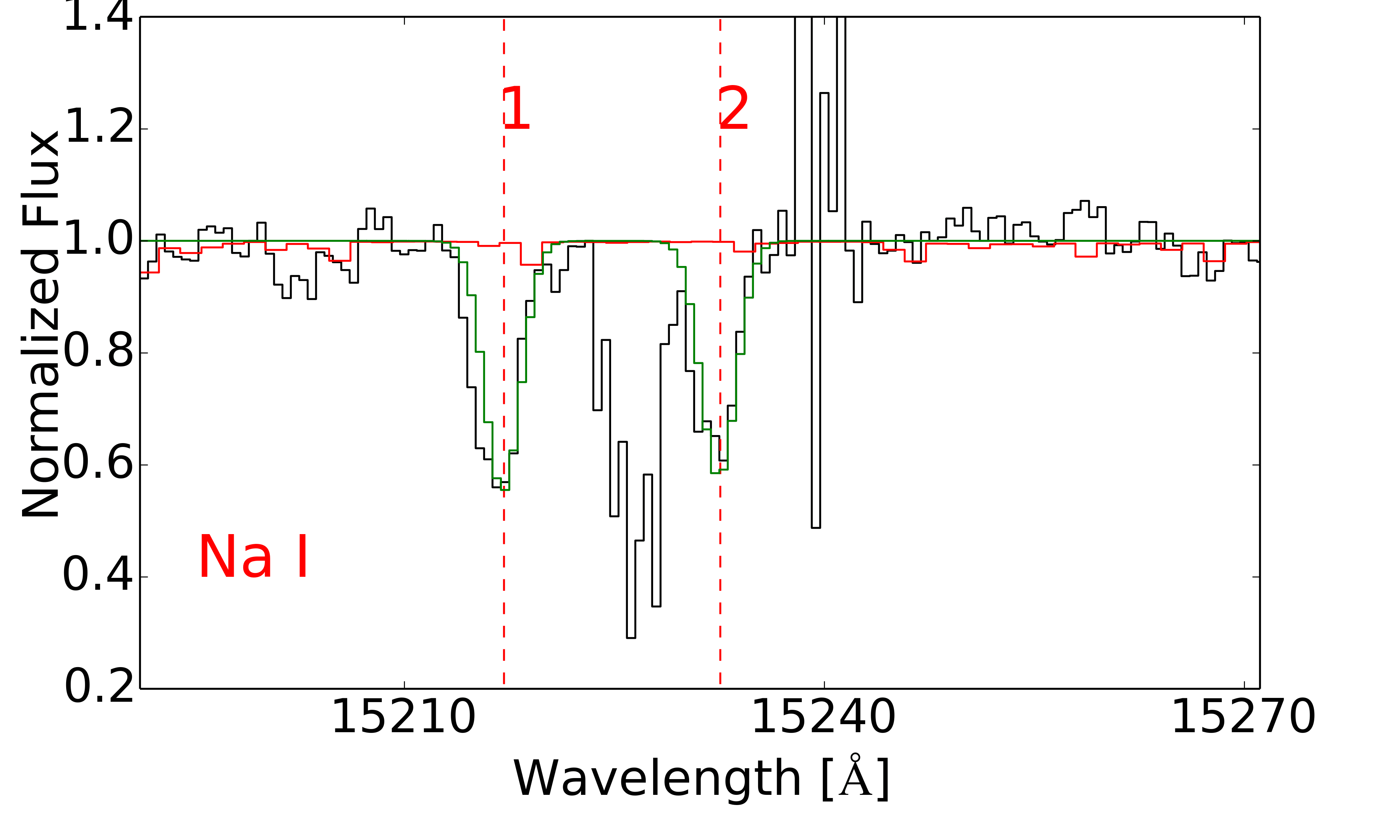}
  \label{1314na}\par\vfill
  \includegraphics[width=6.4cm]{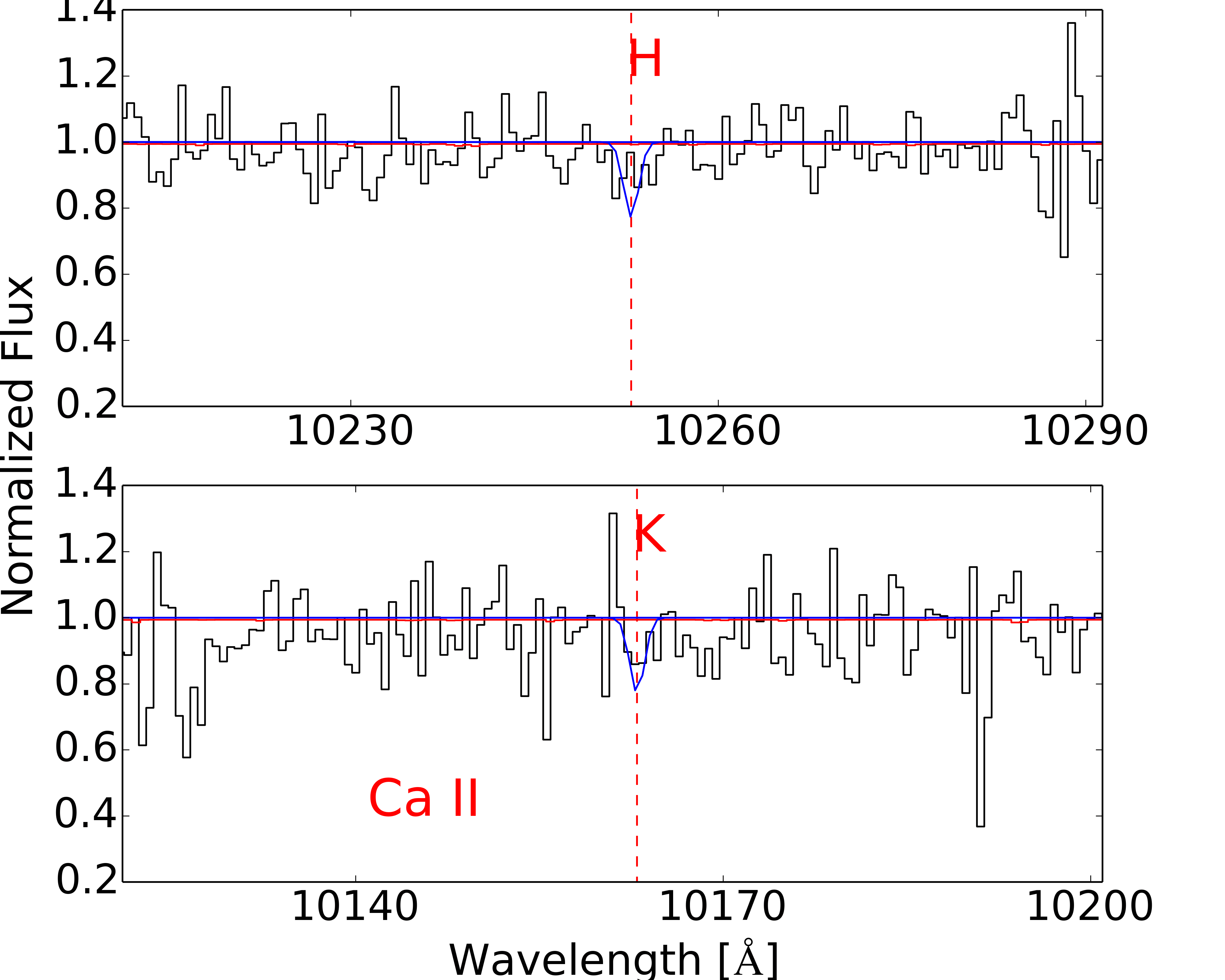}
  \subcaption{}
  \label{1314ca}
\end{minipage}
\caption{J1314+0543 : Same as \ref{0216}.
}
\end{figure*}

\subsection{J1341+1852 $-$ $z_{\rm abs}$ = 1.5442}
This system is peculiar as it turns out to be 
a sub-DLA, with column density log~$N$(HI)~=~18.18$\pm$0.05.
The C~{\sc i} and other metal line absorptions are weak, with C~{\sc i}$\lambda$1560 equivalent width 
$\sim$0.13~\AA~ (well below the mean $EW$ = 0.38~\AA~ in the sample), 
and $W$(Mg~{\sc ii}$\lambda$2798)~=~0.33~\AA~ (for a mean EW~=~2.76~\AA~ in the sample).
The metallicity is high with [OI/HI] = +0.36, relative to solar.
It might be interesting to derive the physical properties of this gas and 
in particular its ionization state. Indeed, it is surprising to detect C~{\sc i} in such a system.
This is, however, out of the scope of the present paper.
There are no obvious detections of either Ca~{\sc ii} H$\&$K or Na~{\sc i} D. 
We give $EW$ upper limits for the lines.

\begin{figure*}
  \begin{minipage}[c][10cm][t]{.5\textwidth}
  \vspace*{\fill}
  \centering
  \includegraphics[width=7cm]{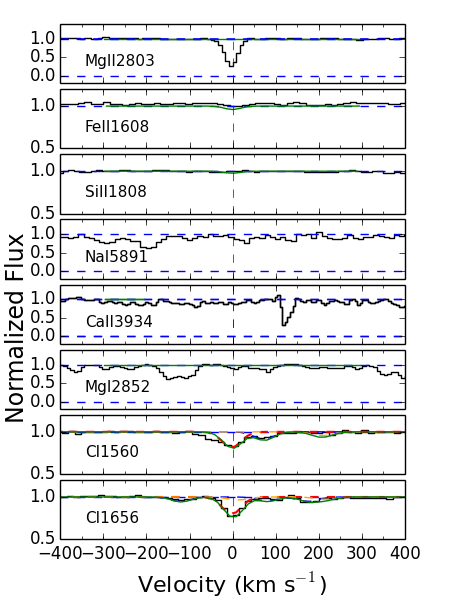}
  \subcaption{}
  \label{1341v}
\end{minipage}%
\begin{minipage}[c][10cm][t]{.5\textwidth}
  \vspace*{\fill}
  \centering
  \includegraphics[width=6.2cm]{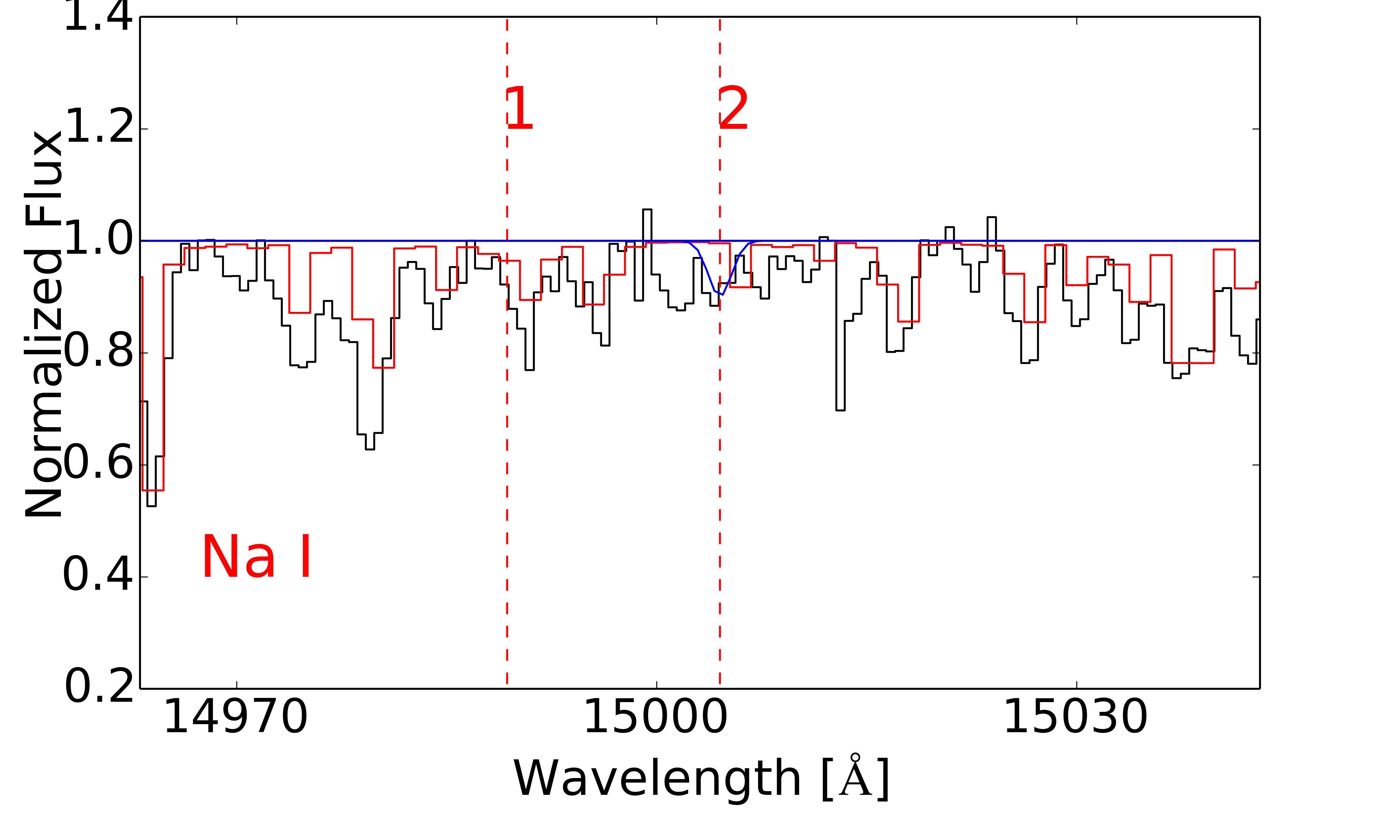}
  \label{1341na}\par\vfill
  \includegraphics[width=6.4cm]{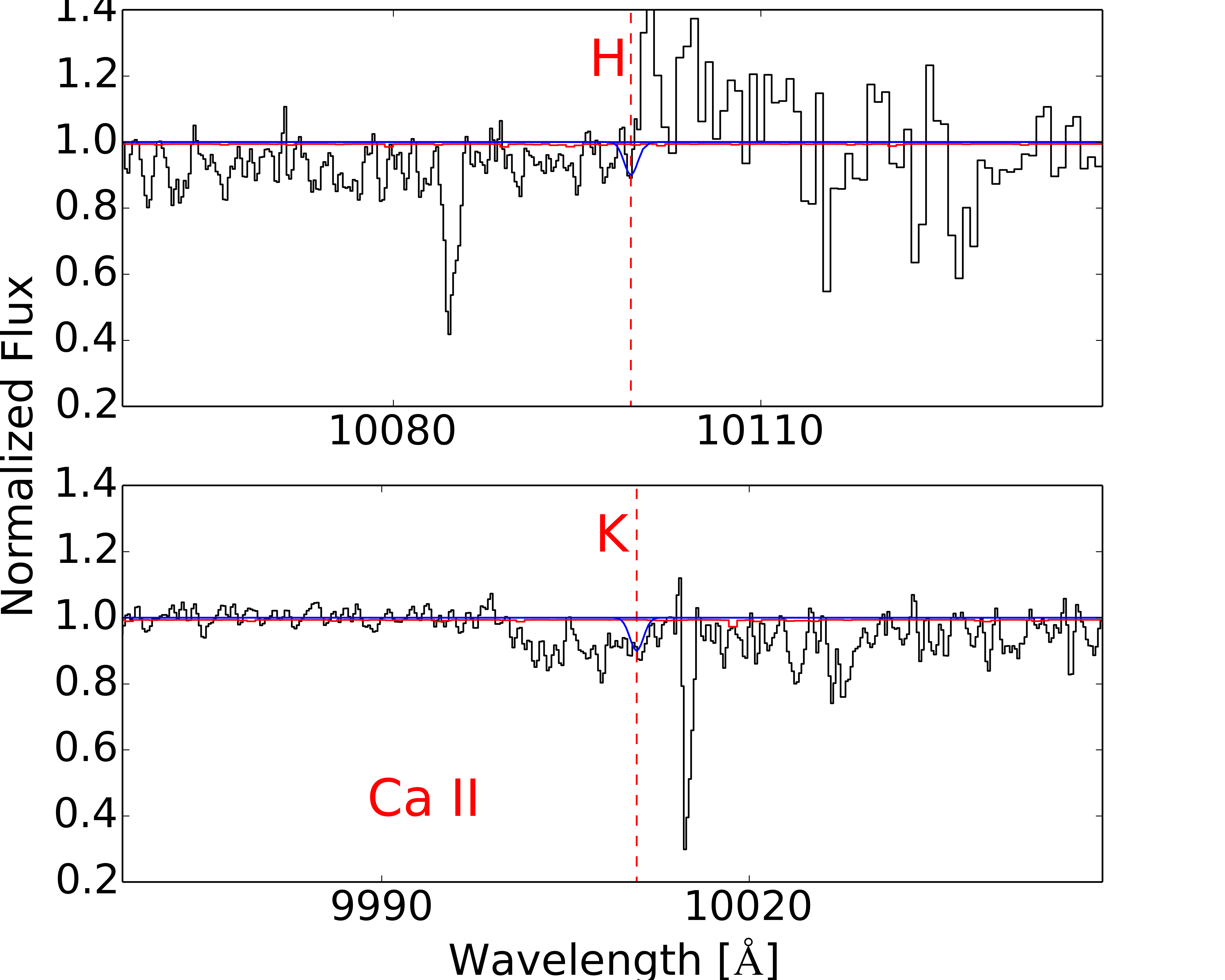}
  \subcaption{}
  \label{1341ca}
\end{minipage}
\caption{J1341+1852 : Same as \ref{0216}.
}
\end{figure*}

\subsection{J1346+0644 $-$ $z_{\rm abs}$ = 1.511938}
The Na~{\sc i} D line is strongly blended with sky features so that it is difficult to measure even
an upper limit. We did not detect Ca~{\sc ii} i. 
The C~{\sc i} absorptions are fitted well with two components at $z = 1.511938$ and 
$z = 1.512393$  ($\Delta v$~$\sim$~50~km~s$^{-1}$). The Mg~{\sc i} absorption is extended over
$\sim$200~km~s$^{-1}$. 
The strongest components of Fe~{\sc ii} and Si~{\sc ii} are
blueshifted by $\sim$100 km$s^{-1}$ relative to the strongest components of C~{\sc i} and Mg~{\sc i}, 
which may indicate there is one more component of C~{\sc i} at $-$100 km$s^{-1}$. 
Since there is basically no flux in the spectrum at wavelengths smaller than 3200~\AA, 
we cannot derive the H~{\sc i} column density. 

\begin{figure*}
  \begin{minipage}[c][10cm][t]{.5\textwidth}
  \vspace*{\fill}
  \centering
  \includegraphics[width=7cm]{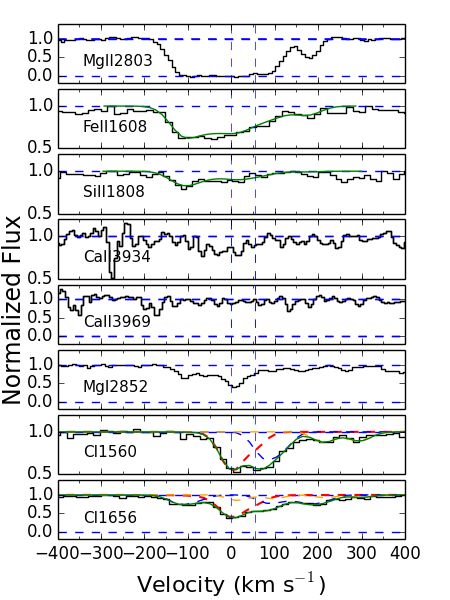}
  \subcaption{}
  \label{1346v}
\end{minipage}%
\begin{minipage}[c][10cm][t]{.5\textwidth}
  \vspace*{\fill}
  \centering
  \includegraphics[width=6.2cm]{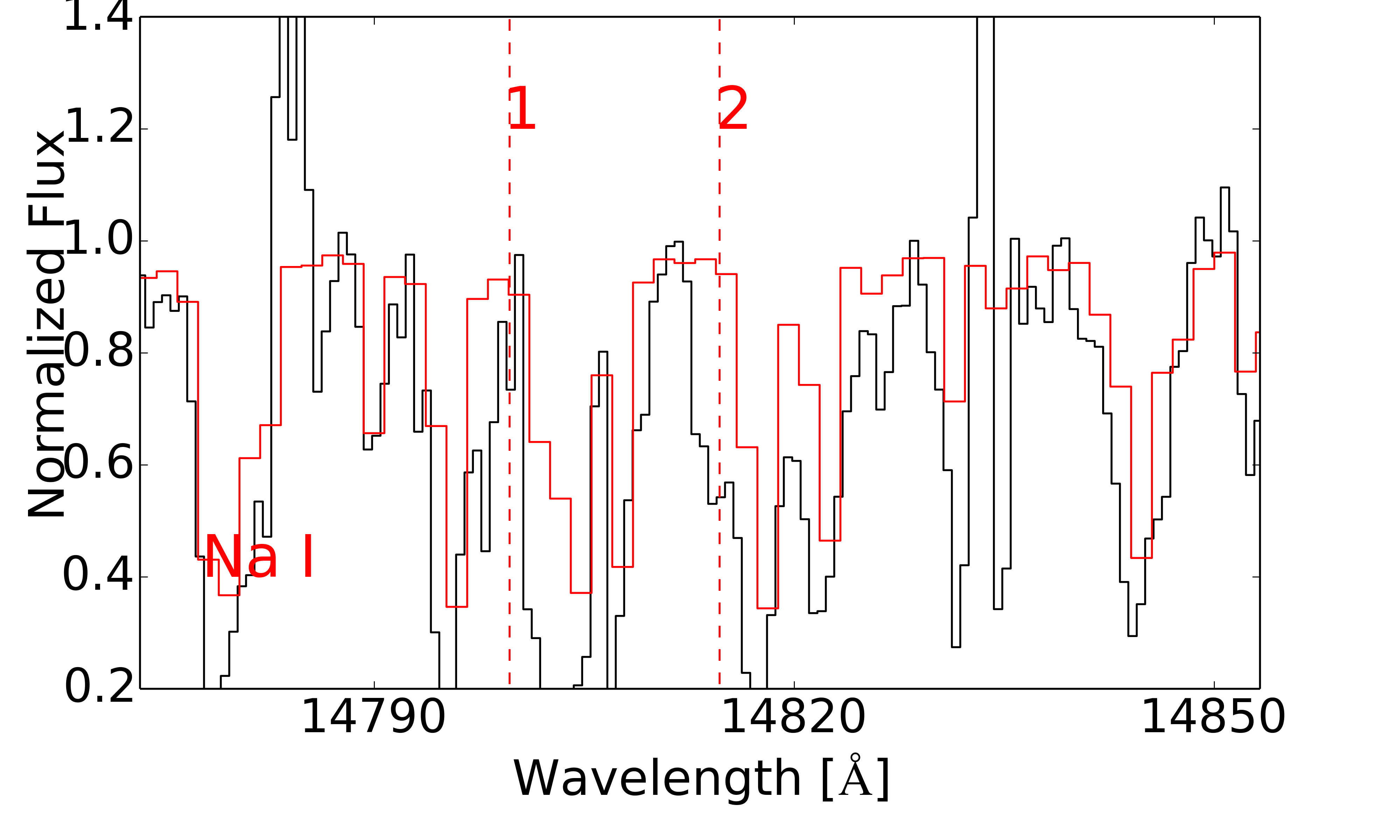}
  \label{1346na}\par\vfill
  \includegraphics[width=6.4cm]{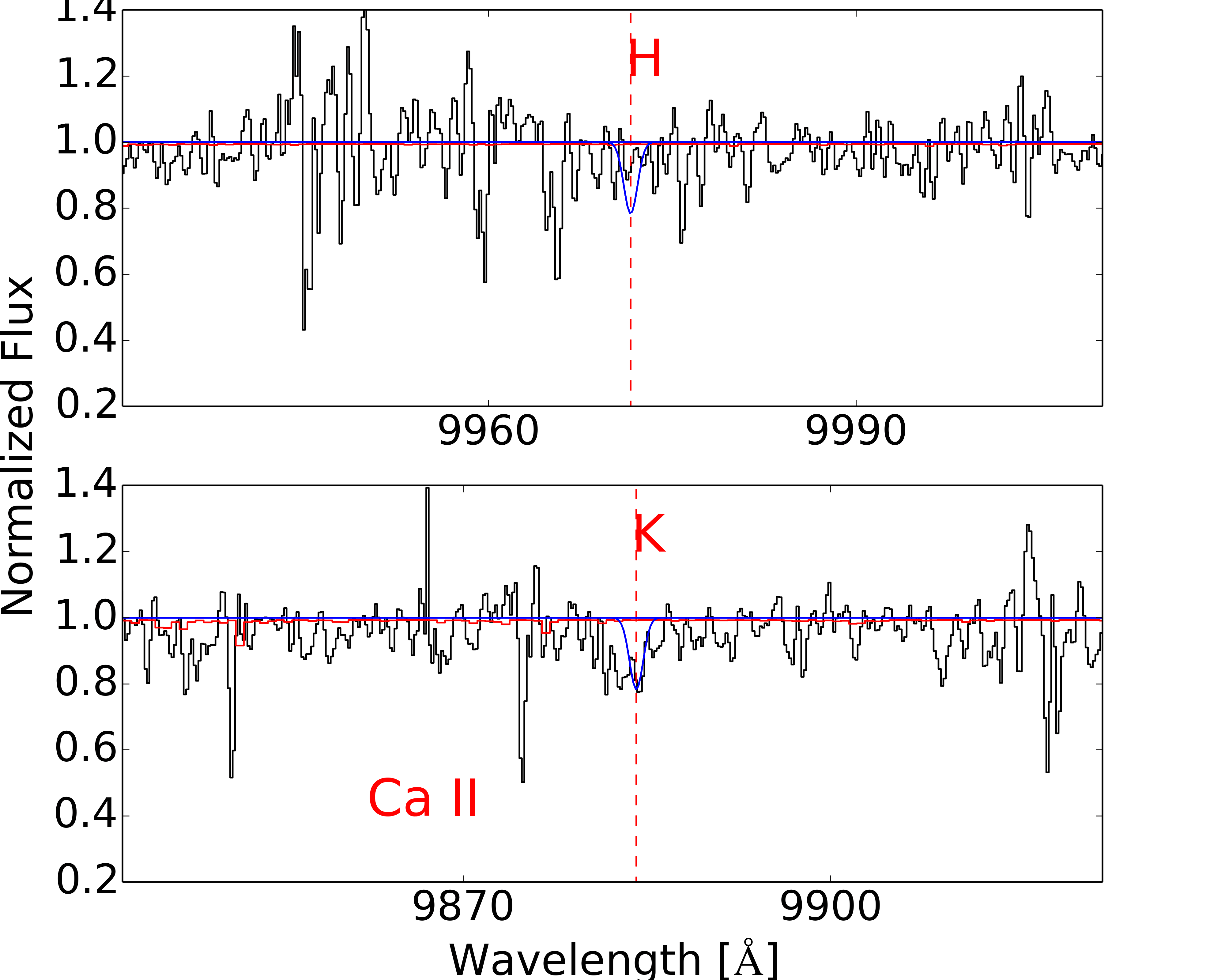}
  \subcaption{ }
  \label{1346ca}
\end{minipage}
\caption{J1346+0644 : Same as \ref{0216}.
}
\end{figure*}

\subsection{J2229+1414 $-$ $z_{\rm abs}$ = 1.585372}
For this system Na~{\sc i} D is clearly detected and strong.
We derive upper limits for the Ca~{\sc ii} lines (0.55 $\AA$). The metal lines 
are relatively weak compared to other systems.
The C~{\sc i} and Mg~{\sc i} main features are located on the blue edge of the Mg~{\sc ii} profile,
which is $\sim$200 km~s$^{-1}$ wide.

\begin{figure*}
  \begin{minipage}[c][10cm][t]{.5\textwidth}
  \vspace*{\fill}
  \centering
  \includegraphics[width=7cm]{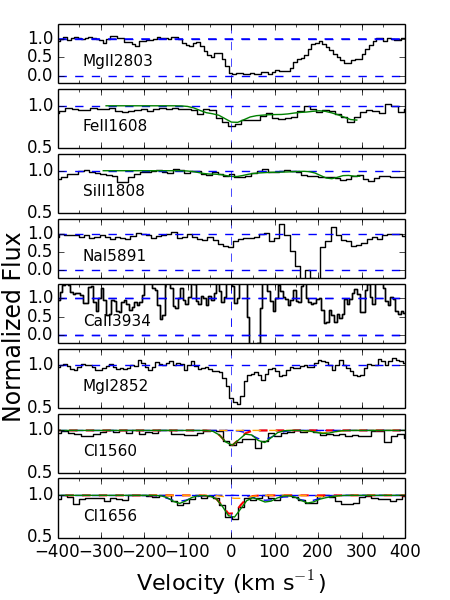}
  \subcaption{}
  \label{2229v}
\end{minipage}%
\begin{minipage}[c][10cm][t]{.5\textwidth}
  \vspace*{\fill}
  \centering
  \includegraphics[width=6.2cm]{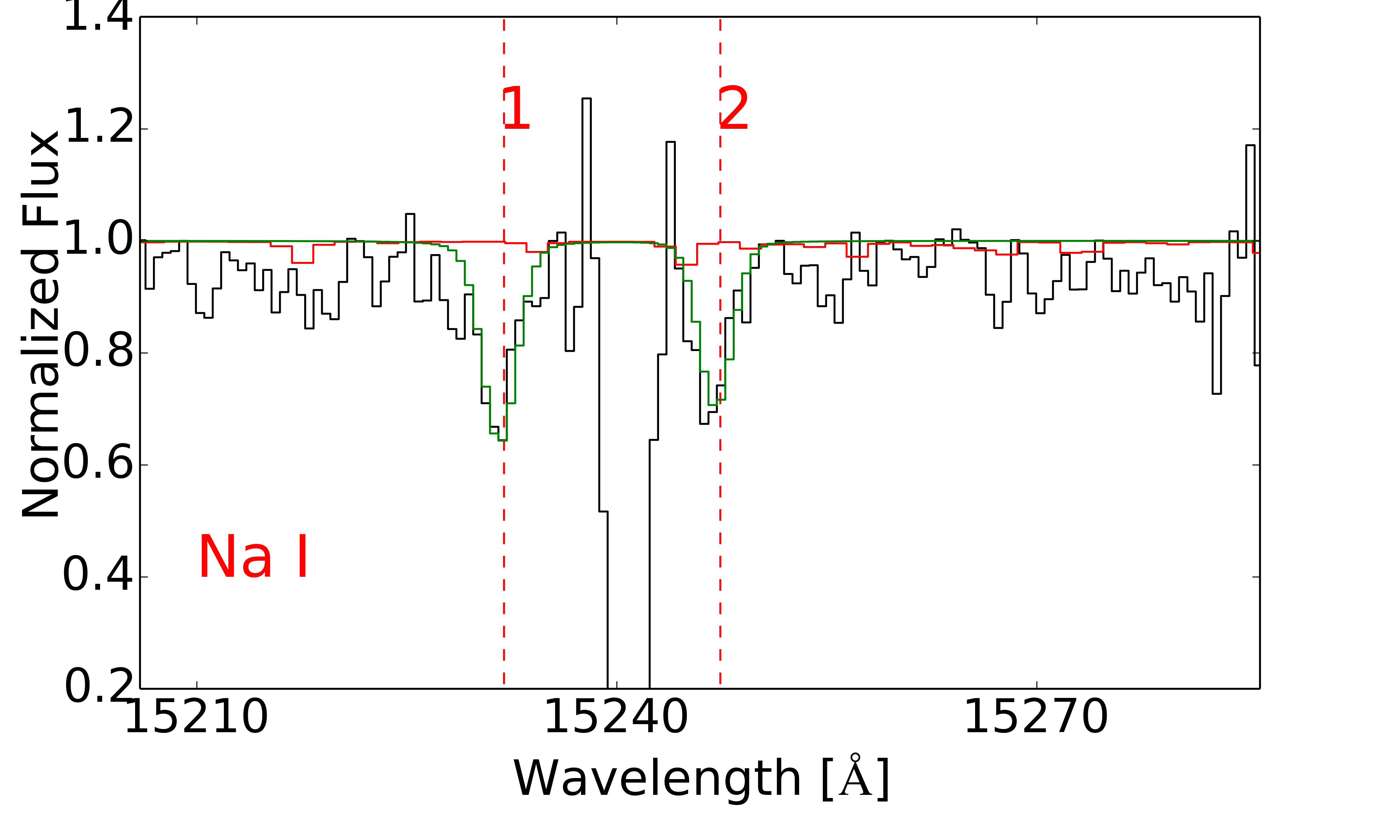}
  \label{2229na}\par\vfill
  \includegraphics[width=6.4cm]{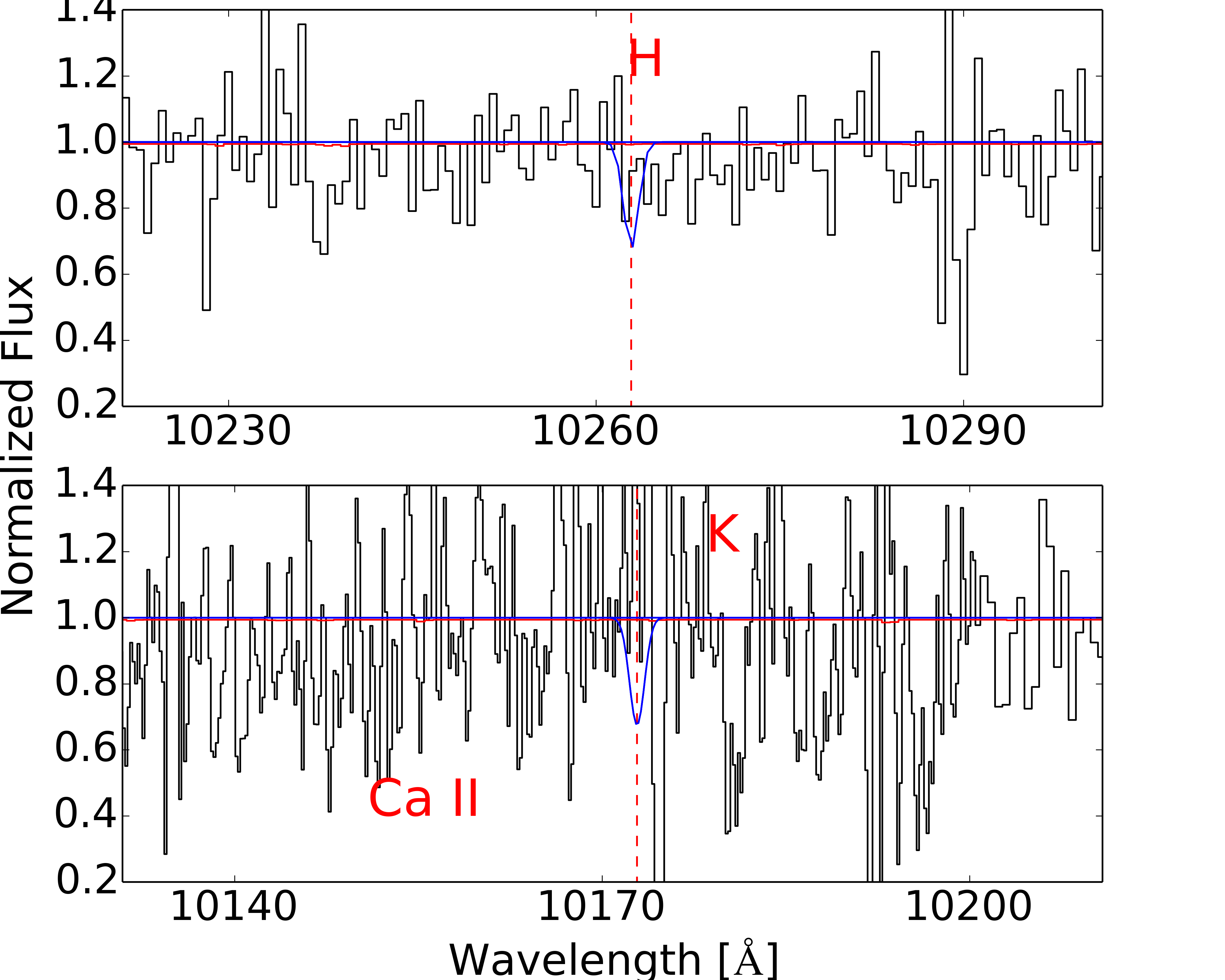}
  \subcaption{}
  \label{2229ca}
\end{minipage}
\caption{J2229+1414 : Same as \ref{0216}.
}
\end{figure*}

\subsection{J2336-1058 $-$ $z_{\rm abs}$=1.828723}
For this system, the Ca~{\sc ii} doublet is lost in strong sky absorption features.
We detect Na~{\sc i} D. The Na~{\sc i}$\lambda$5891 line is slightly blended with a sky absorption, 
which is subtracted to derive $EW$. 
The C~{\sc i} absorptions are not very strong but we can see two components.
The Zn~{\sc ii} line is detected and we derive a depletion factor [Fe/Zn]~=~$-$1.11$\pm$0.10. 

\begin{figure*}
  \begin{minipage}[c][10cm][t]{.5\textwidth}
  \vspace*{\fill}
  \centering
  \includegraphics[width=7cm]{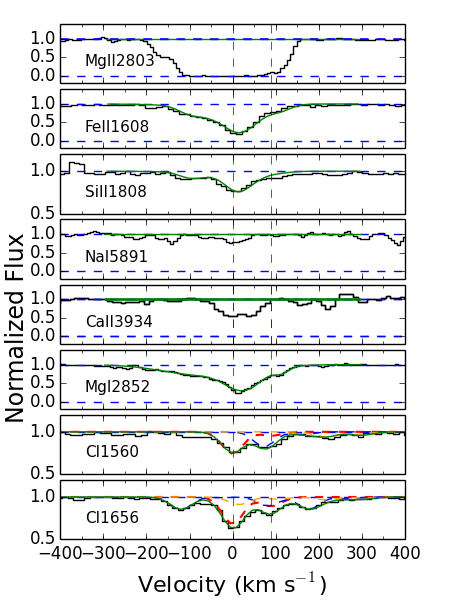}
  \subcaption{Velocity profile}
  \label{2336v}
\end{minipage}%
\begin{minipage}[c][10cm][t]{.5\textwidth}
  \vspace*{\fill}
  \centering
  \includegraphics[width=6.2cm]{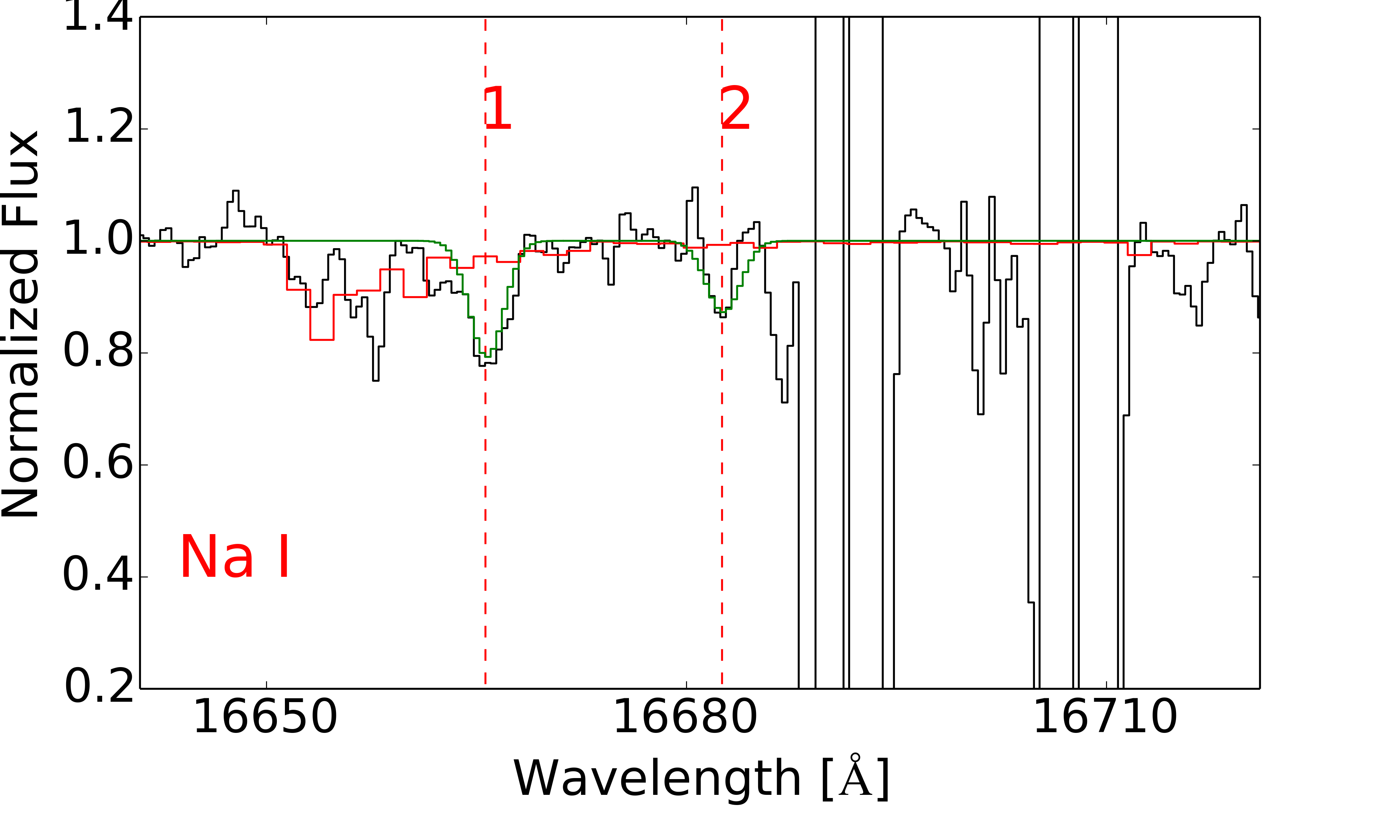}
  \label{2336na}\par\vfill
  \includegraphics[width=6.4cm]{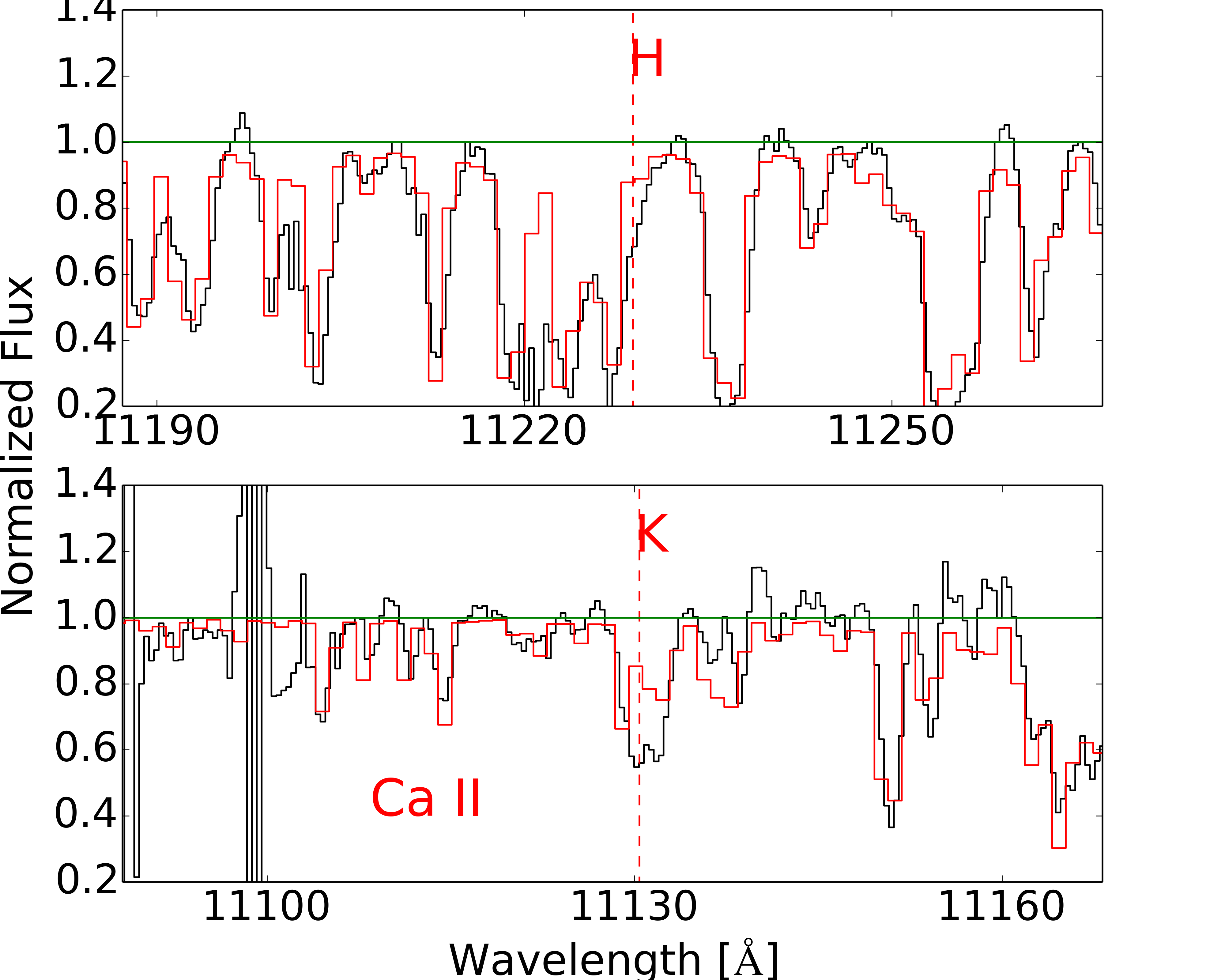}
  \subcaption{}
  \label{2336ca}
\end{minipage}
\caption{J2336-1058 : Same as \ref{0216}.
}
\end{figure*}

\subsection{J2340-0053 $-$ $z_{\rm abs}$ = 2.054643}
For this system Na~{\sc i} D is lost in telluric absorptions.
The C~{\sc i} structure is simple with two components. 
The Ca~{\sc ii} doublet is detected although it is blended with weak sky features. We subtracted
the latter.
H2 is detected in this system. It is mentioned in \cite{boi15}, \citet{bal15}, who 
derived a column density of log~$N$(H$_2$)~=~18.07 $\pm$ 0.06. 

\begin{figure*}
  \begin{minipage}[c][10cm][t]{.5\textwidth}
  \vspace*{\fill}
  \centering
  \includegraphics[width=7cm]{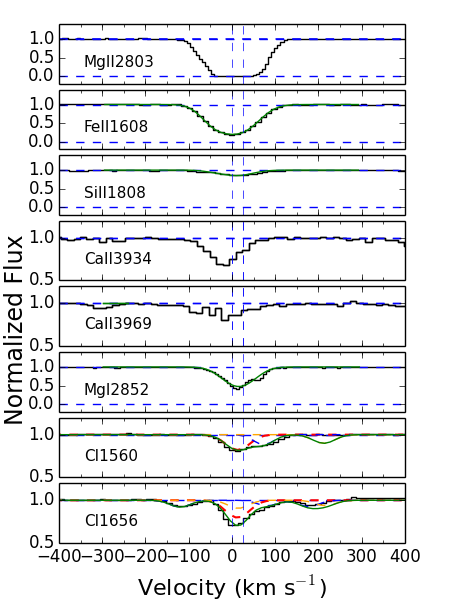}
  \subcaption{}
  \label{2340v}
\end{minipage}%
\begin{minipage}[c][10cm][t]{.5\textwidth}
  \vspace*{\fill}
  \centering
  \includegraphics[width=6.2cm]{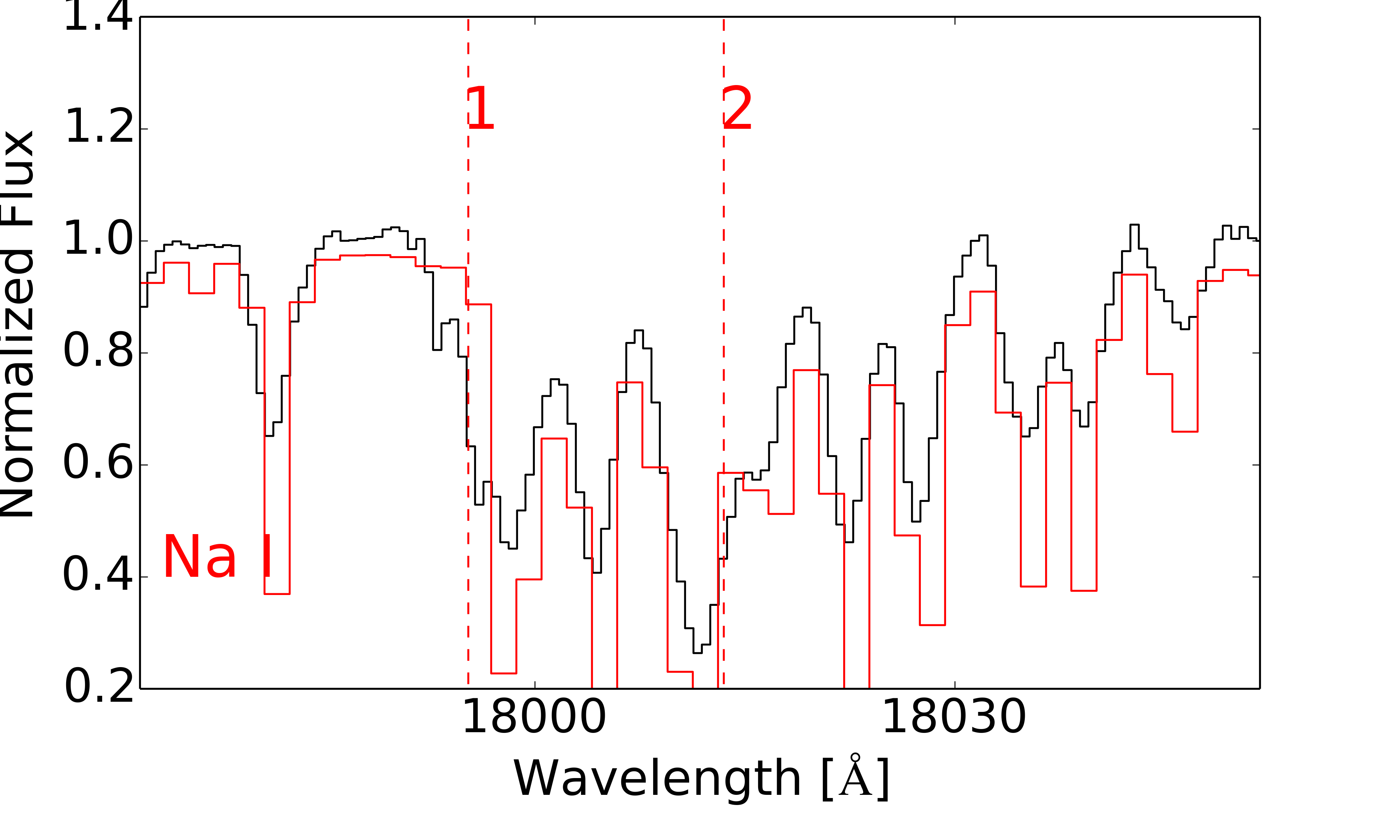}
  \label{2340na}\par\vfill
  \includegraphics[width=6.4cm]{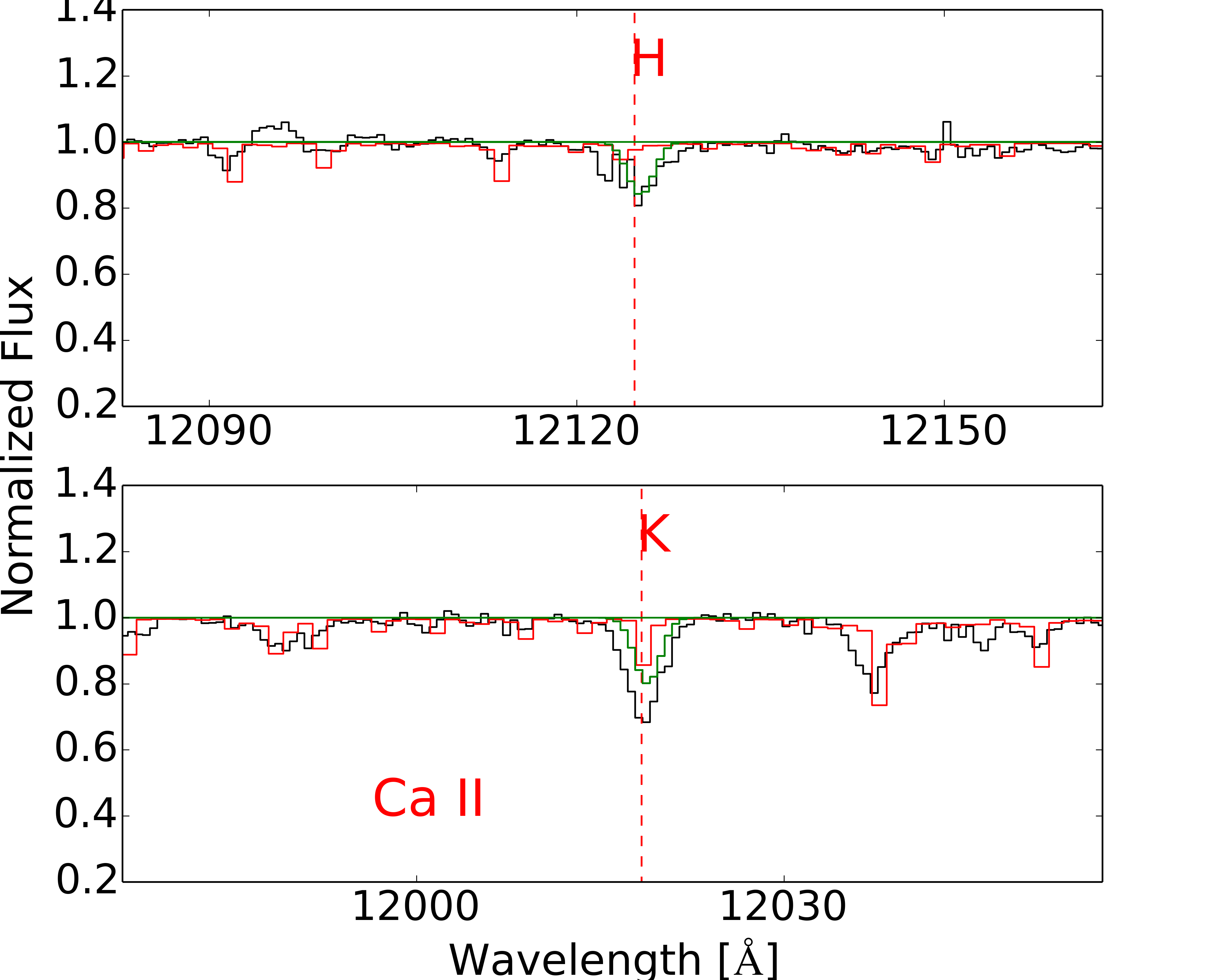}
  \subcaption{}
  \label{2340ca}
\end{minipage}
\caption{J2340-0053 : Same as \ref{0216}.
}
\end{figure*}

\subsection{J2350-0052 $-$ $z_{\rm abs}$ = 2.426475}
Unfortunately both Na~{\sc i} D and Ca~{\sc ii} H$\&$K are lost in telluric absorption features.
The Fe~{\sc ii}$\lambda$1608 and Fe~{\sc ii}$\lambda$1611 lines seem to be blended with other lines. 
There is a $\sim$0.4~\AA~ shift between the UVB and VIS arms that we corrected manually.

\begin{figure*}
  \begin{minipage}[c][10cm][t]{.5\textwidth}
  \vspace*{\fill}
  \centering
  \includegraphics[width=6.6cm]{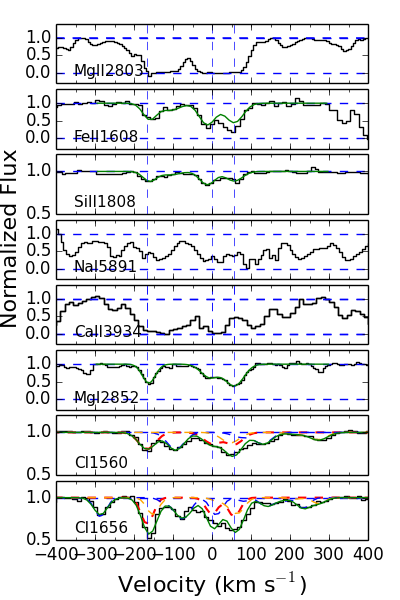}
  \subcaption{}
  \label{2350v}
\end{minipage}%
\begin{minipage}[c][10cm][t]{.5\textwidth}
  \vspace*{\fill}
  \centering
  \includegraphics[width=6.2cm]{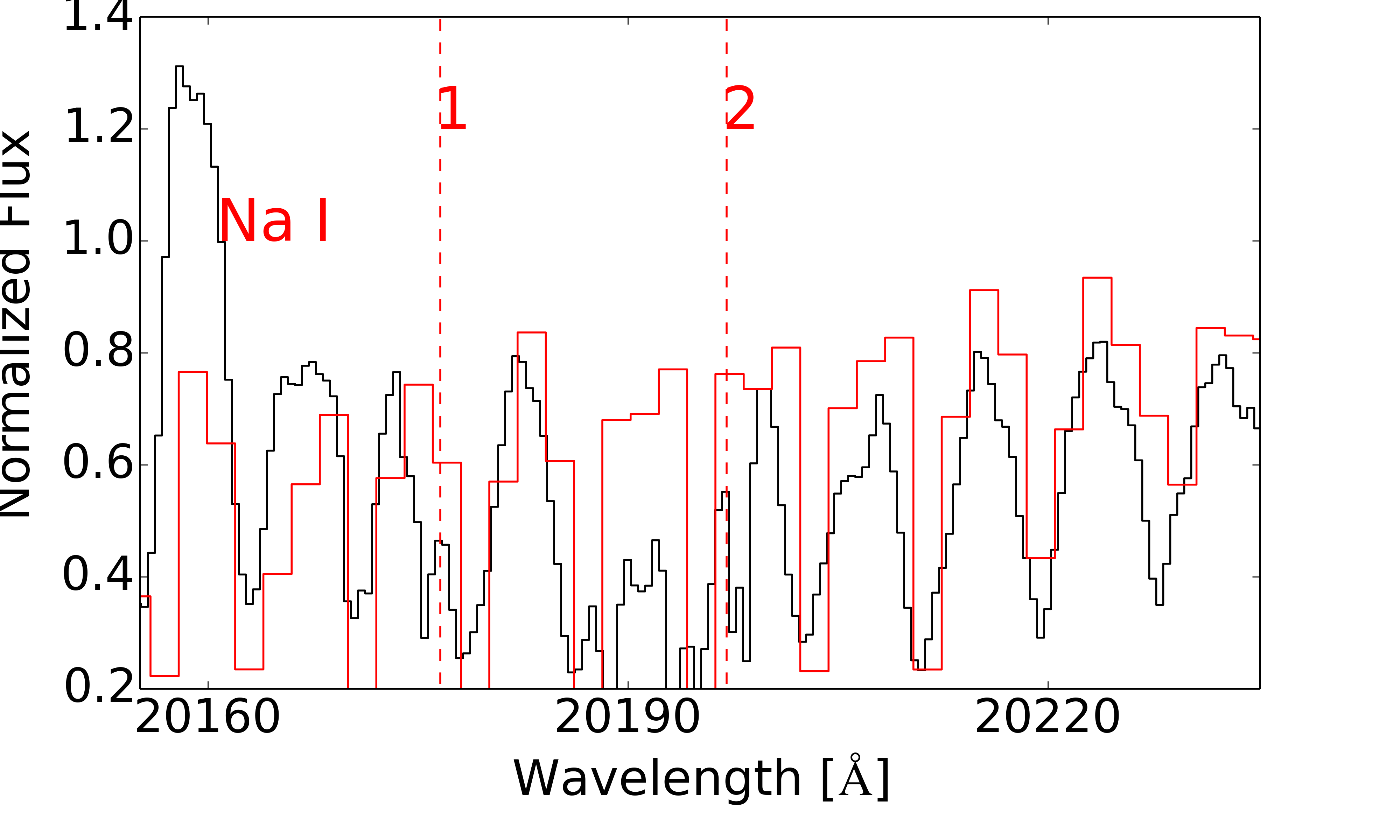}
  \label{2350na}\par\vfill
  \includegraphics[width=6.4cm]{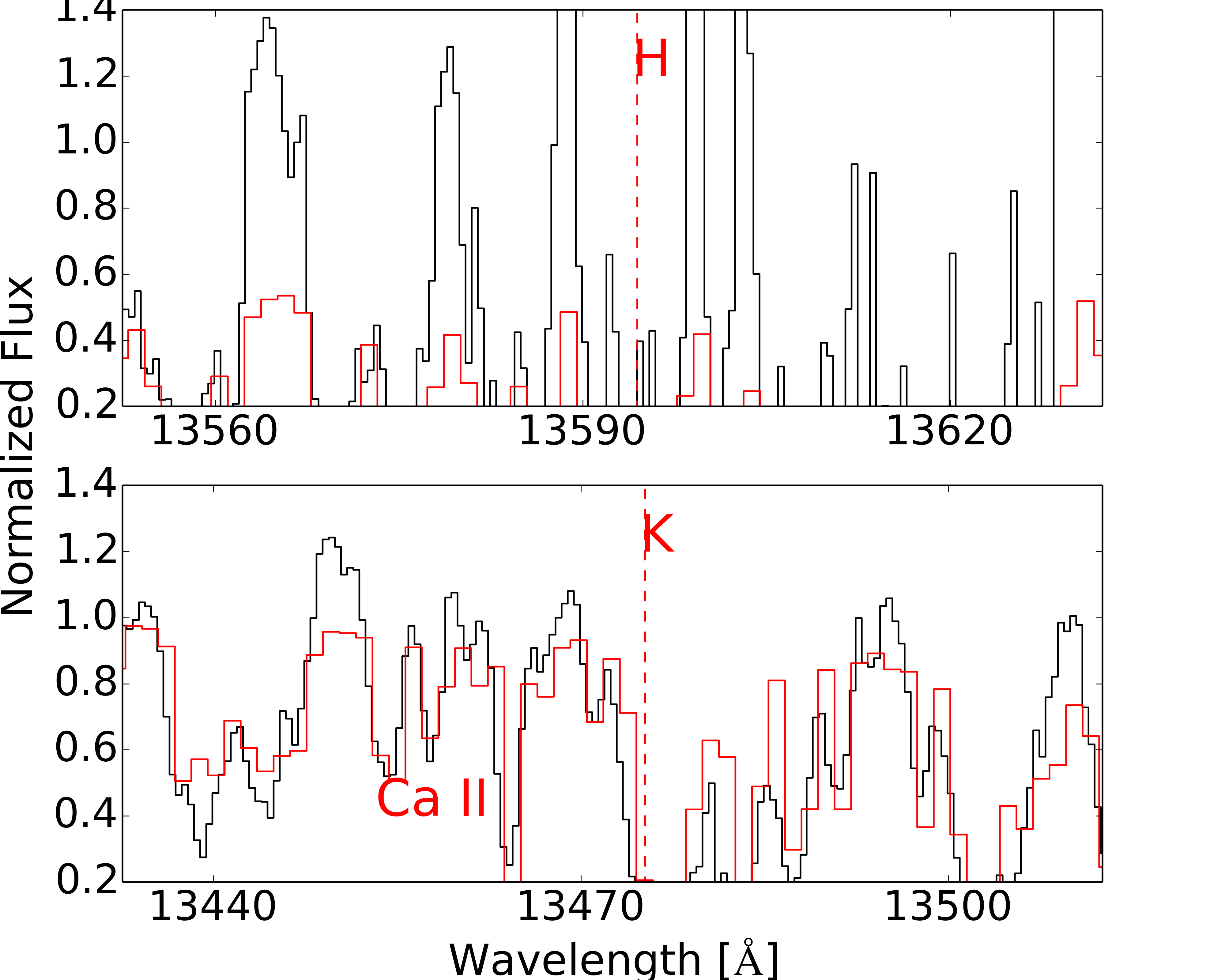}
  \subcaption{}
  \label{2350ca}
\end{minipage}
\caption{J2350-0052 : Same as \ref{0216}
}
\end{figure*}

%


\clearpage

\section{E(B-V)}\label{appendixebv}
X-shooter spectra of quasars in our sample. The red curve corresponds
to the quasar template of \citet{sel16} reddened by different
extinction curves. See Table \ref{tableebv} for results.

%

\begin{figure*}[!htb]
\minipage{0.32\textwidth}
  \includegraphics[width=\linewidth]{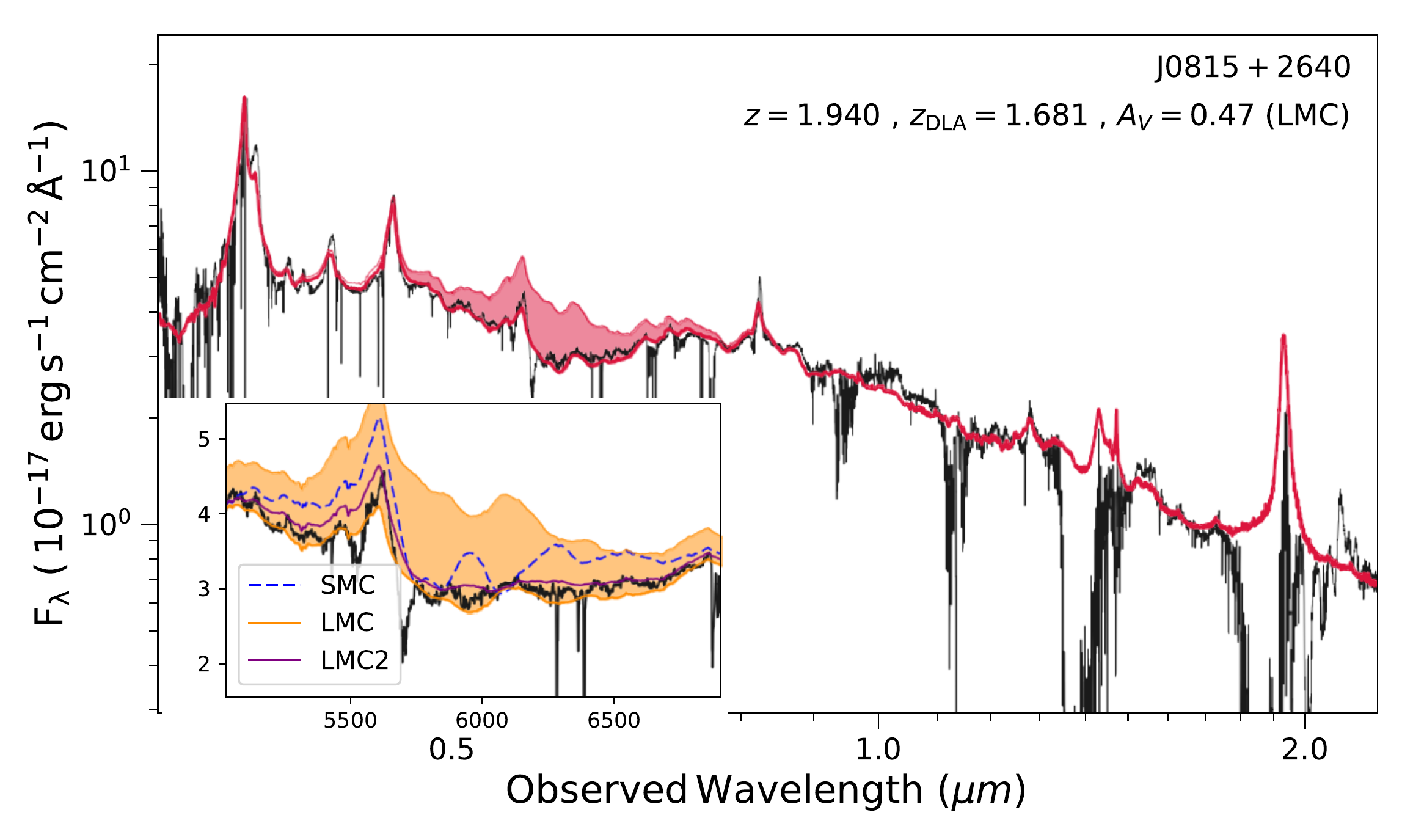}
  \caption{ }\label{J0815ebv}
\endminipage\hfill
\minipage{0.32\textwidth}
  \includegraphics[width=\linewidth]{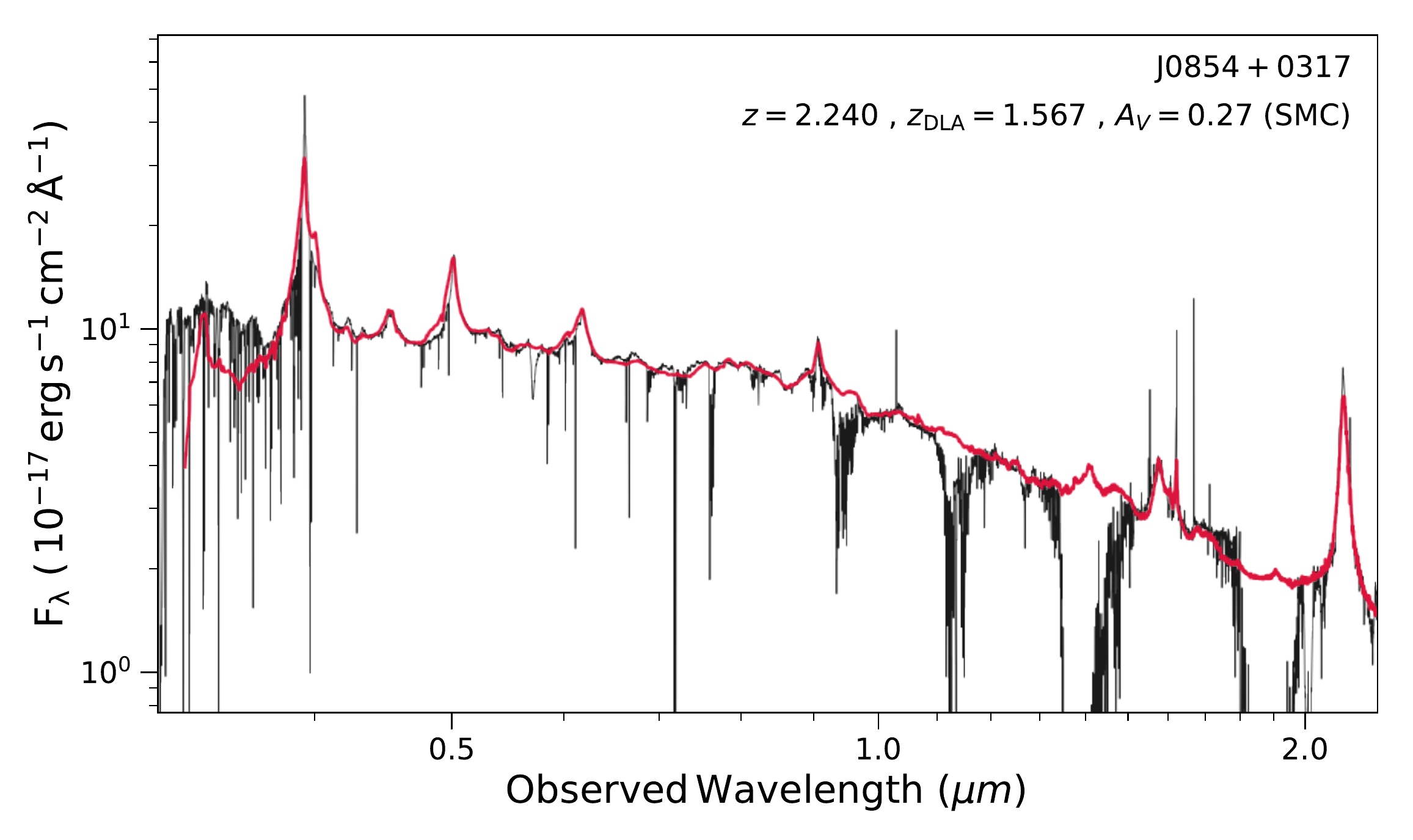}
  \caption{ }\label{J0854ebv}
\endminipage\hfill
\minipage{0.32\textwidth}%
  \includegraphics[width=\linewidth]{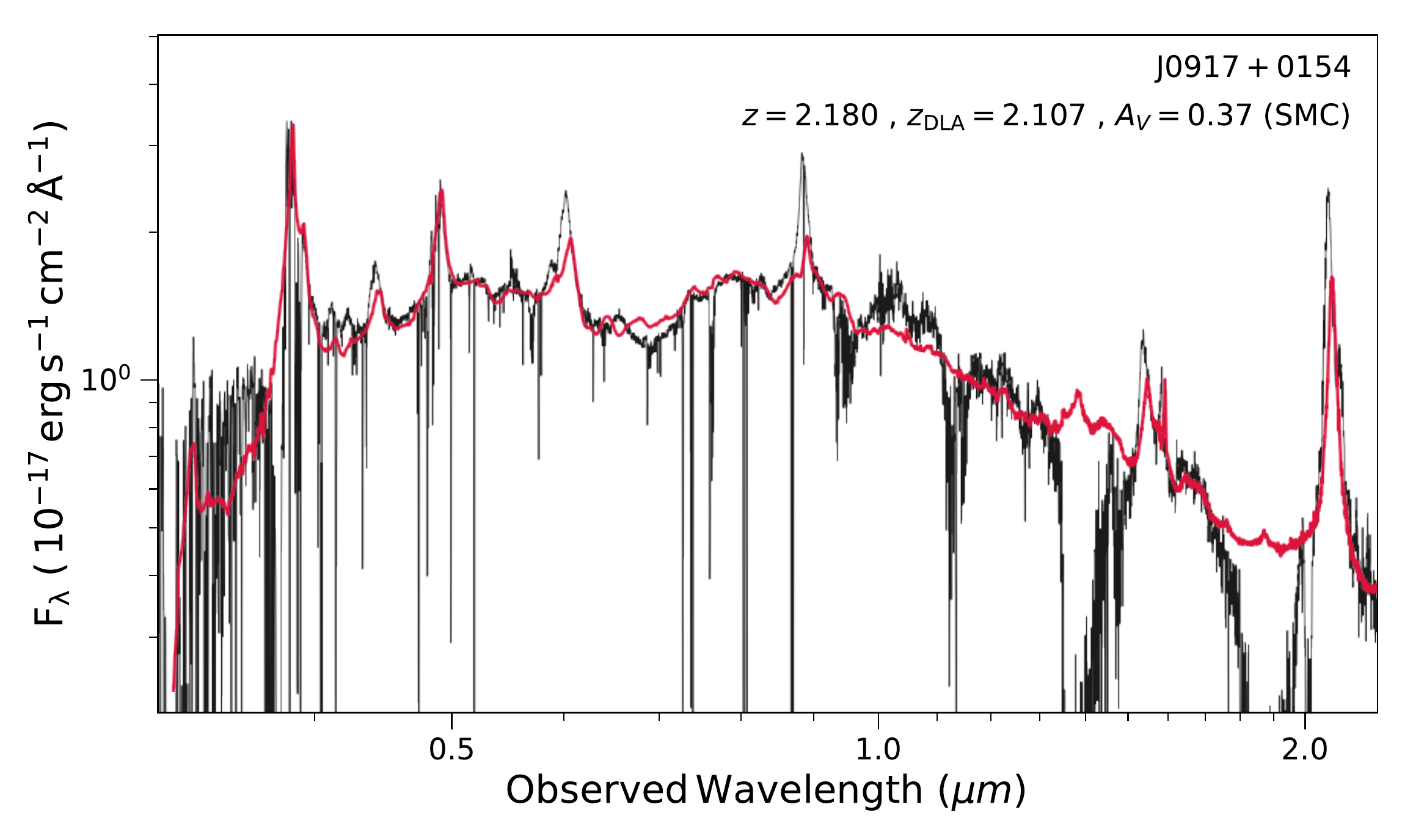}
 \caption{ }\label{J0917ebv}
\endminipage
\end{figure*}

\begin{figure*}[!htb]
\minipage{0.32\textwidth}
  \includegraphics[width=\linewidth]{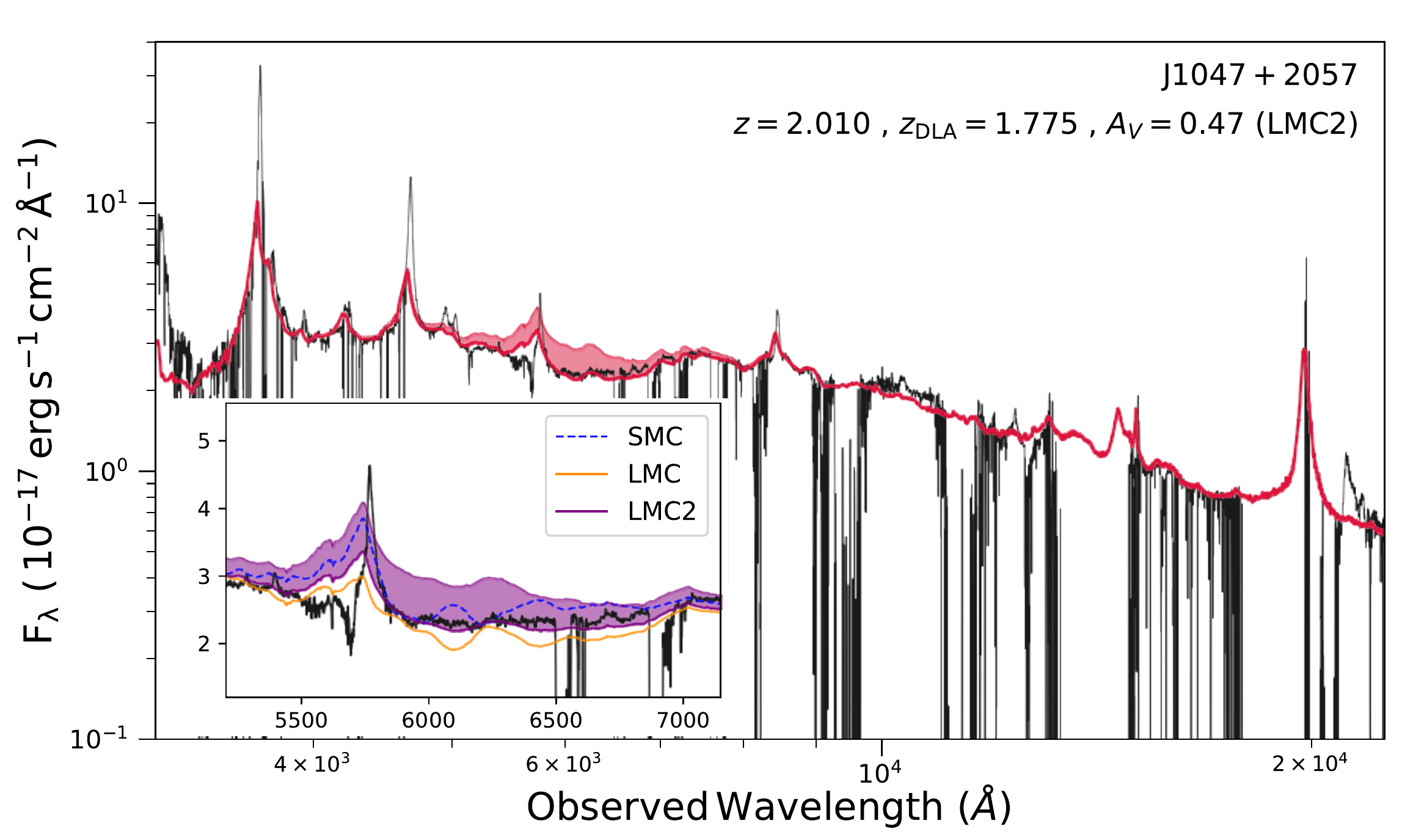}
  \caption{}\label{J1047ebv}
\endminipage\hfill
\minipage{0.32\textwidth}
  \includegraphics[width=\linewidth]{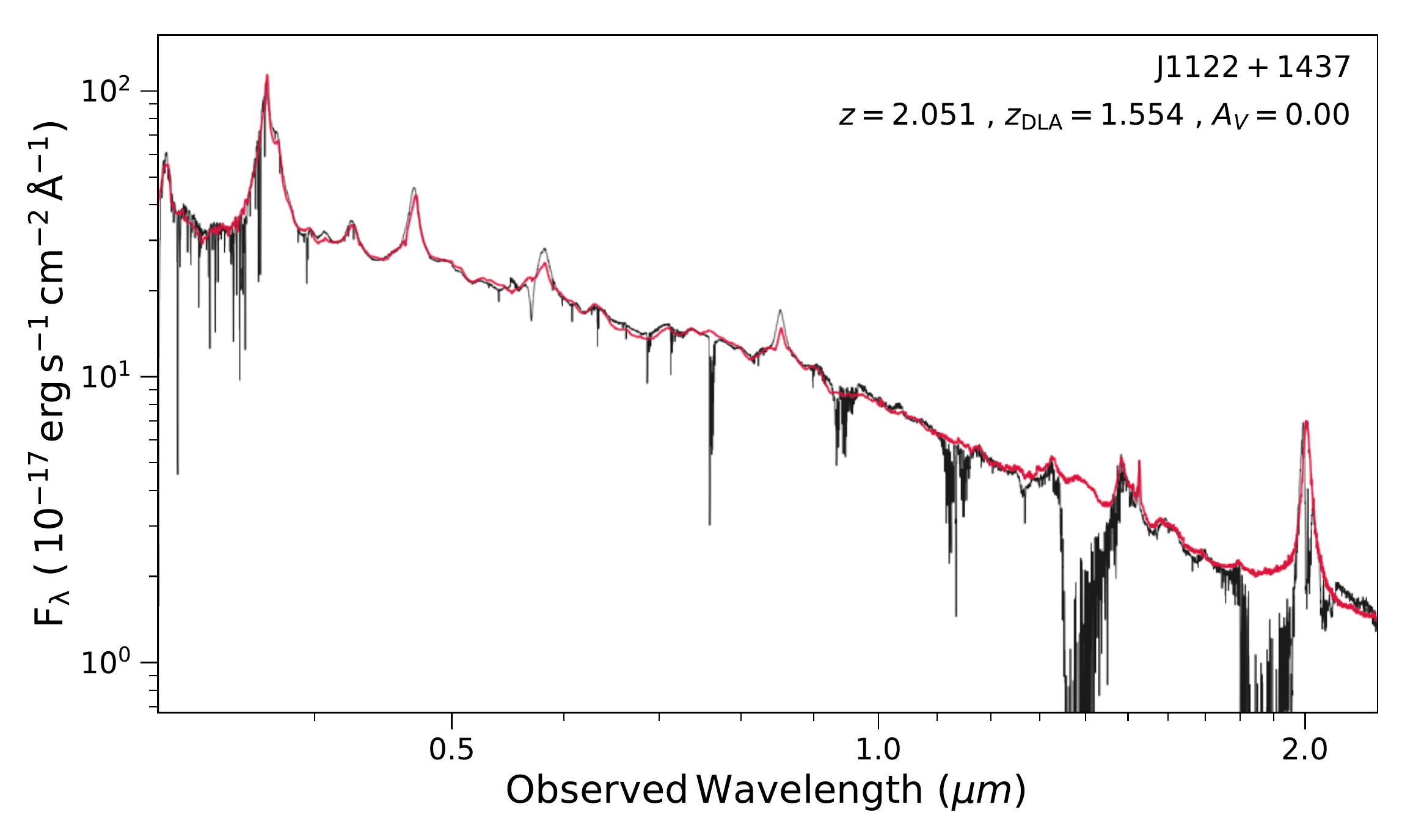}
  \caption{}\label{J1122ebv}
\endminipage\hfill
\minipage{0.32\textwidth}%
  \includegraphics[width=\linewidth]{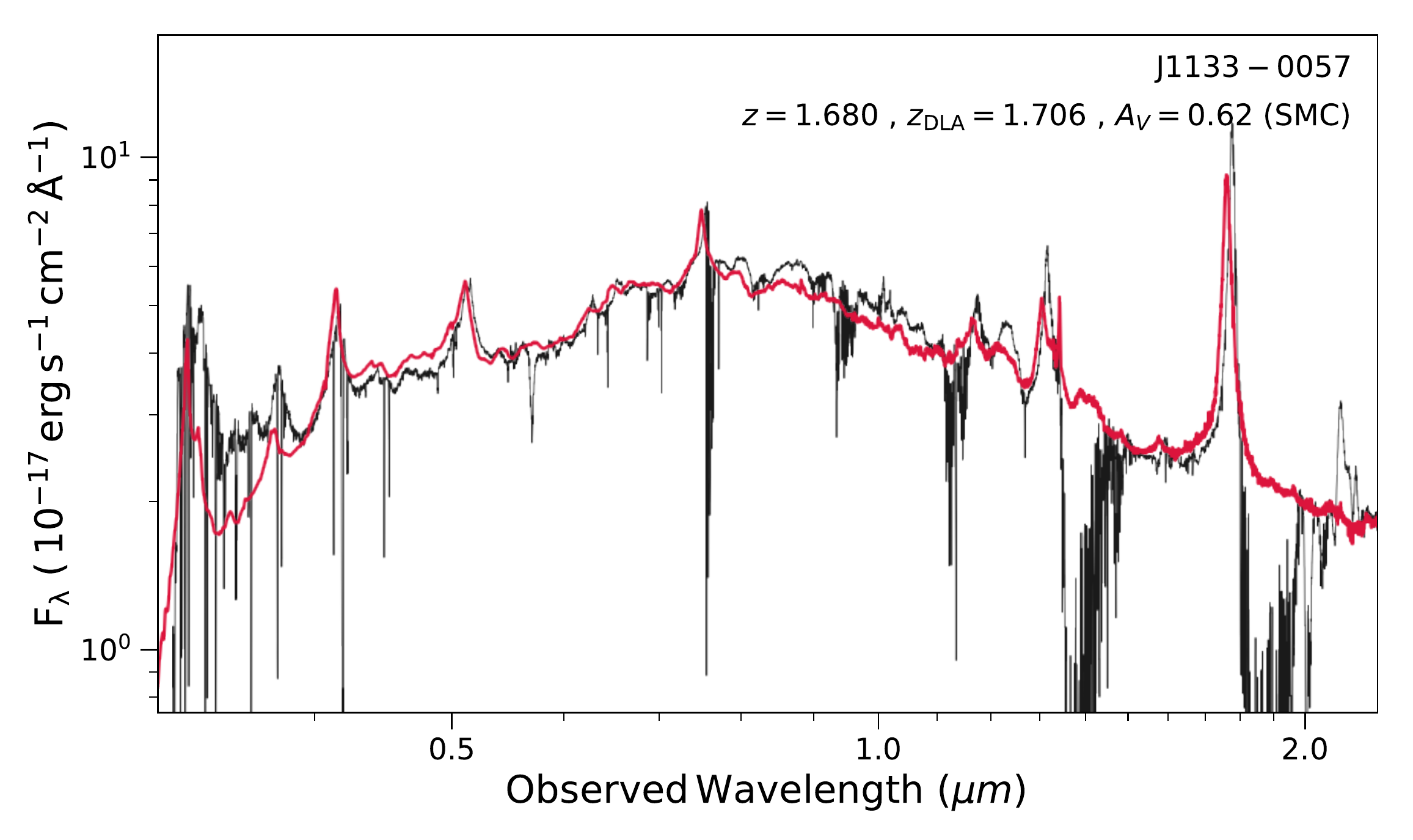}
  \caption{}\label{J1133ebv}
\endminipage
\end{figure*}

\begin{figure*}[!htb]
\minipage{0.32\textwidth}
  \includegraphics[width=\linewidth]{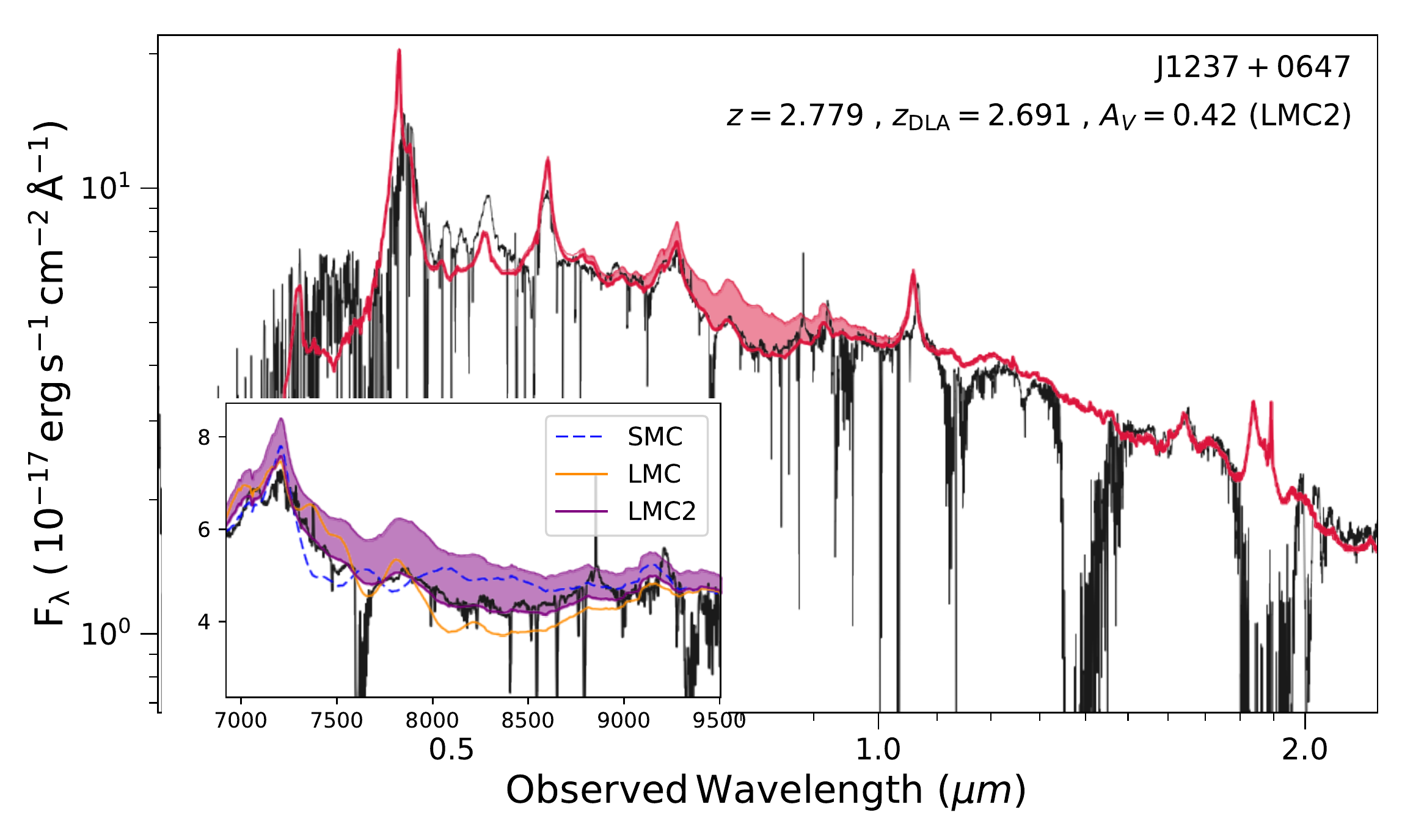}
  \caption{}\label{J1237ebv}
\endminipage\hfill
\minipage{0.32\textwidth}
  \includegraphics[width=\linewidth]{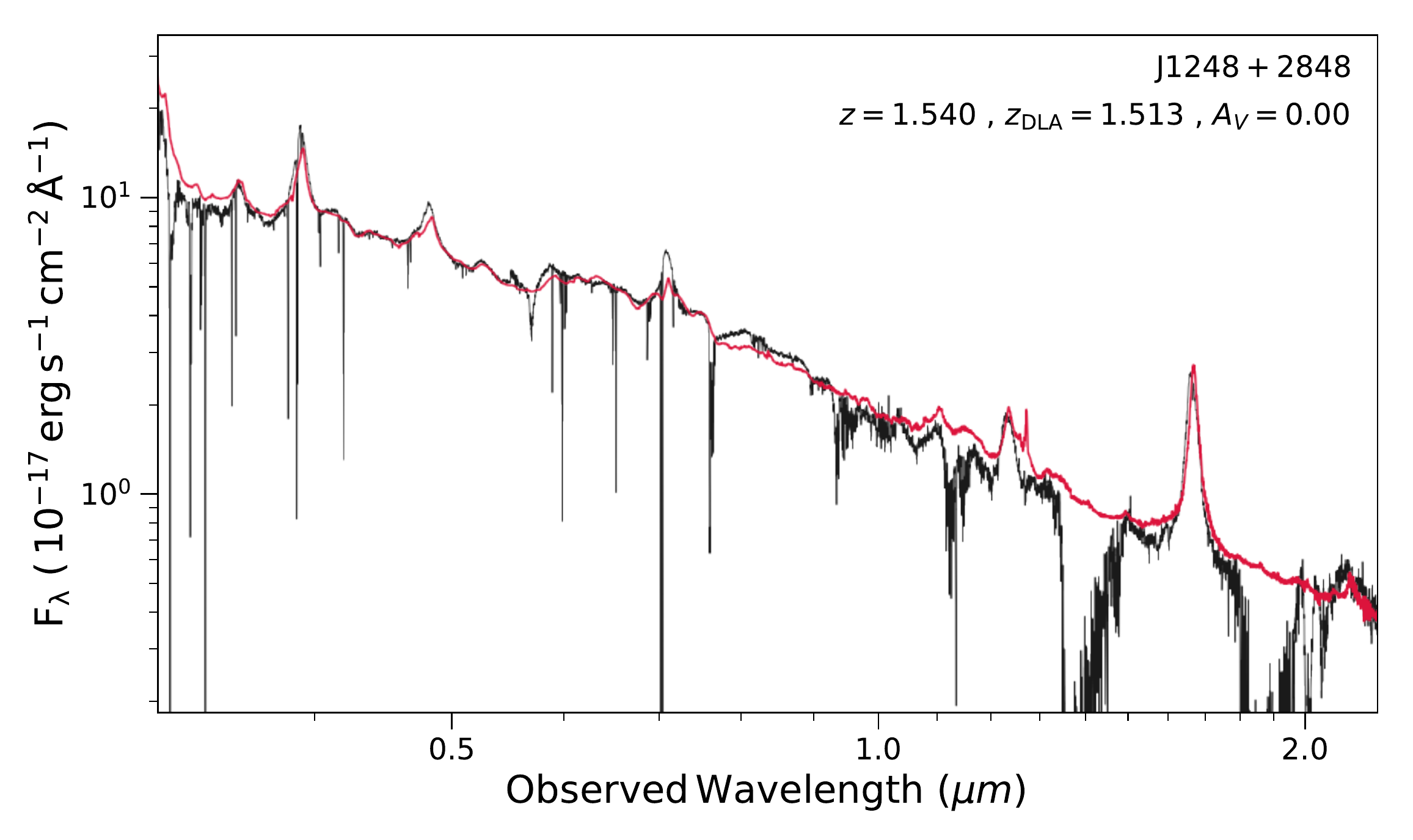}
  \caption{}\label{J1248ebv}
\endminipage\hfill
\minipage{0.32\textwidth}%
  \includegraphics[width=\linewidth]{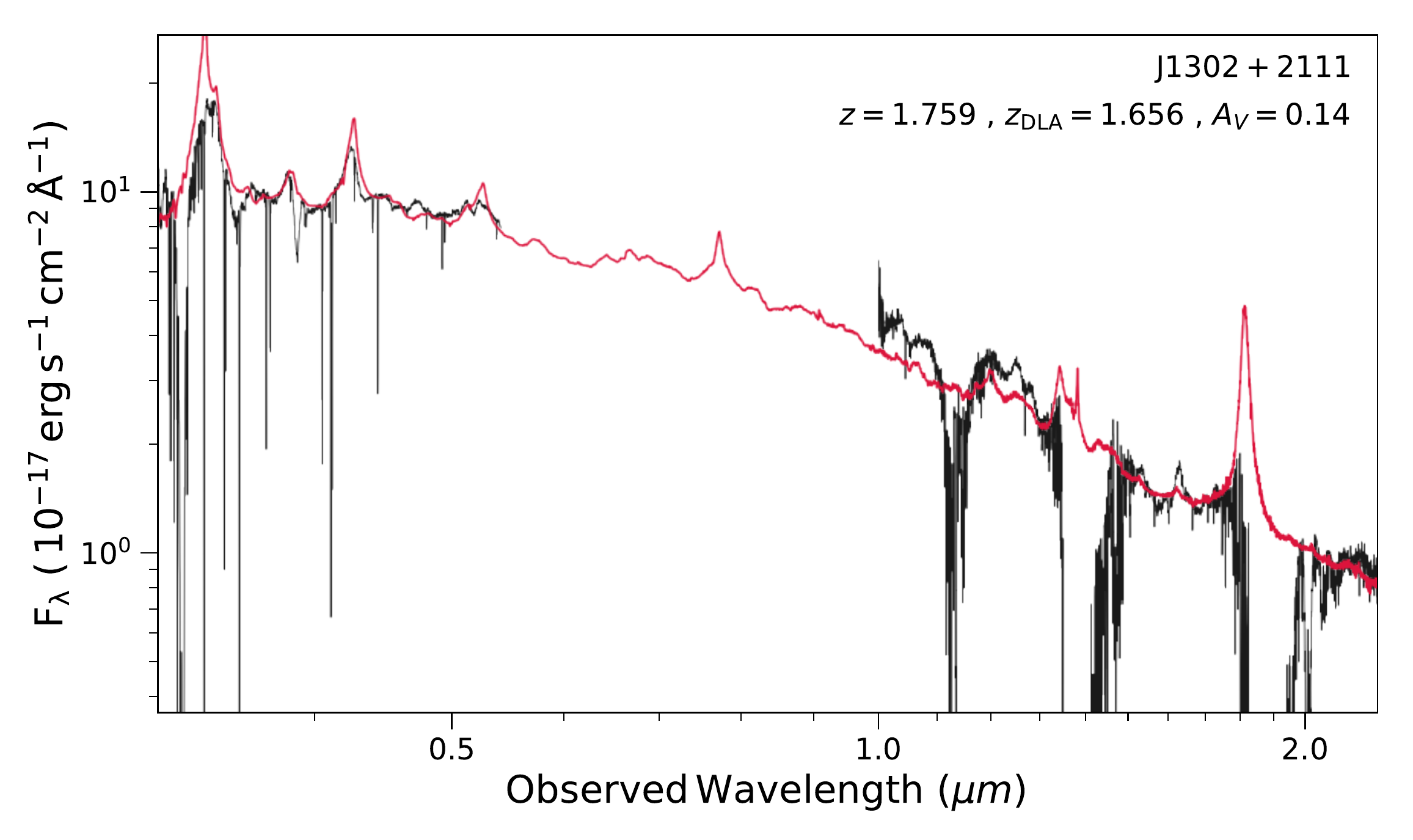}
  \caption{}\label{J1302ebv}
\endminipage
\end{figure*}

\begin{figure*}[!htb]
\minipage{0.32\textwidth}
  \includegraphics[width=\linewidth]{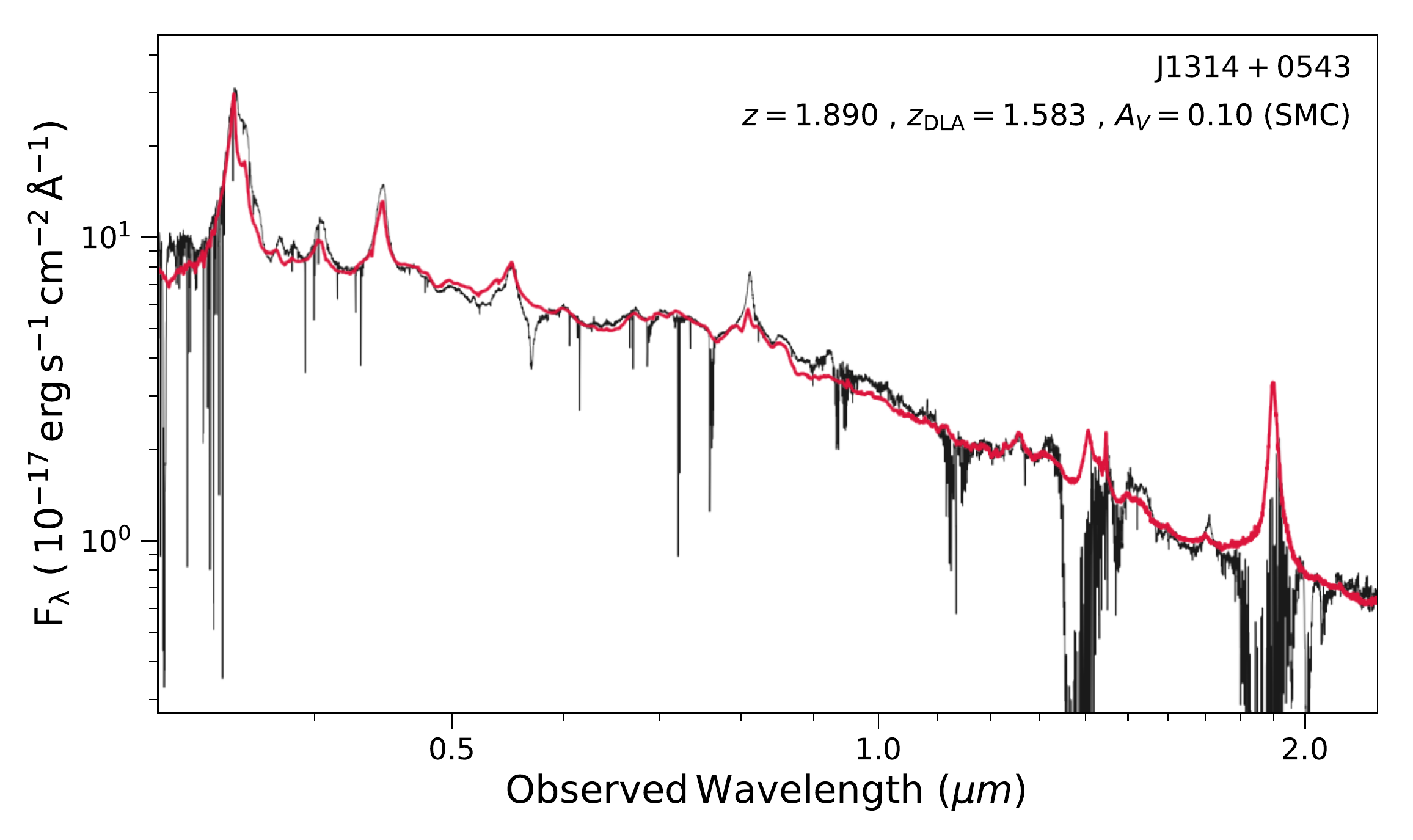}
  \caption{}\label{J1314ebv}
\endminipage\hfill
\minipage{0.32\textwidth}
  \includegraphics[width=\linewidth]{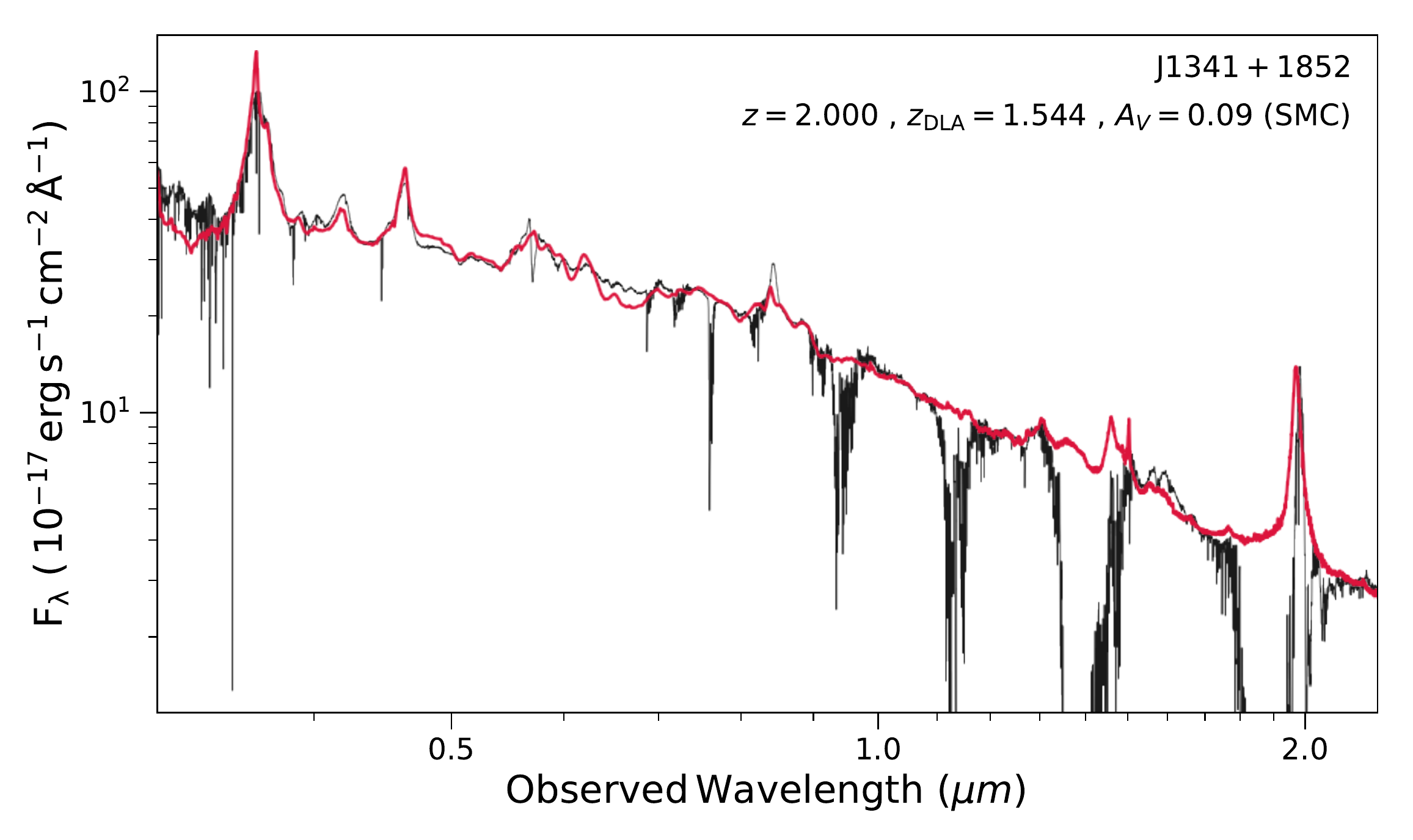}
  \caption{}\label{J1341ebv}
\endminipage\hfill
\minipage{0.32\textwidth}%
  \includegraphics[width=\linewidth]{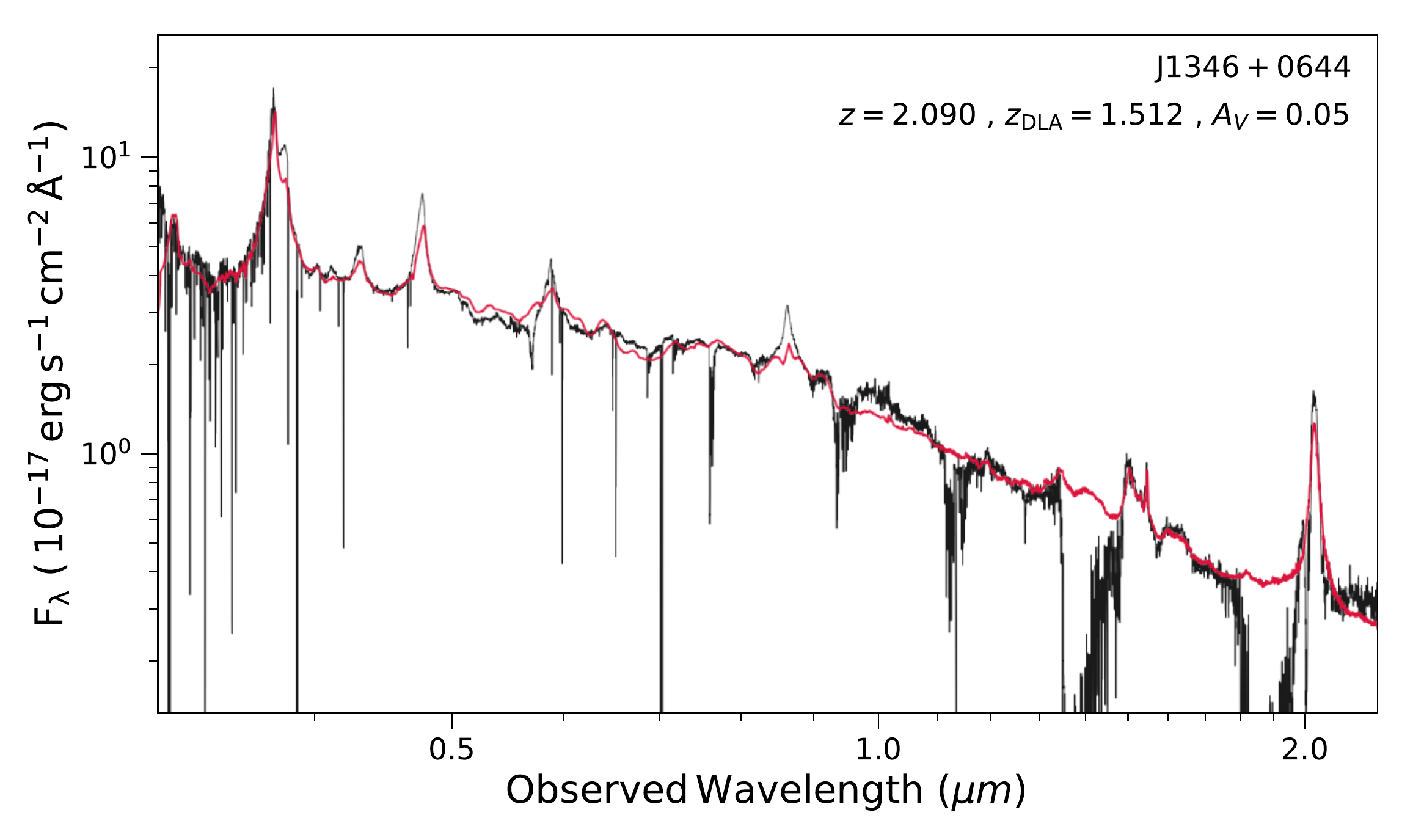}
  \caption{}\label{J1346ebv}
\endminipage
\end{figure*}

\begin{figure*}[!htb]
\minipage{0.32\textwidth}
  \includegraphics[width=\linewidth]{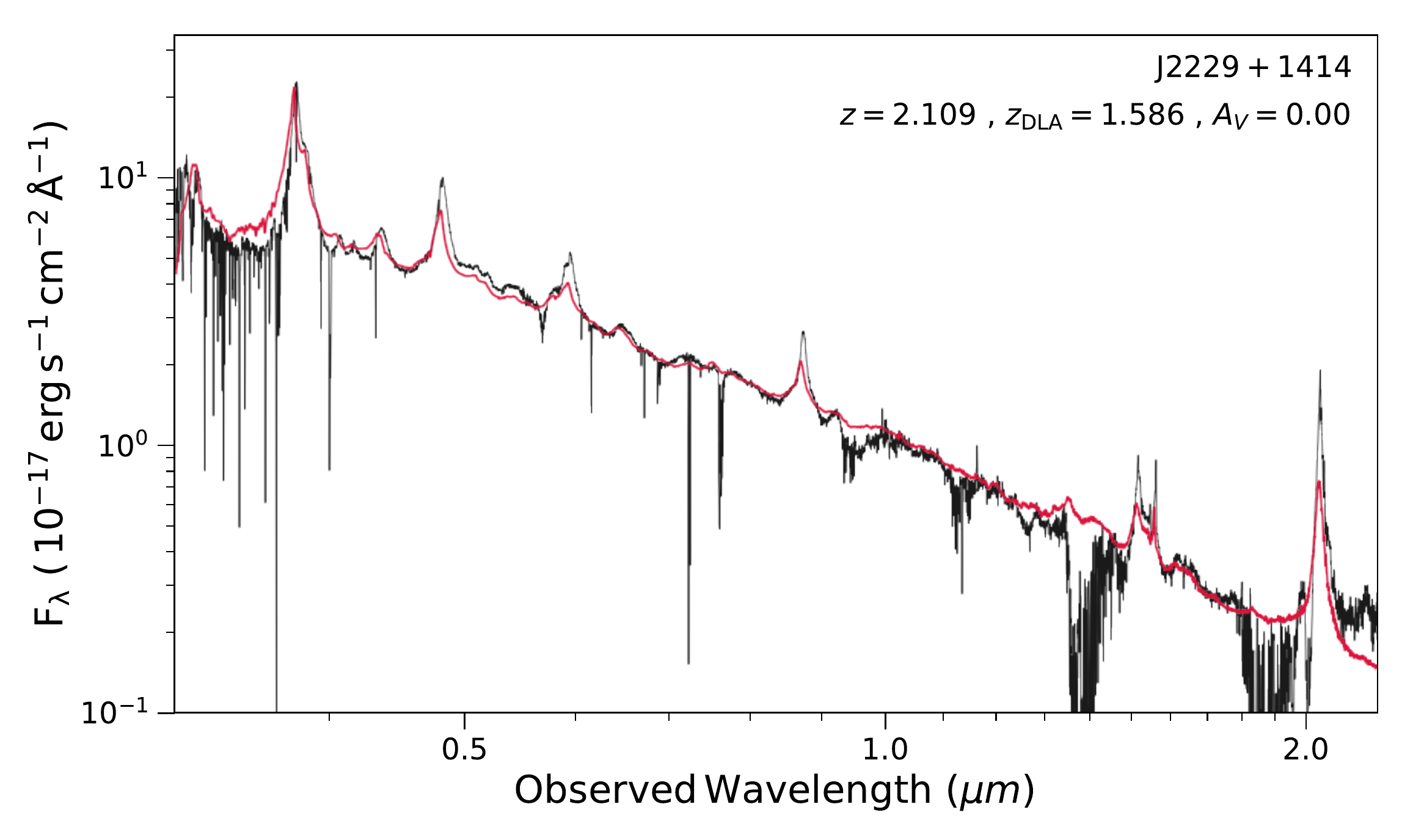}
  \caption{}\label{J2229ebv}
\endminipage\hfill
\minipage{0.32\textwidth}
  \includegraphics[width=\linewidth]{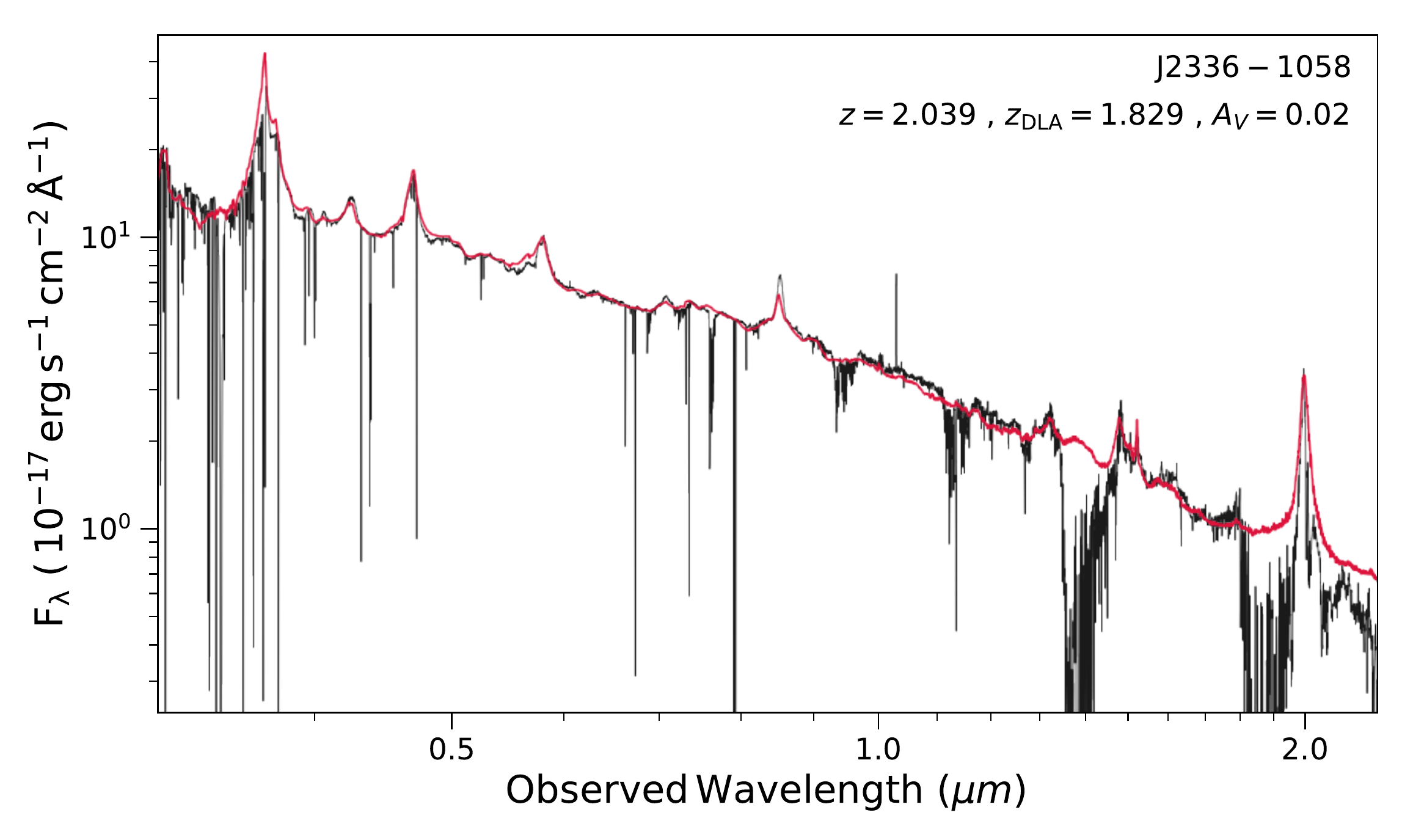}
  \caption{}\label{J2336ebv}
\endminipage\hfill
\minipage{0.32\textwidth}%
  \includegraphics[width=\linewidth]{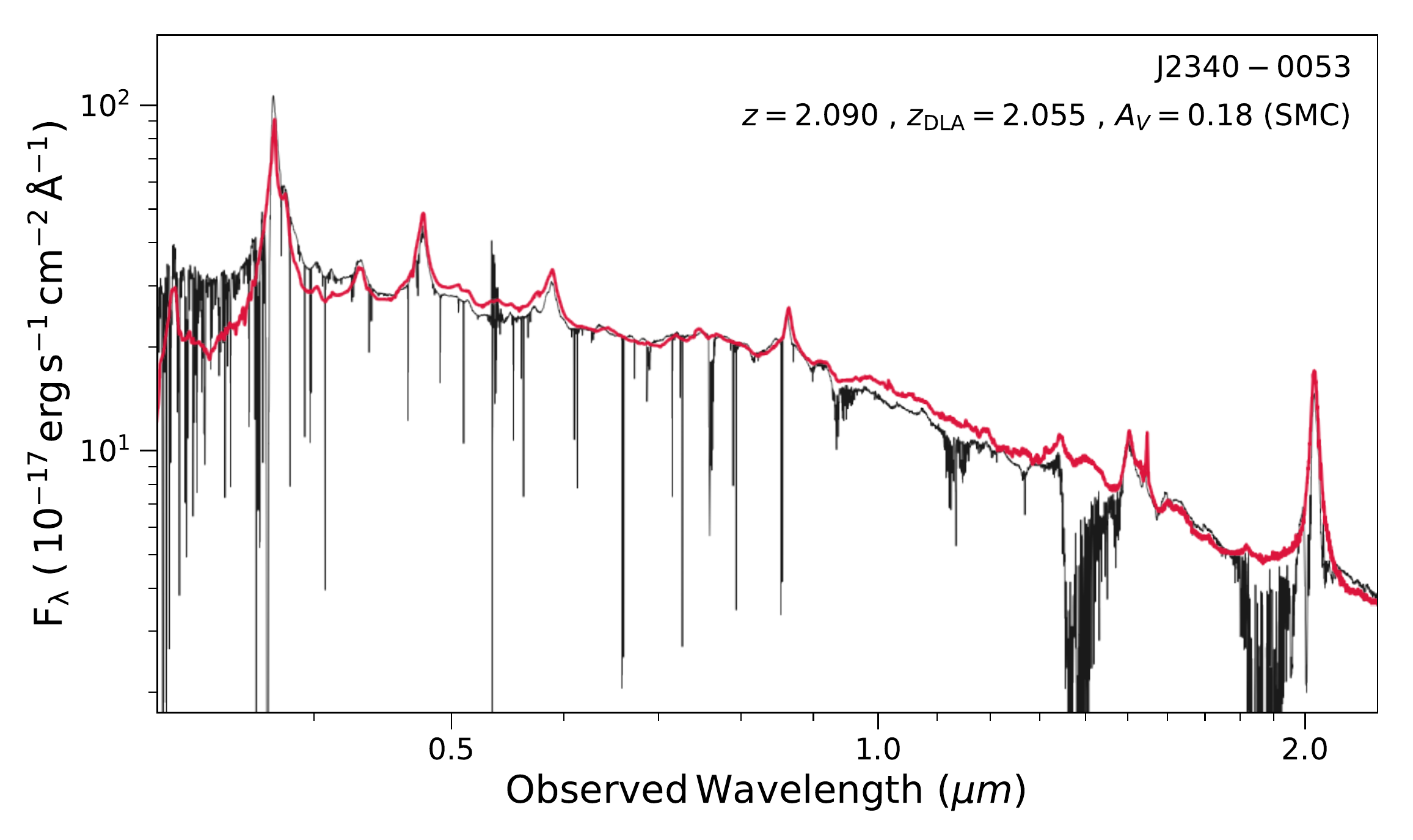}
  \caption{}\label{J2340ebv}
\endminipage
\end{figure*}

\begin{figure*}[!htb]
\minipage{0.32\textwidth}
  \includegraphics[width=\linewidth]{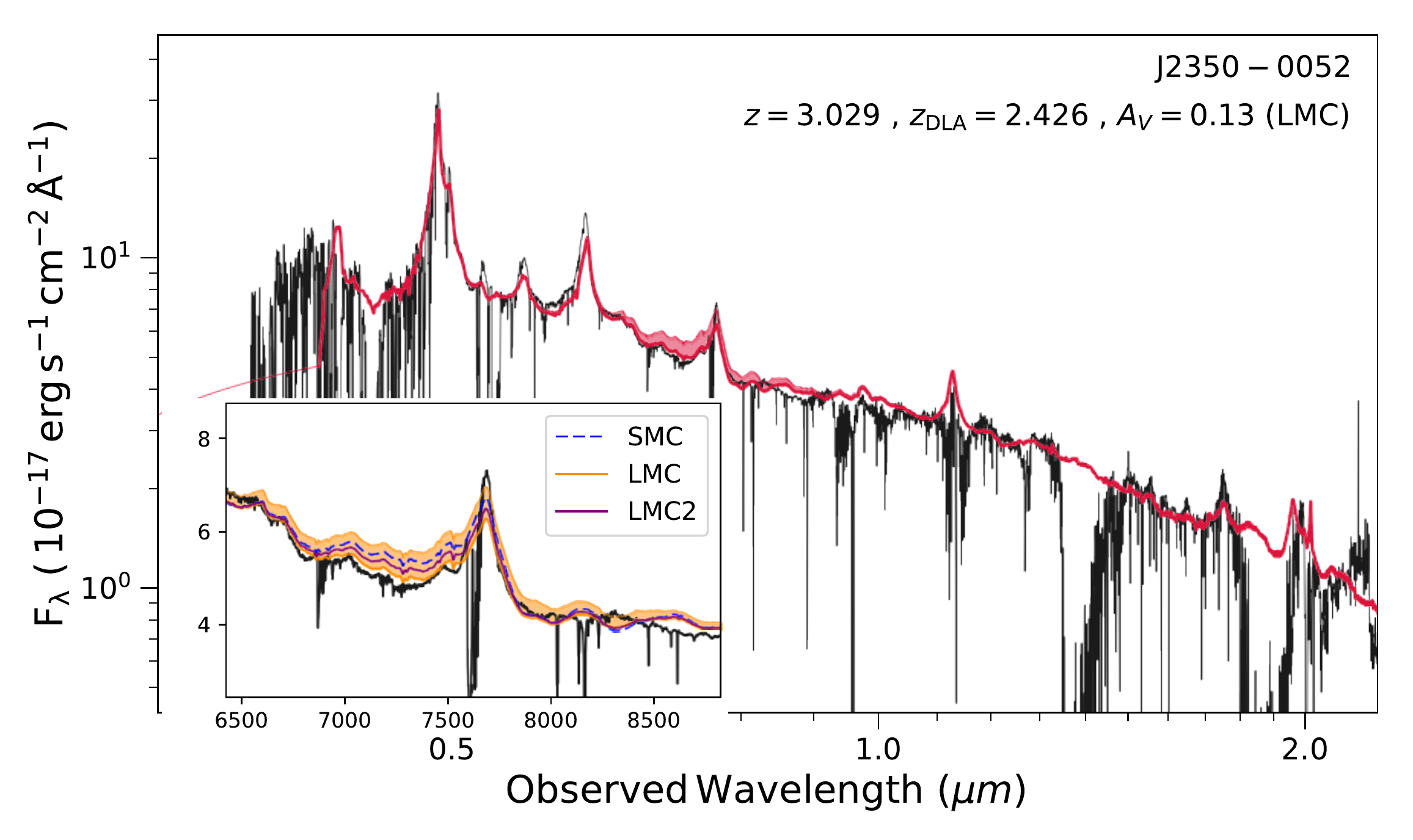}
  \caption{}\label{J2350ebv}
\endminipage\hfill
\end{figure*}

\end{appendix}

\end{document}